\documentclass[
 reprint,
superscriptaddress,
nofootinbib,
 amsmath,amssymb,
 aps,
]{revtex4-2}

\usepackage{graphicx}
\usepackage{dcolumn}
\usepackage{bm}
\usepackage{hyperref}  
\usepackage{caption}
\usepackage{subcaption}
\usepackage{gensymb}
\usepackage{booktabs}
\usepackage{xcolor}

\hypersetup{
colorlinks=true,
linkcolor=blue,
citecolor=blue
}

\newcommand{\herx}[1][ ]{Her X-1#1}
\newcommand{\cw}[1][ ]{continuous gravitational wave#1}
\newcommand{\jstat}[1][ ]{$\mathcal{J}$-statistic#1}
\newcommand{\fstat}[1][ ]{$\mathcal{F}$-statistic#1}
\newcommand{\mufreq}[1][ ]{\mu_{\rm mid}}
\newcommand{\mutemplate}[1][ ]{\mu}
\newcommand{\hUpper}{h_0^{\rm 95\%}}

\begin{document}

\preprint{APS/123-QED}

\title{Pilot search for continuous gravitational waves from Hercules X-1 in O3 LIGO data: enhancements to a hidden Markov model}
\author{Nicholas Kah Yean Low}
\email[]{nlow1@student.unimelb.edu.au}
\author{Andrew Melatos}
\affiliation{School of Physics, University of Melbourne, Parkville, VIC 3010, Australia}
\affiliation{Australian Research Council Centre of Excellence for Gravitational Wave Discovery (OzGrav), University of Melbourne, Parkville, VIC 3010, Australia}

\date{\today}

\begin{abstract}
    A semi-coherent search is conducted for \cw[s]  from stellar oscillations in Hercules X-1 with data from Observing Run 3 of the Laser Interferometer Gravitational-wave Observatory, using an established hidden Markov model algorithm to track stochastic wandering of the signal frequency.
    The primary goal is to implement and test several algorithmic enhancements relevant to future searches, including a second-order projected semi-major axis correction, an orbital eccentricity correction, an adjustment to the width of a search subband, and propagation of the electromagnetically measured orbital parameters with a nonzero orbital period derivative. 
    The enhancements are validated by injecting synthetic signals into Gaussian noise and improve the sensitivity depth by $\approx 14\%$ for a Hercules X-1-like target. 
    In the 60--$200\, {\rm Hz}$ band, the pilot search yields four candidates that survive a five-level veto procedure, whose terminating frequencies are $76.41\, {\rm Hz}$,  $109.47\, {\rm Hz}$, $144.40\, {\rm Hz}$, and $189.24\, {\rm Hz}$. 
    The frequentist upper limit on the maximum detectable wave strain at 95\% confidence is arbitrarily set at 174 Hz and found to be $9.3 \times 10^{-26}$, assuming a uniform prior on polarization.
\end{abstract}

\maketitle

\section{Introduction}

Rotating neutron stars with nonzero mass or current quadrupole moments are sources of \cw[s]. For example, mass quadrupole radiation from a corotating mountain is emitted at $f_*$ and $2f_*$, where $f_*$ equals the rotation frequency, whereas current quadrupole radiation from $r$-modes is emitted at $\sim 4f_*/3$ \cite{caride2019, wette2023, Gittins_2024}. To date, continuous wave searches have targeted neutron stars with $f_* \gtrsim 10 \, {\rm Hz}$ for two reasons: (i) terrestrial detectors such as the Laser Interferometer Gravitational-wave Observatory (LIGO) \cite{ligo}, Virgo \cite{virgo}, and the Kamioka Gravitational Wave Detector (KAGRA) \cite{kagra} lose sensitivity at $\lesssim 10 \, {\rm Hz}$ due to Newtonian and seismic noise \cite{o3LIGOsensitivity,O3LIGOsensitivty2}; and (ii) the characteristic gravitational wave strain obeys $h_0 \propto f_*^2$ for static mountains and $h_0 \propto f_*^3$ for $r$-modes \cite{Riles2023}.


Stellar oscillation modes that emit \cw[s] at frequencies unrelated (or weakly related) to $f_*$ offer an alternative to rotation-driven emission mechanisms. Such modes include $f$-, $p$-, and $g$-modes \cite{McDermott1988,andersson1998, Benhar2004, Gaertig2009, sidery2010, Andersson2021}. Their frequencies are not rational multiples of $f_*$, although they may depend weakly on $f_*$ through rotational splitting \cite{Stergioulas2004splitting, aerts2010asteroseismology, kruger2021splitting}. They may be excited continuously by internal processes, such as starquakes \cite{Baym1971starquake, keer2014starquake, Giliberti2022starquake}, or by external processes, such as clumpy accretion \cite{nagar2004accretion, Romanova2015, dong2024, dong2025}. To the authors' knowledge, no LIGO, Virgo, and KAGRA searches have been published for $f$-, $p$-, and $g$-modes from accreting neutron stars, although many searches have been published that target (i) $f$-, $p$-, and $g$-modes from nonaccreting objects, e.g.\ in the aftermath of a glitch \cite{firstVelaCW, crabVelaCW, knownPulsars2}, and (ii) $r$-modes or static mountains from accreting objects (e.g. \cite{o3_amxp, knownPulsars, knownPulsars2}; see reviews in Refs.\ \cite{wette2023, Riles2023}).

In this paper, we search for \cw[s] from Hercules X-1 (\herx[]), using a hidden Markov model (HMM) algorithm to track stochastic wandering of the signal frequency \cite{HMM, jstat}. \herx is a persistently accreting X-ray pulsar discovered in 1972 \cite{tananbaum1972}, in a binary system with a $2.2 M_\odot$ companion, located $\sim 6 \, {\rm kpc}$ from Earth \cite{herx1CompanionMassAndDistance}. It spins at a frequency of $0.808 \, {\rm Hz}$ \cite{Staubert_2009}, so we do not expect terrestrial detectors to detect \cw emission from mountains and \textit{r}-modes. Instead, we search for signals from other stellar oscillation modes. \herx accretes persistently at $\sim 10\%$ of the Eddington limit \cite{herx1Edd1, herx1Edd2}, so it is plausible energetically that accretion-powered oscillation modes are excited \cite{dong2024}. In addition, \herx presents an opportunity to search regions of the parameter domain that have not been explored in previous HMM \cw searches for binary targets. Specifically, \herx has a relatively large projected semi-major axis $a_0=13 \,$lt-s and a nonzero eccentricity $e=4.2 \times 10^{-4}$ \cite{Staubert_2009}. In comparison, previous HMM \cw searches consider $a_0 \leq 3 \, $lt-s and assume circular orbits \cite{o3_amxp, scox1, psrJ0437andres}\footnote{This parameter domain has been explored in other searches that do not use a HMM, such as in Refs.\ \cite{Ashok_2021, Mirasola2024}, which assume $a_0\lesssim 10^2\, $lt-s and $e\lesssim0.1$.}. \herx['s] larger $a_0$ and nonzero eccentricity imply, that the standard phase model of the signal assumed in Ref.\ \cite{jstat} must be modified.

The paper is structured as follows. In Section \ref{sec:hmm}, we introduce the HMM algorithm and the frequency domain matched filter, called the \jstat[,] used in the search. In Section \ref{sec:search_params}, we define and justify the parameter domain. In Sections \ref{sec:a0^2_correction} and \ref{sec:ecc_correction}, we describe respectively the new $a_0^2$ and $a_0 e$ corrections to the standard phase model. In Section \ref{sec:search_implementation}, we discuss the implementation of the search with respect to frequency binning, orbital templates, detection thresholds, and veto procedures. In Section \ref{sec:results}, we conduct a pilot search by analyzing data from LIGO Observing Run 3 (O3). The pilot search is of interest in its own right, covering the astrophysically relevant band $0.06 \leq f_{\rm GW} / (1 \, {\rm kHz}) \leq 0.2$ for the first time in this target. Importantly, however, it also serves as a  validation and training exercise for more comprehensive generic searches targeting X-ray pulsars in the future.
In Section \ref{sec:upper_limits}, we investigate the improvement in the sensitivity depth $\mathcal{D}$ offered by the new $a_0^2$ and $a_0 e$ corrections.
The astrophysical implications of the results are canvassed briefly in Section \ref{sec:conclusion}. Interestingly, the corrections to the standard phase model discussed in Sections \ref{sec:a0^2_correction}--\ref{sec:search_implementation} increase the computational cost substantially, limiting the maximum frequency of the pilot search to $f_{\rm max} = 0.2 \, {\rm kHz}$ with the computational resources at our disposal, cf.\ $f_{\rm max} \approx 0.5 \, {\rm kHz}$ for previous searches \cite{scox1, psrJ0437andres}.

We emphasize that the main goal of this paper is to validate the methodological enhancements discussed in Sections \ref{sec:a0^2_correction}--\ref{sec:search_implementation} in real data, as a preparation for future searches for continuous gravitational waves from binaries, irrespective of the specific emission mechanism.
Gravitational radiation from stellar oscillations may not be detectable by the current generation of audio-band interferometers, if the oscillation frequencies are too high or the damping timescales are too short.
We refer the reader to Appendix \ref{appendix:stellar_oscillations} for a brief discussion of detectability under various astrophysical scenarios.

\section{HMM algorithm}
\label{sec:hmm}

The \cw search algorithm used in this paper consists of two parts. Section \ref{subsec:hmm_formalism} describes the HMM, which tracks the most likely evolution of the signal frequency during a sequence of time steps. Section \ref{subsec:jstat} describes the frequency domain matched filter, called the \jstat[,] which computes the probability that the signal occupies a particular frequency bin given the data collected during a single time step. 

\subsection{Hidden state formalism}
\label{subsec:hmm_formalism}

A HMM is composed of a discrete hidden state variable $q(t) \in \{q_1, \,\ldots,\,q_{N_Q}\}$ and a discrete observable $o(t) \in \{o_1, \,\ldots,\,o_{N_O}\}$ sampled at discrete time steps $t \in \{t_1, \,\ldots,\,t_{N_T}\}$. For a \cw search, $q(t)$ corresponds to the frequency of the gravitational wave, $f_{\rm GW}(t)$. At each time step, the hidden state $q(t)$ has a probability of jumping from state $q_i$ to state $q_j$, given by the transition matrix $A_{q_j q_i}$. As with previous searches \cite{o3_amxp, scox1, psrJ0437andres}, $A_{q_j q_i}$ is approximated as

\begin{equation}
    \label{eq:transition}
    A_{q_j q_i} = \frac{1}{3} \left( \delta_{q_j q_{i-1}} + \delta_{q_j q_{i}} + \delta_{q_j q_{i+1}} \right),
\end{equation}

\noindent where $\delta_{q_j q_{i}}$ is the Kronecker delta. Eq. (\ref{eq:transition}) corresponds to $f_{\rm GW}(t)$ jumping $-1, 0$ or $ +1$ frequency bins, with equal probability, at each time step.

The observable $o(t)$ is computed from the Fourier transform of the gravitational wave strain measured by the detectors. The dataset, which has duration $T_{\rm obs}$, is divided into $N_T$ coherent chunks of equal duration $T_{\rm drift}$, with $N_T={\rm floor}(T_{\rm obs} / T_{\rm drift})$. The choice of $T_{\rm drift}$, which is justified in Section \ref{subsec:t_drift}, sets the timescale during which $f_{\rm GW}(t)$ wanders by at most one frequency bin. The likelihood of observing state $o_j$, given that the hidden state is $q_i$, equals the emission probability $L_{o_j q_i}$. The \jstat[,] a frequency domain matched filter discussed in Section \ref{subsec:jstat}, is applied to compute $L_{o_j q_i}$ \cite{jstat}.

The probability that the path of hidden states is $Q=\{q(t_0),\,  \ldots , \, q(t_{N_T}) \}$, given the observations $O=\{o(t_0),\,  \ldots , \, o(t_{N_T}) \}$, is 

\begin{align}
    \label{eq:path_likehood}
    {\rm Pr}(Q|O) = &L_{o(t_{N_T})q(t_{N_T})} A_{q(t_{N_T})q(t_{N_T-1})} \, \times \cdots \notag \\
    & \times \, L_{o(t_{1})q(t_{1})} A_{q(t_{1})q(t_{0})} \Pi_{q(t_0)}.
\end{align}

\noindent In Eq.\ (\ref{eq:path_likehood}), $\Pi_{q(t_0)}$ is the prior probability that the starting hidden state is $q(t_0)$. It is assumed to be uniform, i.e.\ $\Pi_{q(t_0)}=1/N_Q$. The Viterbi algorithm \cite{viterbi} searches efficiently for the path $Q^*$ that maximizes Eq.\ (\ref{eq:path_likehood}). The detection statistic adopted in this paper is the log likelihood of the most likely path, i.e.\ $\mathcal{L}= \ln \, {\rm Pr}(Q^*|O)$.

\subsection{$\mathcal{J}$-statistic}
\label{subsec:jstat}

The \jstat \cite{jstat} is a frequency domain matched filter that extends the maximum-likelihood \fstat \cite{fstat} by accounting for Doppler modulation of a binary source through a Jacobi-Anger expansion of the binary orbit. 
The \jstat is computed by first computing the \fstat assuming an isolated target. The Doppler modulation disperses the \fstat power into multiple frequency bins.
The Jacobi-Anger expansion then recombines the dispersed power into one bin.
The \jstat is used as the frequency domain estimator $L_{o_j q_i}$ in this paper, as in previous searches \cite{o3_amxp, scox1, psrJ0437andres}. 

Traditionally, the \jstat assumes that the binary orbit is circular and is described fully by three binary orbital parameters: the projected semi-major axis $a_0$, the orbital period $P$, and the time of ascension $T_{\rm asc}$. The leading-order ($\propto a_0$) phase modulation due to a circular binary is \cite{leaciPrix2015, jstat}

\begin{equation}
    \label{eq:phase_first_order}
    \Phi_{a_0}(t) = -\frac{2\pi f_{\rm GW} a_0}{c} \sin \left [\Omega(t- T_{\rm asc}) \right],
\end{equation}

\noindent where $\Omega=2\pi/P$ is the orbital angular frequency, $f_{\rm GW}$ is the frequency of the gravitational wave, and $c$ is the speed of light. Relaxing the assumption that the binary is circular introduces two extra orbital parameters: the orbital eccentricity $e$ and the argument of periapsis $\omega$. The eccentricity correction is discussed in detail in Section \ref{sec:ecc_correction}. Furthermore, \herx['s] relatively large semi-major axis ($a_0=13 \, $lt-s)  introduces a second-order $(\propto a_0^2)$ correction in Eq.\ (\ref{eq:phase_first_order}), which is discussed in Section \ref{sec:a0^2_correction}.

\section{Her x-1 parameters}
\label{sec:search_params}

\begin{table*}[htbp!]
\caption{\herx search parameters, quoting central values and 1-$\sigma$ uncertainties  in parentheses (referring to the trailing digits). The range of $f_*$ refers to the minimum and maximum measured values during O3. $P_{\rm ref}$, $T_{\rm asc ,\, ref}$ and $\dot{P}$ are measured at the reference time $T_{\rm ref}=182811284.4336$ GPS time, while $P$ and $T_{\rm asc}$ are propagated to the middle of O3, viz.\ $T_{\rm O3, \, mid}=1253718483$ GPS time.}
\label{tab:params}
\begin{tabular}{lcccc}
\hline
Parameter & Symbol & Parameter value & Units & Ref. \\ \hline
Right ascension (J2000) & $\alpha$ & 16h57m49.81s & ${\rm h m s}$ & \cite{herx1_simbad} \\
Declination (J2000) & $\delta$ & +35$\degree$ 20$'$ 32.49$''$ & $\degree\, \, ' \, \, ''$ & \cite{herx1_simbad} \\
Orbital inclination angle & $\iota$ & 80--90 & Degrees & \cite{Leahy_2020_inclination} \\
Projected semi-major axis & $a_0$ & 13.1831(4) & lt-s & \cite{Staubert_2009} \\
Orbital period at $T_{\rm ref}$ & $P_{\rm ref}$ & 146894.4797760(1728) & s & \cite{Staubert_2009} \\
Orbital period propagated to $T_{\rm O3, \, mid}$ & $P$ & 146894.427839(1565) & s & --- \\
Orbital period derivative (constant) & $\Dot{P}$ & $-4.85(13) \times 10^{-11}$ & s s$^{-1}$ & \cite{Staubert_2009} \\
GPS time of ascension at $T_{\rm ref}$ & $T_{\rm asc, \, ref}$ & 182774560.8137(5184) & s & \cite{Staubert_2009} \\ 
GPS time of ascension propagated to $T_{\rm O3, \, mid}$ & $T_{\rm asc}$ & 1253782023.323(6991) & s & --- \\ 
Orbital eccentricity & $e$ & $4.2(8) \times 10^{-4}$ & Unitless & \cite{Staubert_2009} \\
Argument of periapsis & $\omega$ & 96(10) & Degrees & \cite{Staubert_2009}\\
Spin frequency & $f_*$ & 0.807936--0.807943 & Hz & \cite{GBMpulsars, herx1GBM}\\
\hline
\end{tabular}
\end{table*}

The \jstat for an eccentric binary orbit is a function of the right ascension $\alpha$, declination $\delta$, projected semi-major axis $a_0$, orbital period $P$, time of ascension $T_{\rm asc}$, orbital eccentricity $e$ and argument of periapsis $\omega$. Table \ref{tab:params} lists the parameters of \herx measured electromagnetically at $T_{\rm ref}= 182811284.4336\,$ GPS time (21 October 1985), together with their 1-$\sigma$ uncertainties. 
The binary parameters are fitted using X-ray timing observations from 1971 to 2007 \cite{Staubert_2009}.
The orbital reference epoch used in Ref.\ \cite{Staubert_2009} is the epoch of $90^\circ$ mean orbital longitude, $T_{\pi/2}$.%
\footnote{The epoch of superior conjunction of the pulsar (or equivalently the epoch of inferior conjunction of the companion) occurs when the pulsar is behind the companion, which for eccentric orbits, does not necessarily equal $T_{\pi/2}$ \cite{vanderKlis1984, Nieder_2020}.}
The phase model, as described by Eqs.\ (\ref{eq:phase_first_order}), (\ref{eq:phase_second_order}) and (\ref{eq:phase_ecc}), uses the time of ascension $T_{\rm asc}$ instead, as defined in the ELL1 model \cite{Lange2001, leaciPrix2015}.
$T_{\rm asc}$ corresponds to $0^\circ$ mean orbital longitude, and is related to $T_{\pi/2}$ via $T_{\rm asc}= T_{\pi/2} - P/4$ \cite{vanderKlis1984, Lange2001}.
Let us denote the start of O3 as $T_{\rm O3, \,start}=1238166483$ GPS time. 
The quantities $T_{\rm asc, \, ref}$ and $P_{\rm ref}$, which are measured at $T_{\rm ref}$, are propagated to the middle of O3, $T_{\rm O3, \, mid}=T_{\rm O3, \,start} + N_T T_{\rm drift}/2=1253718483$ GPS time, via \cite{tempo2}

\begin{align}
    P &= P_{\rm ref} + \dot{P}\Delta T \label{eq:p_propagated}, \\
    T_{\rm asc} &= T_{\rm asc, \, ref} + \Delta T, \label{eq:t_propagated} \\
    \Delta T &= \frac{P_{\rm ref}}{\dot{P}} \left(1 - \sqrt{1- 2 \dot{P} N_{\rm orb}}\right) \label{eq:delta_t_o3} \\
    N_{\rm orb} &= {\rm round} \left[ \frac{T_{\rm O3, \, mid} - T_{\rm ref}}{P_{\rm ref}} \right. \notag \\ 
    & \quad \left. - \frac{\dot{P}}{2} \left(\frac{T_{\rm O3, \, mid} - T_{\rm ref}}{P_{\rm ref}} \right)^2 \right]. \label{eq:num_orbits}
\end{align}

\noindent In Eqs.\ (\ref{eq:p_propagated}) and (\ref{eq:t_propagated}), $P$ and $T_{\rm asc}$ correspond to the epoch $T_{\rm O3, \, mid}$. In Eqs.\ (\ref{eq:p_propagated}), (\ref{eq:delta_t_o3}) and (\ref{eq:num_orbits}), $\dot{P}$ is measured at $T_{\rm ref}$ and is assumed to stay constant from the reference time to the end of O3 in the phase model. We choose to propagate $T_{\rm asc}$ and $P$ to $T_{\rm O3, \, mid}$ to simplify template placement, which we discuss in detail in Section \ref{subsec:num_templates} and Appendix \ref{appendix:p_t_prop}. 

\begin{figure}[htbp!]
    \centering
    \includegraphics[width=0.95\linewidth]{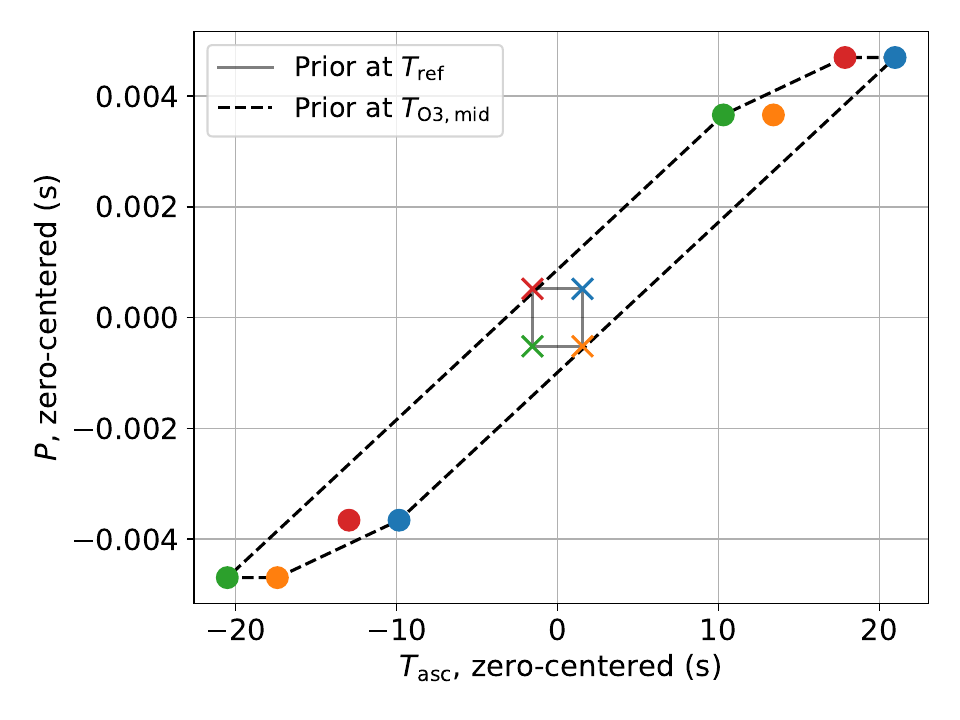}
    \caption{Bounds on the two-dimensional electromagnetic prior of $T_{\rm asc,\, ref}$ and $P_{\rm ref}$ (solid gray curve), measured at $T_{\rm ref}$, as well as $T_{\rm asc}$ and $P$ (dashed black curve), propagated to $T_{\rm O3, \, mid}$. The four corners of the solid gray rectangle are marked with crosses of different colors. Each cross maps to two dots of the same color after propagating to $T_{\rm O3, \, mid}$ assuming $\dot{P} = \bar{\dot{P}}\pm 3\sigma_{\dot{P}}$.}
    \label{fig:uncertainty_prior}
\end{figure}

Fig.\ \ref{fig:uncertainty_prior} illustrates the relative sizes and shapes of the priors on $T_{\rm asc, \,ref}$ and $P_{\rm ref}$, which are measured at $T_{\rm ref}$, and on $T_{\rm asc}$ and $P$, which are measured at $T_{\rm O3, \, mid}$. The priors on $T_{\rm asc,\, ref}$ and $P_{\rm ref}$ are defined as the rectangular region with corners $(\bar{T}_{\rm asc,\, ref} \, \pm 3\sigma_{T_{\rm asc,\, ref}}, \bar{P}_{\rm ref} \, \pm 3\sigma_{P_{\rm ref}})$, as depicted by the solid gray rectangle in Fig.\ \ref{fig:uncertainty_prior}, where $\bar{X}$ and $\sigma_X$ denote the central value and 1-$\sigma$ uncertainty respectively of $X$ from Table \ref{tab:params}. 

The $T_{\rm ref}$ priors are propagated to $T_{\rm O3, \, mid}$ by propagating the four corners, marked by color-coded crosses, using Eqs.\ (\ref{eq:p_propagated})--(\ref{eq:num_orbits}), first assuming $\dot{P}= \bar{\dot{P}} - 3 \sigma_{\dot{P}}$ and then assuming $\dot{P}=\bar{\dot{P}} + 3 \sigma_{\dot{P}}$, to land at the eight matching color-coded dots in Fig.\ \ref{fig:uncertainty_prior}. We join the outermost six dots to form the vertices of a convex polygon to bound the prior on $T_{\rm asc}$ and $P$, as depicted by the dashed black curve in Fig.\ \ref{fig:uncertainty_prior}. We define $T_{\max}$ and $T_{\min}$ as the maximum and minimum $T_{\rm asc}$ values spanned respectively by the dashed black polygon at $T_{\rm O3, \, mid}$, and likewise for $P_{\max}$ and $P_{\rm min}$. Then the symmetrized 3-$\sigma$ uncertainties at $T_{\rm O3, \, mid}$ are given conservatively by 

\begin{align}
    3\sigma_{T_{\rm asc}} &= \max(\bar{T}_{\rm asc}-T_{\min},\, T_{\max} - \bar{T}_{\rm asc}),  \\
    3\sigma_{P} &= \max(\bar{P}-P_{\min},\, P_{\max}-\bar{P}).
\end{align}

To ensure that the dashed black polygon in Fig.\ \ref{fig:uncertainty_prior} brackets the propagated prior, $20^{3}$ points are sampled regularly from the cube with corners $(\bar{T}_{\rm asc,\, ref} \, \pm 3\sigma_{T_{\rm asc,\, ref}}, \bar{P}_{\rm ref} \, \pm 3\sigma_{P_{\rm ref}}, \bar{\dot{P}} \, \pm 3\sigma_{\dot{P}})$ and propagated to $T_{\rm O3, \, mid}$ using Eqs.\ (\ref{eq:p_propagated})--(\ref{eq:num_orbits}). 
Most points fall within the black polygon; the 64 points that do not can be included, if the black curve is magnified by $18\%$ about $(\bar{T}_{\rm asc}, \bar{P})$. 
We opt not to perform the magnification, as the location of the 64 points means that the magnification does not change the number of templates needed.

\section{$a_0^2$ correction to the phase model}
\label{sec:a0^2_correction}

The second-order ($\propto a_0^2$) correction to the signal phase, derived in Appendix \ref{appendix:phase_model}, is given by

\begin{equation}
\label{eq:phase_second_order}
    \Phi_{a_0^2}(t)= \frac{\pi f_{\rm GW} a_0^2 \Omega}{c^2} \sin \left [2\Omega(t- T_{\rm asc}) \right].
\end{equation}

\noindent For \herx[,] which has $a_0/c\sim 10\, {\rm s}$ and $\Omega \sim 10^{-5}\, {\rm s^{-1}}$, the correction at $f_{\rm GW} \sim 10^2\, {\rm Hz}$ has amplitude $f_{\rm GW} \Omega a_0^2/2c^2 \sim 10^{-1}$ cycles.

\begin{figure}[!htbp]
    \centering
    \includegraphics[width=0.95\linewidth]{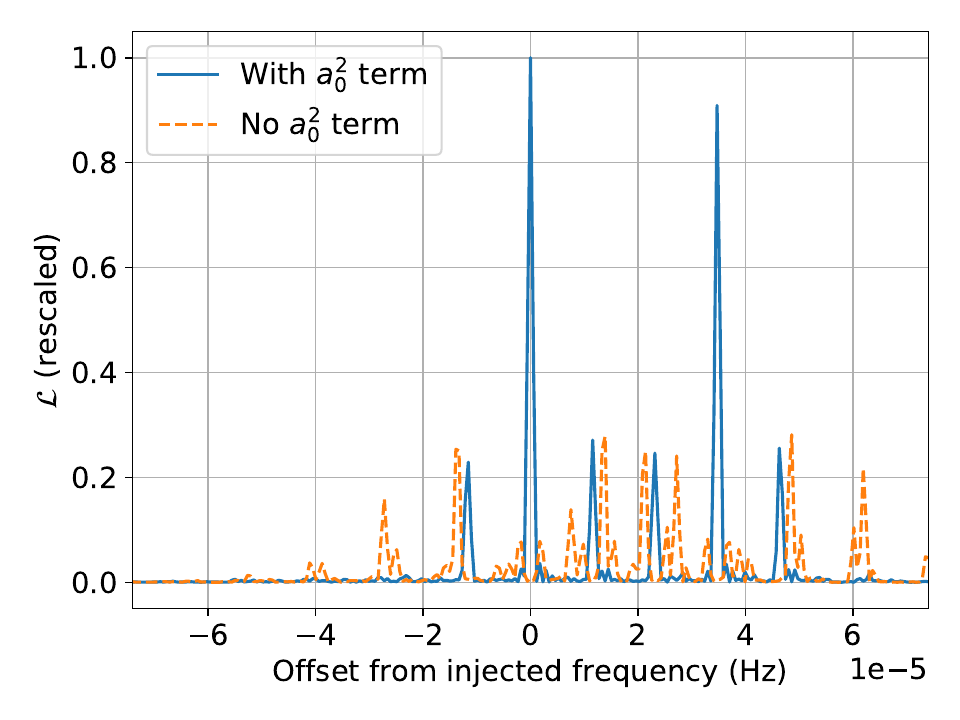}
    \caption{Log likelihood of the most likely path, $\mathcal{L}$, as a function of search frequency offset (units: $10^{-5} \, {\rm Hz}$), when the $a_0^2$ correction in Eq.\ (\ref{eq:phase_second_order}) is included (solid blue curve) and omitted (dashed orange curve). $\mathcal{L}$ has been rescaled to give $\max \mathcal{L}=1$ for the solid blue curve. The injection parameters are $h_0=10^{-24}$, $\sqrt{S_n}=4\times 10^{-24}\, {\rm Hz}^{-1/2}$, $f_{\rm GW}=100\, {\rm Hz}$, $e=0$. All other parameters take the central values in Table \ref{tab:params}.}
    \label{fig:a02_correction}
\end{figure}

To quantify the importance of the second-order correction, a test signal is injected into 10 days of synthetic Gaussian data using the \texttt{lalpulsar\_Makefakedata\_v5} program in the LVK Algorithm Library (LALSuite) \cite{lalsuite}. The signal has gravitational wave strain $h_0=10^{-24}$, one-sided noise amplitude spectral density $\sqrt{S_n}=4\times 10^{-24}\, {\rm Hz}^{-1/2}$, $f_{\rm GW}=100\, {\rm Hz}$ and $e=0$. All other parameters take \herx['s] central values in Table \ref{tab:params}\footnote{At the time of writing, \texttt{Makefakedata} only supports $\dot{P}=0$. A nonzero $\dot{P}$ may be simulated by updating $P$ and $T_{\rm asc}$ manually at each coherent block. As the change in $P$ over O3 is less than the width of a $P$ bin, as discussed in Section \ref{subsec:num_templates}, we opt to inject with constant $P$ for simplicity.}. The HMM algorithm, described in Section \ref{sec:hmm}, is executed with $T_{\rm drift}=10\, {\rm days}$. Fig.\ \ref{fig:a02_correction} displays the results of the test. It plots $\mathcal{L}$ against search frequency for Eq.\ (\ref{eq:phase_first_order}) alone (orange curve) and when Eq.\ (\ref{eq:phase_second_order}) is included (blue curve). Without the $a_0^2$ correction, $\mathcal{L}$ attains its maximum $\approx 4.9 \times 10^{-5}\, {\rm Hz}$ (84 frequency bins) away from the injected frequency, whereas with the $a_0^2$ correction, $\max \mathcal{L}$ falls exactly in the injected frequency bin. Moreover, $\max \mathcal{L}$ is $\approx 3.6$ times higher, when the $a_0^2$ correction is included. In other words, the $a_0^2$ correction leads to appreciable improvements in both accuracy and sensitivity.

Eq.\ (\ref{eq:phase_second_order}) leads to a new factor in the \jstat filter $\propto \sum_{s_2 = -\infty}^\infty J_{s_2}(\pi \bar{f} a_0^2 \Omega / c^2)$, where $J_{s_2}$ is the Bessel function of the first kind of order $s_2$, and $\bar{f}$ is the central frequency of the subband (see Section \ref{subsec:t_drift}). The corresponding Bessel weight for the leading, first-order sum is $J_{s_1}(2\pi \bar{f} a_0/c)$, as in previous HMM implementations \cite{jstat}. The asymptotic properties of $J_{s_1}$ permit truncation of the $s_1$ sum to $2M_1 +1 =2\, {\rm ceil}(2\pi \bar{f} a_0/c) + 1$ terms. Hence a natural choice is to truncate the $s_2$ sum to $2\, {\rm ceil}(\pi \bar{f} a_0^2 \Omega / c^2) + 1$ terms. However, this choice leads sometimes to subtle issues with the placement of orbital templates, described in Appendix \ref{appendix:bessel_sum}, so we opt as a precaution to double the truncation order, viz.\ $2M_2 + 1 = 2\, {\rm ceil}(2\pi \bar{f} a_0^2 \Omega / c^2) + 1$ terms.

\section{$a_0 e$ correction to the phase model}
\label{sec:ecc_correction}

The leading-order ($\propto a_0 e$) eccentricity correction to the signal phase, derived in Appendix \ref{appendix:phase_model}, has amplitude given by \cite{leaciPrix2015}

\begin{equation}
\label{eq:phase_ecc}
    \Phi_{e}(t)=- \frac{\pi f_{\rm GW} a_0 e}{c} \sin \left [2\Omega(t- T_{\rm asc}) - \omega \right],
\end{equation}

\noindent For \herx[,] which has $a_0/c\sim 10\, {\rm s}$ and $e \sim 10^{-4}$, the correction at $f_{\rm GW} \sim 10^2\, {\rm Hz}$ is $f_{\rm GW} a_0 e/2c \sim 10^{-1}$ cycles.

\begin{figure}[!htbp]
    \centering
    \includegraphics[width=0.95\linewidth]{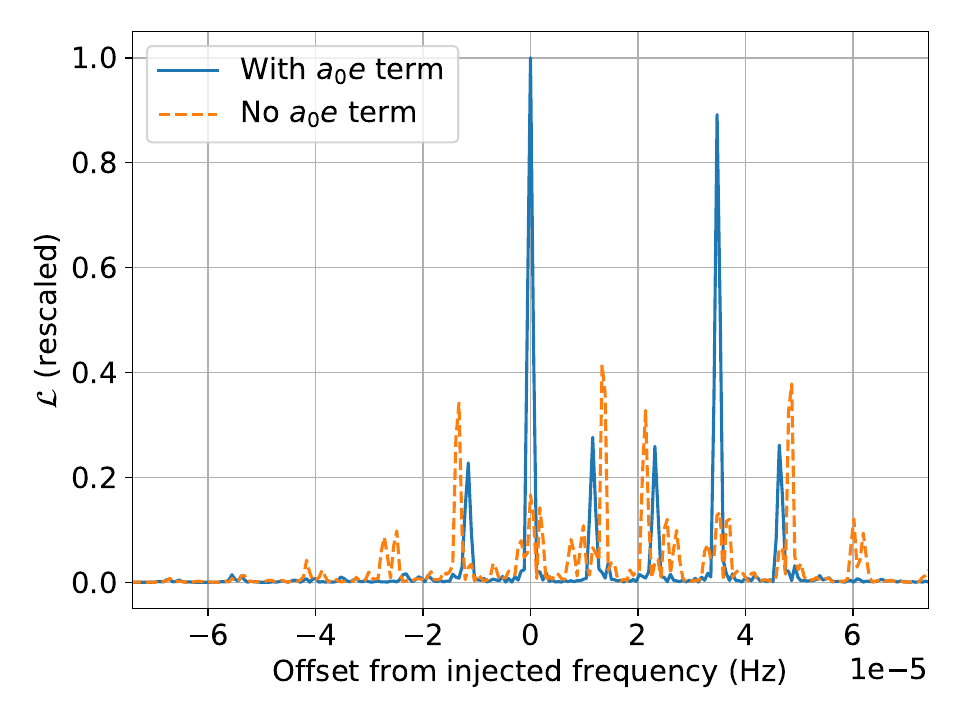}
    \caption{As for Fig.\ \ref{fig:a02_correction}, but with the $a_0 e$ correction in Eq.\ (\ref{eq:phase_ecc}) included (solid blue curve) and omitted (orange dashed curve). The injected test signal has $e=4.2\times 10^{-4}$ instead of $e=0$. The $a_0^2$ correction in Section \ref{sec:a0^2_correction} is included in both the blue and orange curves.}
    \label{fig:ecc_correction}
\end{figure}

We conduct a test like the one in Section \ref{sec:a0^2_correction} to quantify the influence of the first-order eccentricity correction. The injected parameters are identical to Section \ref{sec:a0^2_correction}, except that we set $e=4.2\times 10^{-4}$ instead of $e=0$. Fig.\ \ref{fig:ecc_correction} plots $\mathcal{L}$ against search frequency for Eqs.\ (\ref{eq:phase_first_order}) and (\ref{eq:phase_second_order}) alone (orange curve) and when Eq.\ (\ref{eq:phase_ecc}) is also included (blue curve). In both curves, the $a_0^2$ correction described in Section \ref{sec:a0^2_correction} is included. Without the additional $a_0 e$ correction, $\mathcal{L}$ attains its maximum $\approx 1.3 \times 10^{-5}\, {\rm Hz}$ (23 frequency bins) away from the injected frequency, whereas with the $a_0 e$ correction, $\max \mathcal{L}$ falls exactly in the injected frequency bin. Moreover, $\max \mathcal{L}$ is $\approx 2.4$ times higher, when the $a_0 e$ correction is included. In other words, the $a_0 e$ correction improves the accuracy and sensitivity appreciably for \herx[.]

Similarly to Section \ref{sec:a0^2_correction}, Eq.\ (\ref{eq:phase_ecc}) introduces a new factor $\propto \sum_{s_e=-\infty}^\infty  J_{s_e}(\pi \bar{f} a_0 e/c)$ into the \jstat filter. We truncate the sum to $2 M_e + 1 =2\, {\rm ceil}(2\pi \bar{f} a_0 e/c) + 1$ terms, as justified in Appendix \ref{appendix:bessel_sum}.

\section{Search implementation}
\label{sec:search_implementation}

This section describes the practical implementation of the search. Section \ref{subsec:t_drift} justifies the choice of the coherence time $T_{\rm drift}$ and the width of a search subband. Section \ref{subsec:num_templates} explains how the orbital parameter templates are placed. Section \ref{subsec:threshold} presents the procedure to set detection thresholds for a given false alarm probability. Section \ref{subsec:data} describes the data used in the search.  Section \ref{subsec:vetoes} describes the vetoes applied to candidate signals, that surpass the detection threshold. Section \ref{subsec:real_data} validates briefly the enhancements to the HMM pipeline with real data containing non-Gaussian noise artifacts.

\subsection{$T_{\rm drift}$ and frequency binning}
\label{subsec:t_drift}

The coherence time, $T_{\rm drift}$, bounds the interval during which the time-dependent gravitational wave frequency, $f_{\rm GW}(t)$, wanders by at most one frequency bin of width $\Delta f=1/(2T_{\rm drift})$ \cite{fstat, jstat}. We choose $T_{\rm drift}=10\, $days, partly to facilitate comparisons with previous HMM searches \cite{o3_amxp, scox1, psrJ0437andres}, and partly as a conservative response to the uncertainty regarding the astrophysical mechanisms driving frequency wandering for $f$-, $p$-, and $g$-modes. 

Previous HMM searches partition the search band into subbands that are $N_f=2^{20}$ frequency bins or $\Delta f_{\rm sub}=N_f \Delta f\approx 0.6\, {\rm Hz}$ wide \cite{o3_amxp, scox1, psrJ0437andres}. The number of frequency bins is chosen to be a power of two to speed up the Fourier transforms involved in computing the \jstat \cite{jstat}. In practice, computing the \jstat requires a convolution between the \fstat and a Jacobi-Anger filter, which depends on the phase model, viz.\ Eqs.\ (\ref{eq:phase_first_order}), (\ref{eq:phase_second_order}) and (\ref{eq:phase_ecc}) (see Eq.\ (3) in Ref.\ \cite{scox1}). For computational efficiency, $f_{\rm GW}(t)$ is assumed to coincide with the center $\bar{f}$ of the subband when constructing the Jacobi-Anger filter to avoid recomputing the filter for each frequency bin in the subband \cite{jstat}.

If the true gravitational wave frequency is displaced from $\bar{f}$ by $\delta f$, then Eq.\ (\ref{eq:phase_first_order}) implies that the phase error accumulated during an interval of one cycle equals $\sim a_0 \delta f/c$ cycles. To investigate the effect of $\delta f$ on the detection statistic $\mathcal{L}$, we inject signals into Gaussian noise with $h_0=10^{-24}$, $\sqrt{S_n}=4\times 10^{-24}\, {\rm Hz^{-1/2}}$, $T_{\rm drift}=10\, {\rm days}$, and $T_{\rm obs}=360\, {\rm days}$, such that all other parameters take \herx['s] central value in Table \ref{tab:params}. The signals are injected at $\delta f\in \{-2\times 10^{4} \Delta f, -1.9\times 10^{4} \Delta f, \, \ldots, 2\times 10^{4} \Delta f\}$, which amounts to incrementing $\delta f$ by $\approx 6 \times 10^{-4} \, {\rm Hz}$ per step or $a_0 \delta f/c$ by $\approx 8 \times 10^{-3}$ per step. For each $\delta f$ step, we run the HMM algorithm described in Section \ref{sec:hmm} including both the $a_0^2$ and $a_0 e$ corrections through Eqs.\ (\ref{eq:phase_second_order}) and (\ref{eq:phase_ecc}) respectively. We calculate the mismatch $\mufreq=1-\mathcal{L}/\mathcal{L}_{\rm mid}$, with $\mathcal{L}_{\rm mid} = \mathcal{L}(\delta f = 0)$, to measure how the sensitivity falls away from its maximum value as a function of $\delta f$. 

\begin{figure}[!htbp]
    \centering
    \includegraphics[width=0.95\linewidth]{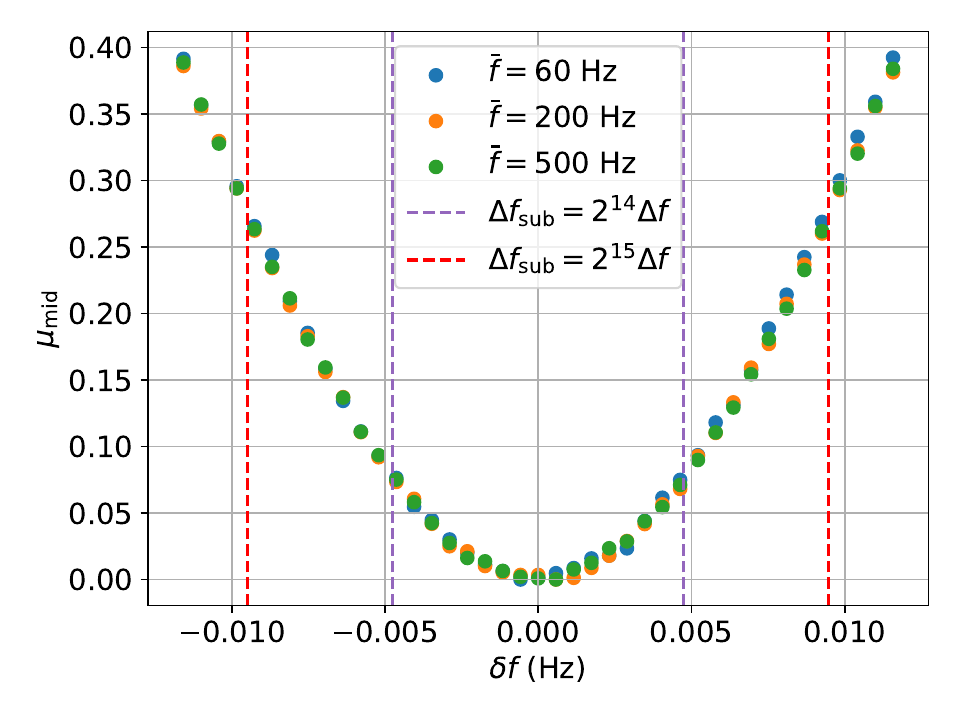}
    \caption{Log likelihood mismatch $\mufreq$ as a function of the displacement $\delta f$ of the injected signal from the central subband frequency $\bar{f}$, for $\bar{f}=60$, 200 and 500 Hz (blue, orange and green dots respectively). Vertical purple and red dashed lines bracket subbands of width $\Delta f_{\rm sub}=2^{14} \Delta f$ and $2^{15} \Delta f$ respectively.}
    \label{fig:mismatch_freq_center}
\end{figure}

Fig.\ \ref{fig:mismatch_freq_center} displays $\mufreq$ as a function of $\delta f$ for $\bar{f}=60$ Hz (blue dots), 200 Hz (orange dots) and 500 Hz (green dots). For reference, we draw dashed vertical purple and red lines to bracket subbands with $N_f=2^{14}$ and $2^{15}$ respectively. We find that the function $\mufreq(\delta f)$ is qualitatively similar for $\bar{f}=60$, 200 and 500 Hz; that is, one can assume to a good approximation, that $\mufreq$ does not depend on $\bar{f}$. Adopting the tolerance $\mufreq \leq 0.1$, in line with previously published searches \cite{scox1, o3_amxp, psrJ0437andres}, one must take $N_f = 2^{14}$ and $\Delta f_{\rm sub} \approx 9.5 \times 10^{-3} \, {\rm Hz}$\footnote{The Jacobi-Anger filter has bandwidth $\approx 4\pi \bar{f} a_0 / P$ \cite{jstat}. When convolving the \fstat with the Jacobi-Anger filter at some frequency $f$, one must evaluate the \fstat in the band $[f-2\pi \bar{f} a_0 / P, f+ 2\pi \bar{f} a_0 / P]$. In other words, for a subband whose width equals $\Delta f_{\rm sub}$, the \fstat must be evaluated in the band $[\bar{f}-\Delta f_{\rm sub}/2-2\pi \bar{f} a_0 / P, \bar{f}+\Delta f_{\rm sub}/2+2\pi \bar{f} a_0 / P]$.}. In Appendix \ref{appendix:mismatch_center_band}, we show that $\mufreq=0.1$ corresponds to $|a_0 \delta f/c| \approx 0.07$.

In this paper, we search $f_{\rm GW} \in [60, 200]\, {\rm Hz}$, where the LIGO detectors are most sensitive \cite{o3LIGOsensitivity,O3LIGOsensitivty2}. The search is narrower than previous broadband HMM searches using the \jstat \cite{scox1, psrJ0437andres}, which select $f_{\rm GW} \in [60, 500]\, {\rm Hz}$, due to the increased computational cost incurred from  the $a_0$ and $a_0 e$ corrections in Sections \ref{sec:a0^2_correction} and \ref{sec:ecc_correction}. The $a_0 e$ correction, discussed in Section \ref{sec:ecc_correction}, lifts the number of orbital parameters from three to five. Moreover, by selecting $\Delta f_{\rm sub}=2^{14} \Delta f$ (see above) instead of $\Delta f_{\rm sub}=2^{20} \Delta f$ as in previous searches \cite{o3_amxp, scox1, psrJ0437andres}, we have $2^6$ times more Jacobi-Anger filters to compute. 

\subsection{Placing orbital templates}
\label{subsec:num_templates}

In previous HMM searches using the \jstat \cite{scox1, o3_amxp, psrJ0437andres}, the parameter space is covered by a rectangular grid that spans $[\bar{X} - 3\sigma_X, \bar{X} + 3\sigma_X]$, where $\bar{X}$ and $\sigma_X$ are the central value and 1-$\sigma$ uncertainty respectively of parameter $X$. The spacing $\Delta X$ between consecutive points in this rectangular grid for parameter $X$ is given by \cite{leaciPrix2015}

\begin{align}
    \Delta a_0 &= \frac{c\sqrt{2\mu_{\rm max}}}{\pi f_{\rm GW}}, \label{eq:template_spacing_first} \\
    \Delta T_{\rm asc} &= \frac{cP\sqrt{2\mu_{\rm max}}}{2\pi^2 a_0 f_{\rm GW}}, \label{eq:template_spacing_t}\\
    \Delta P &= \frac{cP^2 \sqrt{6\mu_{\rm max}}}{\pi^2 N_T T_{\rm drift}a_0 f_{\rm GW}}, \label{eq:template_spacing_p} \\
    \Delta e &= \frac{2c \sqrt{2\mu_{\rm max}}}{\pi a_0 f_{\rm GW}}, \\
    \Delta \omega &= \frac{2 c\sqrt{2\mu_{\rm max}}}{\pi e  a_0 f_{\rm GW}}. \label{eq:template_spacing_last}
\end{align}

\noindent In Eqs.\ (\ref{eq:template_spacing_first})--(\ref{eq:template_spacing_last}), $\mu_{\rm max}$ defines the maximum tolerable mismatch $1-\mathcal{L_{\rm grid}/\mathcal{L}_{\rm true}}$, where $\mathcal{L}_{\rm grid}$ and $\mathcal{L}_{\rm true}$ are the log likelihoods obtained from executing the search at the nearest grid point and the true parameters respectively. To be conservative when evaluating Eqs.\ (\ref{eq:template_spacing_first})--(\ref{eq:template_spacing_last}), we use $f_{\rm GW}=\bar{f}+\Delta f_{\rm sub}/2$, $P=\bar{P}-3\sigma_P$, $a_0=\bar{a_0}+3\sigma_{a_0}$, and $e=\bar{e}+3\sigma_e$. We have $\dot{P}T_{\rm obs}<\Delta P/2$ for $f_{\rm GW} \in [ 60,200]\, {\rm Hz}$, so we assume that $P$ remains in the same bin throughout O3.

\begin{figure}[htbp!]
    \centering
    \includegraphics[width=0.95\linewidth]{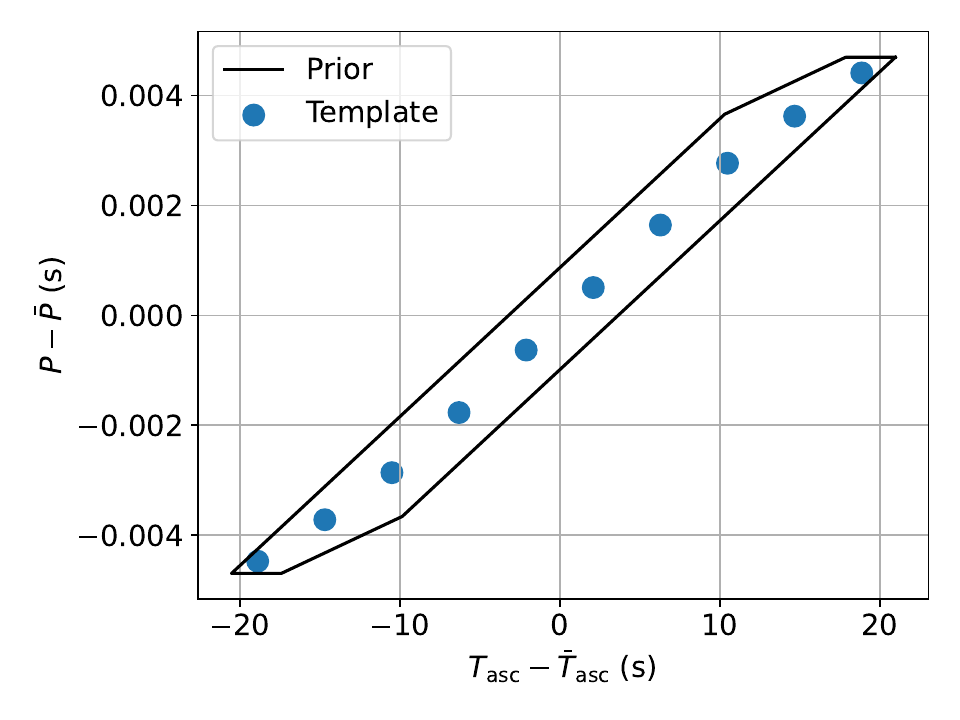}
    \caption{Template sites in the two-dimensional electromagnetic prior of $T_{\rm asc}$ and $P$, shown as blue dots. We obtain $\Delta P \gtrsim 0.02 \, {\rm s}>6\sigma _P$ for $60 \,{\rm Hz} \leq f_{\rm GW} \leq 200 \,{\rm Hz}$, so we only have one ``diagonal'' row of templates in the $T_{\rm asc}$-$P$ plane.}
    \label{fig:template_place_p_t}
\end{figure}

To construct the rectangular grid, we first calculate the number of grid points for parameter $X$ as $N_X={\rm ceil}(6\sigma_X/\Delta X)$, which yields an adjusted spacing $\Delta X'=6\sigma_X/N_X$. For \herx[,] we have $N_P=1$ and $N_{X \neq P}>1$ for $60 \,{\rm Hz} \leq f_{\rm GW} \leq 200 \,{\rm Hz}$. We construct the rectangular grid with $N_X$ points for each parameter $X$, where consecutive points are separated by $\Delta X'$. Additionally, because the bounds on the two-dimensional electromagnetic prior of $T_{\rm asc}$ and $P$ have a different shape from the other orbital parameters, as shown in Fig.\ \ref{fig:uncertainty_prior}, we place grid points along the diagonal of the two-dimensional prior in the $T_{\rm asc}$-$P$ plane. To be precise, for each $T_{\rm asc}$ grid point, we place $P$ at the midpoint of the corresponding bound on $P$ implied by the two-dimensional prior of $T_{\rm asc}$ and $P$. Fig.\ \ref{fig:template_place_p_t} illustrates how we place templates\footnote{There are other ways to account for the correlation in $T_{\rm asc}$ and $P$, e.g. Ref.\ \cite{Wagner_2022} discusses a shear transformation.} for $T_{\rm asc}$ and $P$.

\begin{figure}[htbp!]
    \centering
    \includegraphics[width=0.95\linewidth]{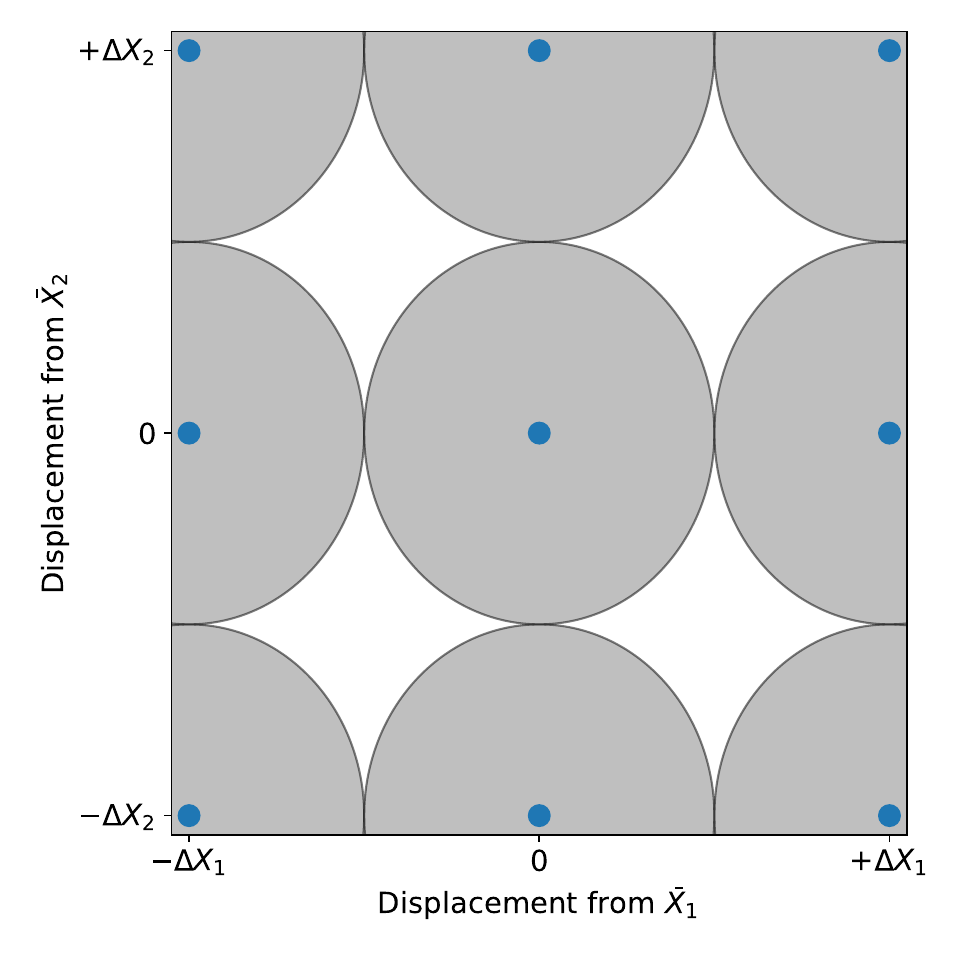}
    \caption{An idealized schematic of the parameter space covered by two uncorrelated parameters $X_1$ and $X_2$. The blue dots mark a rectangular lattice with grid points separated by $(\Delta X_1,\Delta X_2)$, the gray shaded regions satisfy $1 - \mathcal{L}_{\rm grid} / \mathcal{L}_{\rm true} \leq \mu_{\rm max}$ centered on the blue dots, and the white unshaded regions (``holes'') satisfy $1 - \mathcal{L}_{\rm grid} / \mathcal{L}_{\rm true} > \mu_{\rm max}$.}
    \label{fig:template_place}
\end{figure}

A purely rectangular grid with spacing given by Eqs.\ (\ref{eq:template_spacing_first})--(\ref{eq:template_spacing_last}) does not achieve the desired $\mu_{\rm max}$, because there are ``holes'' between the five-dimensional ellipsoids satisfying $1 - \mathcal{L}_{\rm grid}/\mathcal{L}_{\rm true}$ around every grid point \cite{leaciPrix2015}. Fig.\ \ref{fig:template_place}  illustrates schematically a two-dimensional cross-section of the parameter space covered by a rectangular grid, for two uncorrelated parameters $X_1$ and $X_2$. The blue dots mark where the templates are centered, the gray shaded regions satisfy $1 - \mathcal{L}_{\rm grid}/\mathcal{L}_{\rm true} \leq \mu_{\rm max}$, and the white unshaded regions are holes which satisfy $1 - \mathcal{L}_{\rm grid} / \mathcal{L}_{\rm true} > \mu_{\rm max}$.
The mismatch at the center of the holes is $\sim \mu_{\rm max}N_d$, where $N_d=5$ is the number of dimensions of the parameter space.
Hence, the separation between grid points per dimension should be $\Delta X / \sqrt{N_d}$ to maintain $1 - \mathcal{L}_{\rm grid}/\mathcal{L}_{\rm true} \leq \mu_{\rm max}$ \cite{Prix_template}.
However, a rectangular grid is inefficient, in terms of minimizing the number of templates.
There are known efficient methods to place templates to guarantee $1 - \mathcal{L}_{\rm grid}/\mathcal{L}_{\rm true} \leq \mu_{\rm max}$ across the entire parameter domain \cite{Prix_template, Wette_template, Wagner_2022}, but we defer their use to future work; the overhead is unjustified for the pilot search in this paper.
In this paper, we emulate previous HMM searches by using a grid separation of $\Delta X'$, as described above \cite{scox1, o3_amxp, psrJ0437andres}.

The correlation between $T_{\rm asc}$ and $P$ shears the ellipse $1 - \mathcal{L}_{\rm grid} / \mathcal{L}_{\rm true} \leq \mu_{\rm max}$. The shear, for a single coherent time step, is proportional to $T_{\rm asc} - T_{\rm mid}$, where $T_{\rm mid}$ is the midpoint of the coherent timestep \cite{leaciPrix2015}. Hence, we opt to propagate $T_{\rm asc}$ and $P$ to the middle of O3, so that we average out the shear across the $N_T$ coherent blocks. In Appendix \ref{appendix:p_t_prop}, we show, through injections in Gaussian noise, that propagating $T_{\rm asc}$ and $P$ to the midpoint of O3 yields an unsheared ellipse. This strategy does not change the number of templates, because a shear preserves area, but it allow us to apply the same template placement strategy to $a_0$, $e$, and $\omega$, which are uncorrelated, as to $T_{\rm asc}$ and $P$. Previous HMM searches using the \jstat propagate the orbital parameters to the start of the dataset \cite{o3_amxp, scox1, psrJ0437andres}, but other searches propagate to the midpoint \cite{CrossCorrScox1, Whelan_2023}. The secondary question of how the mismatch $\mu_{\rm mid}$ (see Section \ref{subsec:t_drift}) and $1-\mathcal{L}_{\rm grid}/\mathcal{L}_{\rm true}$ compound is deferred to future work.

\subsection{Detection threshold and false alarm probability}
\label{subsec:threshold}

For each subband, the HMM algorithm described in Section \ref{sec:hmm} outputs $N_{\rm tot}=N_f N_{\rm B}$ paths and their associated log likelihoods $\mathcal{L}$, where $N_{\rm B}$ is the number of binary orbital parameters used in that subband. 
We consider a subband to contain a candidate, if $\max(\mathcal{L})$ exceeds a threshold $\mathcal{L}_{\rm th}$, which depends on a user-selected false alarm probability per subband, $\alpha_{N_{\rm tot}}$. 
We estimate ${\cal L}_{\rm th}$ using off-target noise realizations and the percentile method described in Appendix A in Ref.\ \cite{o3_amxp}. 
Setting $\mathcal{L}_{\rm th}$ at every subband is computationally expensive.
Instead, we opt to set $\mathcal{L}_{\rm th}$ at 29 subbands that are not contaminated by known lines \cite{O3lines}, and interpolate linearly between them.
We present the interpolation in Appendix \ref{appendix:threshold}.

In this paper, we take $N_{\rm sub}={\rm ceil} \left[ (200-60) \, {\rm Hz} /\Delta f_{\rm sub} \right]=14766$ subbands.
Previous HMM searches typically take $0.01 \leq \alpha_{N_{\rm tot}} \leq 0.3$ to produce $7 \leq N_{\rm sub} \alpha_{N_{\rm tot}}  \leq 18$ \cite{viterbiO1_scox1, viterbiO2_scox1, o3_amxp, scox1, psrJ0437andres}.
We arbitrarily choose $\alpha_{N_{\rm tot}}=5 \times 10^{-4}$ to yield $N_{\rm sub} \alpha_{N_{\rm tot}} \approx 7$.
To estimate the expected number of false alarms, we note that the thresholds are set in subbands that are not contaminated by known lines.
Looking ahead to the known lines veto in Section \ref{subsec:vetoes}, we find that 5884 subbands are contaminated, in the sense that they would be vetoed if they contain an above-threshold candidate.
Hence, we expect $8882 \alpha_{N_{\rm tot}} \approx 4$ false alarms after applying the known lines veto but before applying the other vetoes.

\subsection{O3 data}
\label{subsec:data}

We use the whole O3 dataset collected by the LIGO Hanford and Livingston detectors, which spans from April 1, 2019 15:00 UTC to March 27, 2020, 17:00 UTC \cite{O3_open_data}. 
There is a month-long commissioning break from 1 October 2019 to 1 November 2019, during which no data were collected. 
The data ingested by the HMM algorithm described in Section \ref{sec:hmm} are short Fourier transforms (SFTs) of duration $T_{\rm SFT}=1800 \, {\rm s}$ generated from the ``CO1 calibrated self-gated 60 Hz subtracted'' dataset, which subtracts loud transient glitches and the 60Hz mains power noise line \cite{O3segments, CalibrationO3, CalibrationO3_60Hz}.
Coherent blocks which contain no SFTs due to the commissioning break are given uniform log likelihoods across all frequency bins, to allow for stochastic wandering of the signal frequency during these blocks. 
When generating Gaussian realizations in Figs.\ \ref{fig:mismatch_freq_center} and \ref{fig:ellipse}, we do not generate SFTs during the commissioning break to emulate O3.

The \jstat depends on the \fstat computed assuming an isolated target, as described in Section \ref{subsec:jstat}.
In practice, the LALSuite \cite{lalsuite} implementation of the \fstat assumes a constant signal frequency during each SFT for convenience.
The assumption implies a maximum allowed SFT duration \cite{leaciPrix2015}

\begin{equation}
    T_{\rm SFT}^2 \leq \frac{6\sqrt{5\mu_{\rm SFT}}}{\pi (a_0/c)f_{\rm GW}\Omega^2}, \label{eq:max_sft}
\end{equation}

\noindent where $\mu_{\rm SFT}=0.05$ is the maximum mismatch in the constant frequency approximation.
Because we compute the \fstat assuming an isolated target, the values of $a_0$ and $\Omega$ in Eq.\ (\ref{eq:max_sft}) are set predominantly by the Earth's sidereal motion plus smaller binary corrections, i.e.\ $a_0/c \approx 0.02\, {\rm s}$ and $\Omega \approx 2\pi/(1 \, {\rm day})$ \cite{lalsuite}. 
Hence, for the \jstat[,] the maximum frequency allowed with $T_{\rm SFT}=1800 \, {\rm s}$ is $f_{\rm GW} \approx 2.8 \, {\rm kHz}$, cf.\ $f_{\rm GW}\approx12\, {\rm Hz}$ for the binary \fstat and the values in Table \ref{tab:params}.
In Appendix \ref{appendix:binary_fstat}, we verify, with injections into Gaussian noise, that the above approach produces qualitatively similar $\mathcal{L}$ for both the \jstat and the binary \fstat[.]

\subsection{Vetoes}
\label{subsec:vetoes}

We employ a sequence of five standard veto procedures, as done in previous work \cite{o3_amxp, scox1, psrJ0437andres, viterbiO2_AMXP, viterbiO2_Fomalhaut_b, viterbiO3_supernova, viterbiO2_scox1}, to determine if a candidate is likely to be non-astrophysical. 

(1) Narrowband noise features (instrumental lines) in the interferometers may artificially increase $\mathcal{L}$. We veto any candidate whose optimal path $f_{\rm GW}(t)$, for any epoch $t$ of the search, satisfies

\begin{equation}
\label{eq:line}
    |f_{\rm GW}(t) - f_{\rm line}| < \frac{2\pi a_0 f_{\rm line}}{cP},
\end{equation}

\noindent where $f_{\rm line}$ is the frequency of any noise line  found in the reference list published in Ref.\ \cite{O3lines}. We refer to this veto as the known lines veto.

(2) Both the LIGO Hanford and Livingston detectors have comparable sensitivity for $60 \,{\rm Hz} \leq f_{\rm GW} \leq 200 \,{\rm Hz}$ \cite{o3LIGOsensitivity, O3LIGOsensitivty2}. If a candidate is only seen in one detector, then the candidate likely originates from a noise artifact in that detector. Let $ \mathcal{L}_a$, $\mathcal{L}_b < \mathcal{L}_a$ and $\mathcal{L}_\cup$ denote the maximum log likelihood obtained from detector $a$ only, detector $b$ only and both detectors respectively, with associated optimal paths $f_a(t)$, $f_b(t)$ and $f_\cup(t)$. We veto candidates that satisfy $\mathcal{L}_a > \mathcal{L}_\cup$, $\mathcal{L}_b < \mathcal{L}_{\rm th}$ and

\begin{equation}
    \label{eq:oneIFO}
    |f_\cup (t) - f_a (t)| < \frac{2\pi a_0 f_\cup (t)}{cP},
\end{equation}

\noindent for any epoch $t$ of the search. 
Both $\mathcal{L}_a > \mathcal{L}_\cup$ and $\mathcal{L}_b < \mathcal{L}_{\rm th}$ hold, if we seek to veto noise artifacts that are present in only one detector. 
If instead one finds $\mathcal{L}_a < \mathcal{L}_\cup$ and $\mathcal{L}_b < \mathcal{L}_{\rm th}$, the candidate could be a faint astrophysical signal which requires both interferometers to be detectable.
If one finds $\mathcal{L}_a > \mathcal{L}_\cup$ and $\mathcal{L}_b > \mathcal{L}_{\rm th}$, the candidate could be a strong astrophysical signal. 
Eq.\ (\ref{eq:oneIFO}) checks if the frequency extent of the \jstat filter used in computing $\mathcal{L}_\cup$ contains $f_a(t)$.
We refer to this veto as the single interferometer veto. It too has been employed in many previous analyses \cite{o3_amxp, scox1, psrJ0437andres}.

(3) We search an off-target sky position using the same orbital parameters as the candidate. If the off-target $\mathcal{L}$ exceeds $\mathcal{L}_{\rm th}$, then we veto the candidate. To obtain the off-target sky position, we arbitrarily shift the right ascension by $-40$ minutes from the value listed in Table \ref{tab:params}. We refer to this veto as the off-target veto \cite{viterbiO2_AMXP, viterbiO2_Fomalhaut_b, o3_amxp, viterbiO3_supernova}.

(4) We turn off the correction from the Doppler modulation due to the Earth's motion. Without the Doppler correction, an astrophysical signal should become undetectable. We veto candidates, if $\mathcal{L}$ without the Doppler correction exceeds $\mathcal{L}_\cup$. We refer to this veto as the DM-off veto \cite{dm_off_veto, viterbiO2_Fomalhaut_b, viterbiO3_supernova, Jones_2022}.

(5) An astrophysical continuous-wave signal should persist throughout O3, although one can imagine exceptions in principle, e.g.\ a transient stellar oscillation which lasts for days (say) but not months \cite{zhao2022}.  Hence $\mathcal{L}$ should be similar in the first and second halves of the O3 data. Accordingly, we split the dataset into the first 17 and last 19 coherent steps, to account for the two coherent steps that contain no SFTs. Let $\mathcal{L}_1$ and $\mathcal{L}_2$ be the log likelihoods of the first and second halves respectively. To remove the log likelihood gain of the two coherent steps without SFTs, we adjust $\mathcal{L}_2$ to $\mathcal{L}_2'=\mathcal{L}_2-2\log(1/3)$ and similarly $\mathcal{L}_\cup'$ and $\mathcal{L}_{\rm th}'$. Define $\mathcal{L}_a = \max(\mathcal{L}_1, \mathcal{L}_2')$ and $\mathcal{L}_b = \min(\mathcal{L}_1,\mathcal{L}_2')$. We veto candidates that satisfy $\mathcal{L}_a > \mathcal{L}_\cup '/2$, $\mathcal{L}_b < \mathcal{L}_{\rm th}' / 2$ and Eq.\ (\ref{eq:oneIFO}). Injections into real O3 data yield $1.0 < \mathcal{L}_1/\mathcal{L}_2'<1.2$ for a strong signal ($h_0=10^{-24}$), so we additionally require $\mathcal{L}_1 > 1.2\mathcal{L}_2'$ or $\mathcal{L}_2'> \mathcal{L}_1$ to veto a candidate. We refer to this veto as the $T_{\rm obs}/2$ veto. This veto has been previously employed using z-score normalized log likelihoods
\cite{viterbiO2_AMXP, viterbiO2_scox1, viterbiO2_Fomalhaut_b}. Here we apply it equivalently to log likelihoods without a z-score for the first time.

\subsection{Real detector noise}
\label{subsec:real_data}

The modifications to the HMM pipeline described in Sections \ref{sec:a0^2_correction}, \ref{sec:ecc_correction}, \ref{subsec:t_drift} and \ref{subsec:num_templates} are validated in Gaussian noise. However, noise in the real detectors is generally not Gaussian \cite{Davis_2021}. Previous authors have tested several aspects of the performance of the HMM in real detector noise, including the false dismissal rates of the veto procedures with the \jstat \cite{viterbiO2_scox1} and Bessel-weighted \fstat \cite{viterbiO1_scox1},  the DM-off veto and effective point spread function in real O2 data \cite{Jones_2022}, and hardware injections in real O2 data  \cite{viterbiO2_supernova}.

\begin{figure}[!htbp]
    \centering
    \includegraphics[width=\linewidth]{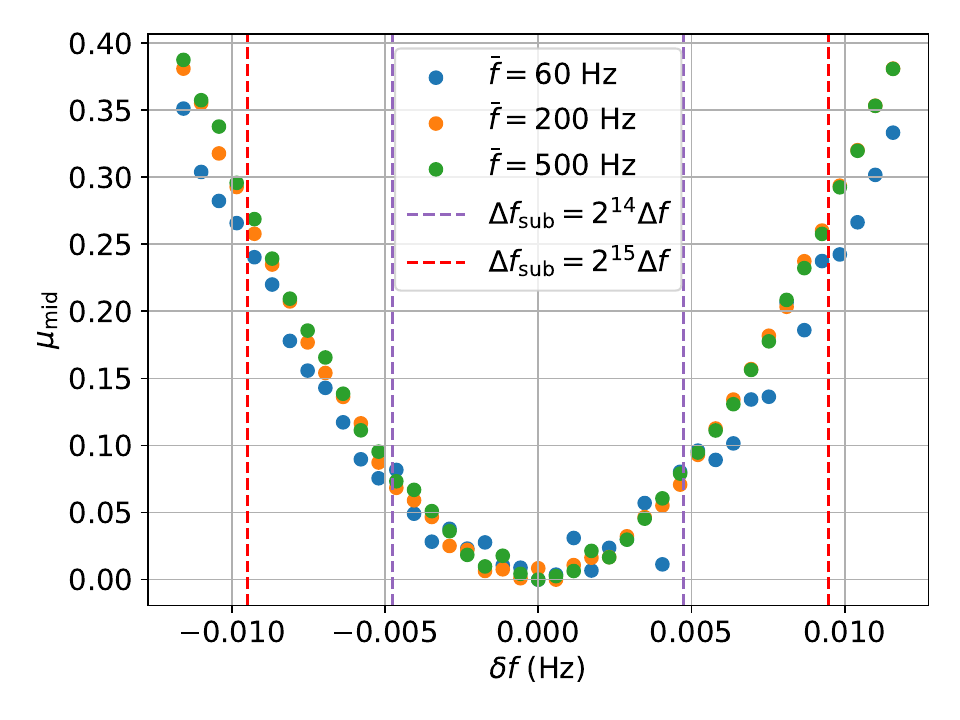}
    \caption{Same as Fig.\ \ref{fig:mismatch_freq_center}, except the signal is injected into real O3 SFTs.}
    \label{fig:mismatch_freq_center_real}
\end{figure}

In this section, we check briefly that previous validations with non-Gaussian noise continue to hold, after the modifications in Sections \ref{sec:a0^2_correction}, \ref{sec:ecc_correction}, \ref{subsec:t_drift} and \ref{subsec:num_templates} are made. To this end, we repeat the tests in Figs.\ \ref{fig:a02_correction}, \ref{fig:ecc_correction} and \ref{fig:mismatch_freq_center} by injecting the same synthetic signals into real O3 SFTs drawn from the interval $[T_{\rm O3, \,start},\, T_{\rm O3, \,start} + N_T T_{\rm drift}]$. As one example, Fig.\ \ref{fig:mismatch_freq_center_real}  is the counterpart of Fig.\ \ref{fig:mismatch_freq_center}. The results in Figs.\ \ref{fig:mismatch_freq_center} and \ref{fig:mismatch_freq_center_real} are qualitatively similar. A subband containing $N_f=2^{14}$ bins is required to maintain $\mufreq \leq 0.1$ in both figures. The corresponding injection tests into real O3 SFTs for Figs.\ \ref{fig:a02_correction} and \ref{fig:ecc_correction} show qualitatively similar improvements, when the $a_0^2$ and $a_0 e$ corrections respectively are applied, and are omitted for brevity.

\section{O3 Analysis}
\label{sec:results}

In this section, we present the output of the pilot search and associated veto procedures described in Sections \ref{sec:hmm}--\ref{sec:search_implementation}.

The initial scan through the full search band $f_{\rm GW} \in [60,200] \, {\rm Hz}$ returns 90 candidates above the threshold set according to Section \ref{subsec:threshold}. 
The candidates span frequencies from $\approx 60 \, {\rm Hz}$ to $\approx 189 \, {\rm Hz}$. 
The loudest and faintest candidates have $\mathcal{L}-\mathcal{L_{\rm th}} =$ $2081.10$ and $0.10$ respectively. 

\begin{table*}[htbp!]
\caption{The terminating frequency of each above-threshold candidate that survive the known lines veto, and if the candidate survives ($\checkmark$) or fails ($\times$) the subsequent four vetoes in Section \ref{subsec:vetoes}. The orbital parameters, $\max \mathcal{L}$ and $\mathcal{L}_{\rm th}$ are recorded in Table \ref{tab_app:candidate_params} in Appendix \ref{appendix:candidate_params} for the sake of reproducibility.}
\label{tab:candidates}
\begin{tabular}{ccccc}
\toprule
Terminating frequency (Hz) & Single interferometer &   Off target &       DM-off & Half $T_{\rm obs}$ \\
\midrule
64.23285822 & $\times$ & -- & -- & -- \\
64.33650174 & $\times$ & -- & -- & -- \\
64.36863600 & $\times$ & -- & -- & -- \\
64.60612674 & $\times$ & -- & -- & -- \\
64.67312153 & $\times$ & -- & -- & -- \\
64.67559201 & $\times$ & -- & -- & -- \\
76.40570197 & $\checkmark$ & $\checkmark$ & $\checkmark$ & $\checkmark$ \\
109.47137095 & $\checkmark$ & $\checkmark$ & $\checkmark$ & $\checkmark$ \\
114.70074190 & $\checkmark$ & $\checkmark$ & $\checkmark$ & $\times$ \\
144.39718171 & $\checkmark$ & $\checkmark$ & $\checkmark$ & $\checkmark$ \\
189.23836111 & $\checkmark$ & $\checkmark$ & $\checkmark$ & $\checkmark$ \\ \midrule
Number of survivors &       5 &      5 &      5 &      4  \\
\bottomrule
\end{tabular}
\end{table*}

Most of the above-threshold candidates coincide with narrowband instrumental noise artifacts. 
Upon comparing with the known lines list published in Ref.\ \cite{O3lines}, we find that 11 candidates survive the known lines veto. 
Table \ref{tab:candidates} summarizes the terminating frequencies of the optimal path of each candidate, as well as the outcome of the vetoes described in Section \ref{subsec:vetoes}. 
Six candidates are eliminated by the single interferometer veto. 
The off-target and DM-off vetoes do not eliminate any candidates. 
Finally, the $114.70\, {\rm Hz}$ candidate is eliminated by the $T_{\rm obs}/2$ veto for satisfying $\mathcal{L}_2' > \mathcal{L}_\cup' / 2$ and $\mathcal{L}_1 < \mathcal{L}_{\rm th}' / 2$. 
We are left with four survivors, whose terminating frequencies are 76.41, 109.47, 144.40 and $189.24\,  {\rm Hz}$.
As an indicative follow-up, we inspect visually four versions of the time-averaged power spectral density (PSD) near the survivors, constructed from the O3a Hanford, O3b Hanford, O3a Livingston and O3b Livingston SFTs. 
The O3b Hanford PSD near the $109.47\, {\rm Hz}$ candidate appears contaminated. 
We display the PSDs near each candidate in Appendix \ref{appendix:psd_inspection} for reference. 
For the sake of reproducibility, we list the orbital parameters of the 11 above-threshold candidates in Table \ref{tab_app:candidate_params} of Appendix \ref{appendix:candidate_params}.

\begin{figure}[htbp!]
    \centering
    \includegraphics[width=\linewidth]{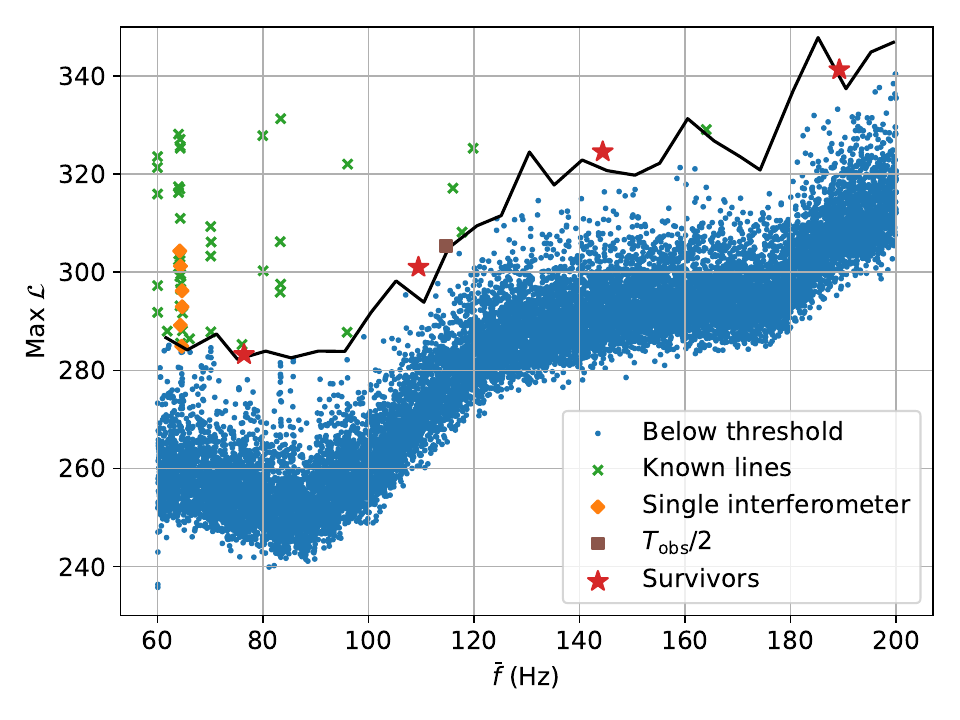}
    \caption{Maximum likelihood per subband as a function of the subband central frequency $\bar{f}$. 
    The solid black curve represents the threshold set according to Section \ref{subsec:threshold}, while the blue dots are below-threshold subbands. 
    Candidates eliminated by the known lines, single interferometer, and $T_{\rm obs}/2$ vetoes are coded as green crosses, orange diamonds, and brown squares respectively. 
    Survivors are coded as red stars. 
    We omit the 35 above-threshold candidates with $\max \mathcal{L} > 350$ that fail the known lines veto because they rise well above the top of the vertical scale.}
    \label{fig:max_ll}
\end{figure}

Fig.\ \ref{fig:max_ll} visualizes $\max \mathcal{L}$ in each subband. 
The solid black line marks $\mathcal{L}_{\rm th}$, as calculated in Section \ref{subsec:threshold}. 
The blue dots represent below-threshold subbands. 
The 79, six, and one candidates that fail the known lines, single interferometer, and $T_{\rm obs}/2$ vetoes are marked as green crosses, orange diamonds, and brown squares respectively. 
The four survivors are marked as red stars. 
The six candidates that fail the single interferometer veto all have terminating frequencies between 64 Hz and 65 Hz and satisfy $\mathcal{L}_{\rm H} > \mathcal{L}_\cup$, $\mathcal{L}_{\rm L} < \mathcal{L}_{\rm th}$, where $\mathcal{L}_{\rm H}$ and $\mathcal{L}_{\rm L}$ are the log likelihoods obtained from the Hanford and Livingston detectors respectively. 
These candidates are  $\leq 0.06\, {\rm Hz}$ away from satisfying Eq.\ (\ref{eq:line}), i.e.\ they are close to a known line. 
It is likely that these six candidates originate from a common noise source, and should be treated as a single candidate for the purpose of comparing against the number of expected false alarms.
Hence, we obtain six known line veto survivors effectively, consistent with the expected number of roughly four.
The $114.70\, {\rm Hz}$ candidate, which is eliminated by the $T_{\rm obs}/2$ veto, is located $0.01\, {\rm Hz}$ away from satisfying Eq.\ (\ref{eq:line}) and is close to a known line.
In comparison, among the four veto survivors, the $189.24\, {\rm Hz}$ survivor is $0.11\, {\rm Hz}$ from being eliminated by the known lines veto, while the other three survivors are $\geq 0.24\, {\rm Hz}$ away.

\begin{figure*}[htbp!]
    \centering
    \includegraphics[width=0.25\linewidth]{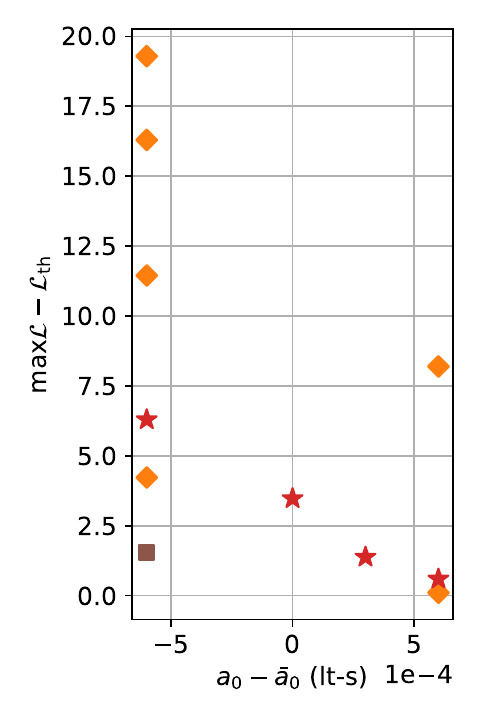}%
    \includegraphics[width=0.25\linewidth]{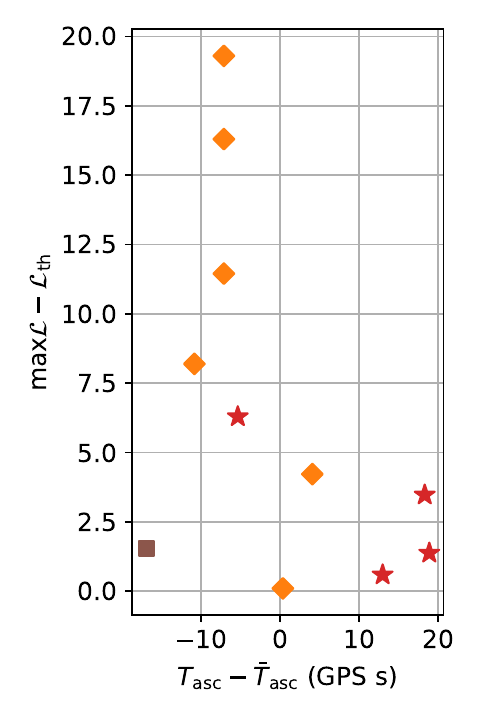}%
    \includegraphics[width=0.25\linewidth]{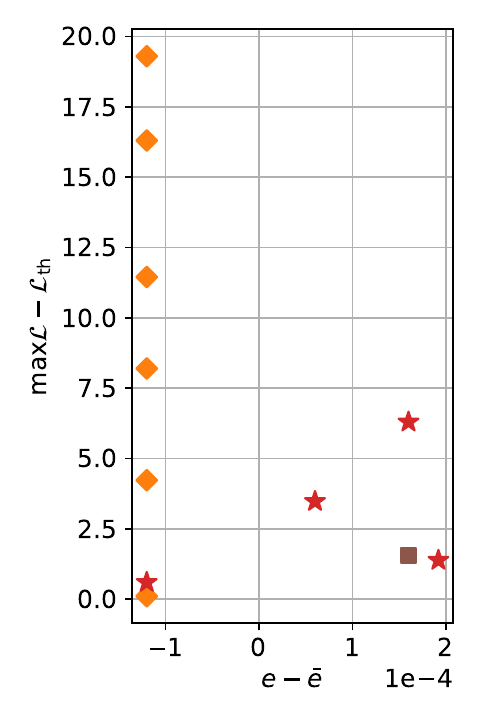}%
    \includegraphics[width=0.25\linewidth]{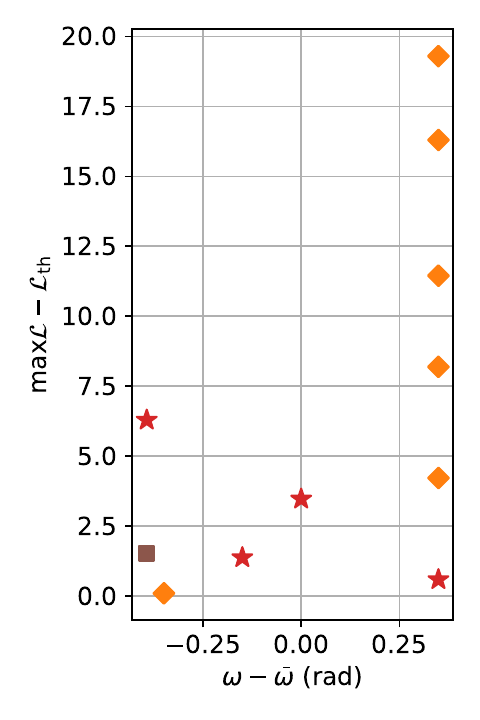}%
    \caption{Maximum likelihood excess $\max\mathcal{L} - \mathcal{L}_{\rm th}$ of the above-threshold survivors of the known lines veto, as a function of the orbital template parameters. Every unique $T_{\rm asc}$ value corresponds to a unique $P$ value (see Fig.\ \ref{fig:template_place_p_t}), so we do not show the corresponding plot for $P$. Candidates eliminated by the single interferometer, and $T_{\rm obs}/2$ vetoes are coded as orange diamonds, and brown squares respectively to match Fig.\ \ref{fig:max_ll}. Survivors are coded as red stars.}
    \label{fig:param_ll}
\end{figure*}

Fig.\ \ref{fig:param_ll} plots $\max\mathcal{L} - \mathcal{L}_{\rm th}$ as a function of the orbital template parameters.
The candidates are coded in the same manner as Fig.\ \ref{fig:max_ll}. 
We do not show the corresponding plot for $P$, as every unique $T_{\rm asc}$ value corresponds to a unique $P$ value.
The bimodality in the $a_0$ plot is caused by having $N_{a_0}=2$ for all candidates except the two with the highest $\bar{f}$.

\section{Representative frequentist upper limits}
\label{sec:upper_limits}

In this section, we quantify the improvement in sensitivity of the second-order ($\propto a_0^2$) and leading-order eccentricity ($\propto a_0 e$) corrections introduced in Sections \ref{sec:a0^2_correction} and \ref{sec:ecc_correction} respectively by calculating the sensitivity depth $\mathcal{D}$ at a representative subband.

We elect arbitrarily to calculate $\mathcal{D}$ at $\bar{f}=174.2281481 \, {\rm Hz}$ because we do not find an above-threshold candidate there, and the subband is one of the 29 subbands where we explicitly set $\mathcal{L}_{\rm th}$, as explained in Section \ref{subsec:threshold}.
The dimensionless sensitivity depth $\mathcal{D}$ is defined as \cite{BehnkeEtAl2015,DreissigackerEtAl2018,wette2023}

\begin{equation}
    \label{eq:depth}\mathcal{D}=\frac{\sqrt{S_n / ({\rm 1\, Hz^{-1}})}}{\hUpper}
\end{equation}

\noindent where $S_n$ is one-sided power spectral density and $\hUpper$ is the frequentist upper limit on the detectable wave strain set at 95\% confidence.
We follow Ref.\ \cite{o3_amxp} to compute $\hUpper$.
We draw 100 sets of random injection parameters.
For each of the five binary parameters $X \in \{P, T_{\rm asc}, a_0, e, \omega\}$ in Table \ref{tab:params}, we draw uniformly from the range $[\bar{X} \pm 3\sigma_X]$, where $\sigma_X$ is the 1-$\sigma$ uncertainty of $X$ as measured electromagnetically. 
We ensure that $T_{\rm asc}$ and $P$ fall within the two-dimensional prior in Fig.\ \ref{fig:uncertainty_prior}.
The cosine of the inclination angle $\cos \iota$ and polarization angle $\psi$ are drawn uniformly from the ranges $[-1,1]$ and $[0,2\pi]$ respectively.
The signal is always injected at $\bar{f}$ with zero spin wandering.
We vary the injected $h_0$ and record $\eta(h_0)$, the percentage of the 100 injections that yield $\max \mathcal{L} > \mathcal{L}_{\rm th}$.
$\hUpper$ is obtained by fitting a sigmoid to $\eta(h_0)$, and by solving $\eta(\hUpper)=0.95$.
$\hUpper$ is computed for two scenarios: (a) when only the leading-order phase term is included, as in Eq.\ (\ref{eq:phase_first_order}); and (b) when the $a_0^2$ and $a_0 e$ corrections are included through Eqs.\ (\ref{eq:phase_second_order}) and (\ref{eq:phase_ecc}) respectively.

\begin{figure}[htbp!]
    \centering
    \includegraphics[width=0.99\linewidth]{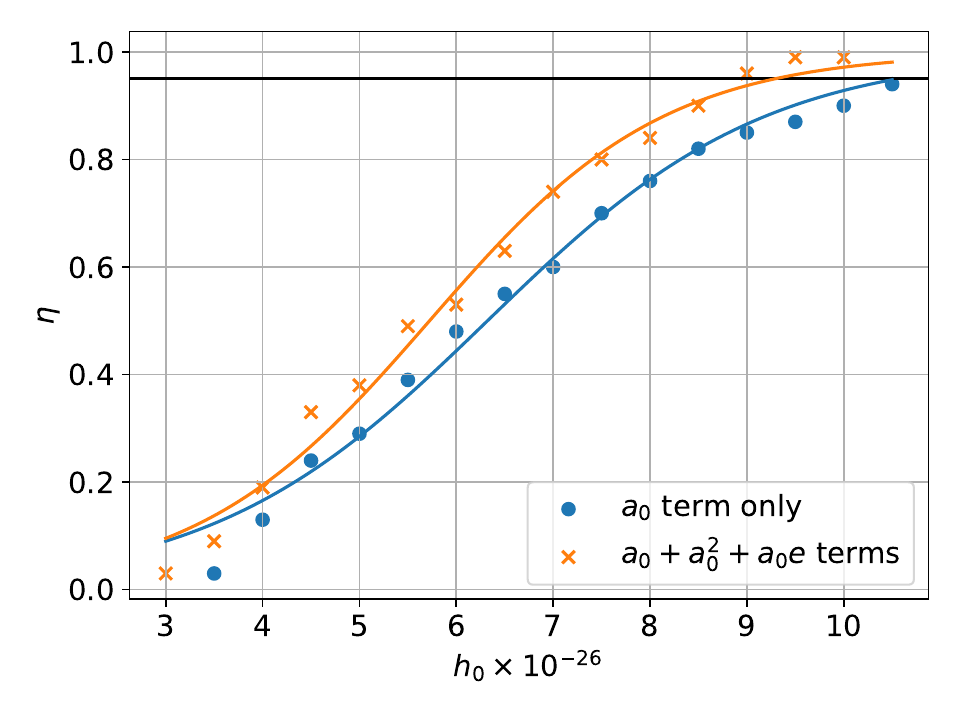}
    \caption{Percentage of above-threshold injections, $\eta$, as a function of the injected $h_0$. 
    The blue dots mark the discrete trial values, while the blue curve is the sigmoid fit, for Eq.\ (\ref{eq:phase_first_order}) alone.
    The orange crosses and curve are similar, except we also include Eqs.\ (\ref{eq:phase_second_order}) and (\ref{eq:phase_ecc}).
    The black solid horizontal line marks $\eta=0.95$.}
    \label{fig:sigmoid}
\end{figure}

Fig.\ \ref{fig:sigmoid} displays $\eta(h_0)$ and the corresponding sigmoid fits. 
We find $h_0^{95\%}/10^{-25}=1.06$ and 0.93 for scenarios (a) and (b) respectively.
We compute $\sqrt{S_n / ({\rm 1\, Hz^{-1}})}=4.78 \times 10^{-24}$ by invoking \texttt{lalpulsar\_ComputePSD --PSDmthopSFTs=harmsum --PSDmthopIFOs=harmsum --PSDnormByTotalNumSFTs=true} in \texttt{LALSuite} \cite{lalsuite}. 
From Eq.\ (\ref{eq:depth}), we obtain $\mathcal{D}=45.3$ and 51.4 for scenarios (a) and (b) respectively.
That is, $\mathcal{D}$ increases by $\approx 14\%$.

\section{Conclusion}
\label{sec:conclusion}

In this paper, we conduct a pilot search for \cw[s] from \herx[] in LIGO O3 data to prepare for other generic X-ray pulsar searches in the future. 
The 60--$200 \, {\rm Hz}$ search band is relevant astrophysically; it may contain signals from stellar oscillations, for example, although it is not limited to stellar oscillations. 
The frequencies and damping timescales of neutron star oscillations are reviewed briefly in Appendix \ref{appendix:stellar_oscillations}.
Several changes to the \jstat HMM search implementation are introduced in this pilot search. (1) We introduce a second-order ($\propto a_0^2$) and a leading-order eccentricity ($\propto a_0 e$) correction to the signal phase model (see Sections \ref{sec:a0^2_correction}, \ref{sec:ecc_correction} and Appendix \ref{appendix:bessel_sum}). These corrections raise the log likelihood by factors of $\approx 3.6$ and $\approx 2.4$ corrections respectively.
Together, both corrections improve the sensitivity depth $\mathcal{D}$ by $\approx 14\%$.
(2) We quantify the log likelihood mismatch as a function of any displacement $\delta f$ from the center of the subband, and find that $|a_0 \delta f/c|<0.07$ must be satisfied to maintain a maximum mismatch of $10\%$ (see Section \ref{subsec:t_drift} and Appendix \ref{appendix:mismatch_center_band}). Accordingly, we limit the width of a subband to $\Delta f_{\rm sub} \approx 9.5 \times 10^{-3}\, {\rm Hz}$ in this paper, cf.\ $\Delta f_{\rm sub} \approx 0.6 \, {\rm Hz}$ in previously published searches \cite{o3_amxp, psrJ0437andres, scox1}. (3) We propagate the orbital reference time $T_{\rm asc}$ to the middle of the dataset to minimize the correlation of the log likelihood mismatch between $P$ and $T_{\rm asc}$, copying other \cw searches \cite{CrossCorrScox1, Whelan_2023} (see Section \ref{subsec:num_templates} and Appendix \ref{appendix:p_t_prop}). (4) We propagate the electromagnetic prior in the $T_{\rm asc}$-$P$ plane, when $\dot{P}$ exceeds a minimum value (see Section \ref{sec:search_params}). 

We search for \cw[s] from \herx[] in the frequency range 60--$200\, {\rm Hz}$. 
We find 90 above-threshold candidates, 79 of which fail the known lines veto. 
Among the 11 survivors, six fail the single interferometer veto. 
The $T_{\rm obs}/2$ veto further eliminates one candidate.
We are left with four survivors of the five vetoes described in Section \ref{subsec:vetoes}.
One key goal for future work is to implement known efficient ways to place orbital templates to cover the entire parameter domain \cite{Prix_template, Wette_template}.

We obtain a frequentist strain upper limit at 95\% confidence given by $\hUpper = 9.3 \times 10^{-26}$ at $\bar{f}\approx 174.23 \, {\rm Hz}$.
As a comparison, we calculate some astrophysical theoretical upper limits relevant to the stellar oscillations scenario. Firstly, by assuming that all accretion power $L_{\rm acc}=GM_*\dot{M}_*/R_*$ is converted without losses into gravitational radiation, we obtain \cite{Ju_2000}

\begin{align}
    h_0^{\rm acc} \lesssim \, & 2.9 \times 10^{-26}  
\left( \frac{M_*}{1.4 M_{\odot}} \right)^{1/2} \left( \frac{\dot{M}_*}{10^{-9}\, M_{\odot}\, {\rm yr}^{-1}}  \right)^{1/2}  \notag \\
&  \times \left( \frac{10 \,\mathrm{km}}{R_*} \right)^{1/2} 
\left( \frac{100 \,\mathrm{Hz}}{f_{\mathrm{GW}}} \right) \left( \frac{6 \,\mathrm{kpc}}{D} \right), \label{eq:h0_acc}
\end{align}

\noindent where $R_*$, $M_*$, $\dot{M}_*$ and $D$ are the radius, mass, accretion rate and distance to the neutron star respectively. Secondly, by assuming that all of the observed X-ray flux $F_X$ is powered by accretion, and \herx[] is in torque equilibrium, i.e.\ the spin-up torque from accretion equals the spin-down torque from gravitational radiation, we obtain the torque balance upper limit \cite{Zhang_2021} 


\begin{align}
    h_{0}^{\mathrm{torque}} \lesssim & 4.3 \times 10^{-27} 
\left( \frac{R_*}{10 \,\mathrm{km}} \right)^{1/2} 
\left( \frac{1.4 M_{\odot}}{M_*} \right)^{1/4} 
\left( \frac{r_{m}}{10 \,\mathrm{km}} \right)^{1/4} \notag \\
&  \times\left( \frac{F_{X}}{10^{-9} \,\mathrm{erg}\,\mathrm{cm}^{-2}\,\mathrm{s}^{-1}} \right)^{1/2} 
\left( \frac{100 \,\mathrm{Hz}}{f_{\mathrm{GW}}} \right)^{1/2}. \label{eq:h0_tb}
\end{align}

\noindent In Eq.\ (\ref{eq:h0_tb}), $r_m$ is the torque lever arm. Eqs.\ (\ref{eq:h0_acc}) and (\ref{eq:h0_tb}) are comparable informally with
the value of $\hUpper$ obtained in this paper, as well as
the strain upper limits at 95\% confidence achieved in previous HMM searches, e.g.\ $h_0^{95\%} \approx 6\times 10^{-26}$ for Scorpius X-1 at $256.06\, {\rm Hz}$ in O3, assuming an orbital inclination angle of $44\degree$ \cite{scox1},
or $4 \leq \hUpper / 10^{-26} \leq 24$ for the 20 accreting millisecond X-ray pulsars in O3 \cite{o3_amxp}.




\section{Acknowledgments}

This research has made use of data or software obtained from the Gravitational Wave Open Science Center (gwosc.org), a service of the LIGO Scientific Collaboration, the Virgo Collaboration, and KAGRA. This material is based upon work supported by NSF's LIGO Laboratory which is a major facility fully funded by the National Science Foundation, as well as the Science and Technology Facilities Council (STFC) of the United Kingdom, the Max-Planck-Society (MPS), and the State of Niedersachsen/Germany for support of the construction of Advanced LIGO and construction and operation of the GEO600 detector. Additional support for Advanced LIGO was provided by the Australian Research Council. Virgo is funded, through the European Gravitational Observatory (EGO), by the French Centre National de Recherche Scientifique (CNRS), the Italian Istituto Nazionale di Fisica Nucleare (INFN) and the Dutch Nikhef, with contributions by institutions from Belgium, Germany, Greece, Hungary, Ireland, Japan, Monaco, Poland, Portugal, Spain. KAGRA is supported by Ministry of Education, Culture, Sports, Science and Technology (MEXT), Japan Society for the Promotion of Science (JSPS) in Japan; National Research Foundation (NRF) and Ministry of Science and ICT (MSIT) in Korea; Academia Sinica (AS) and National Science and Technology Council (NSTC) in Taiwan.

The computations in this paper were performed on the OzSTAR national facility at Swinburne University of Technology. The OzSTAR program receives funding in part from the Astronomy National Collaborative Research Infrastructure Strategy (NCRIS) allocation provided by the Australian Government, and from the Victorian Higher Education State Investment Fund (VHESIF) provided by the Victorian Government. Parts of this research are supported by the University of Melbourne Research Scholarship, the Robert D Hill Scholarship, and the Australian Research Council (ARC) Centre of Excellence for Gravitational Wave Discovery (OzGrav) (project number CE230100016). We would like to thank Liam Dunn for his assistance in deriving the second-order phase correction, Eq.(\ref{eq:phase_second_order}), and for incorporating Eqs. (\ref{eq:phase_second_order}) and (\ref{eq:phase_ecc}) into the pre-existing HMM search pipeline. We also thank Andr\'es F. Vargas and Liam Dunn for advice regarding the workings of the pipeline, Yutong Bu and Thippayawis (Tong) Cheunchitra for discussions, and Yi Shuen (Christine) Lee for wording and visualization suggestions regarding Fig. \ref{fig:uncertainty_prior}. We thank Karl Wette for discussions regarding Section \ref{subsec:num_templates}. We thank Eric Dong for discussions about the theoretical strain upper limits. 
We thank the anonymous reviewer for their constructive comments.

\appendix

\section{Frequencies and damping timescales of stellar oscillation modes}
\label{appendix:stellar_oscillations}

In this paper, we search for continuous gravitational waves from \herx in the band $0.06 \leq f_{\rm GW} / (1 \, {\rm kHz}) \leq 0.2$. 
The search does not assume a particular emission mechanism. 
However, if the emission arises from stellar oscillation modes, it may not be detectable by current terrestrial interferometers, if the mode oscillates at a frequency outside the search band, or if the damping timescale of the mode is too short.
In this appendix, we briefly review the frequencies and damping timescales of neutron star oscillation modes and comment on the implications for semi-coherent \herx[-]type searches.

Oscillations like $f$- and $p$-modes emit at $f_{\rm GW} \gtrsim 1 \, {\rm kHz}$ for many plausible equations of state (EOS); see for example Fig.\ 4 in Ref.\ \cite{sidery2010} and Table 2 in Ref.\ \cite{McDermott1988}.
$g$-modes, which exist if the adiabatic speed of sound exceeds the equilibrium speed of sound \cite{Reisenegger1992, Gusakov2013}, typically oscillate in the range $10^{-3} \lesssim f_{\rm GW} / (1\, {\rm kHz}) \lesssim 1$, depending on the stellar composition \cite{McDermott1988, Reisenegger1992, Lai1999, Kantor2014, zhao2022, Lozano2022}.
The search band 60--200 Hz in this paper therefore covers some $g$-modes, but not $f$- or $p$- modes.

Theoretical calculations suggest that $f$- and $p$-modes are damped over a timescale in the range $10^{-2} \lesssim \tau / (1 \, {\rm s}) \lesssim 10$ \cite{Reisenegger1992, Benhar2004, zhao2022}. $g$-modes are damped more slowly, with $10^{-1} \lesssim \tau / (1 \, {\rm s}) \lesssim 10^{10}$ \cite{Reisenegger1992, Lai1999, Wei_2020, Zhao_2025}.
In the regime $\tau < T_{\rm drift}$,  the gravitational wave signal is likely to be stochastic (e.g.\ excited continuously by random accretion fluctuations \cite{Kulkarni2009, Romanova2015, dong2025}), like microseismic vibrations on Earth \cite{besedina2023}, rather than coherent.
In the regime $\tau > T_{\rm drift}$, which may apply to $g$-modes under certain circumstances, the signal is likely to be semi-coherent in a continuous-wave sense, as assumed in this paper. 
A third possibility is that gravitational radiation from \herx[-]type binaries may be generated by some mechanism other than stellar oscillations, which has not been modeled theoretically to date, e.g.\ related to superfluid vorticity \cite{Melatos2015,Cheunchitra2024}.

Irrespective of the emission mechanism, the algorithmic enhancements in Sections \ref{sec:a0^2_correction}--\ref{sec:search_implementation} are relevant to any semi-coherent search targeting wide binaries.
The higher-order phase corrections discussed in Sections \ref{sec:a0^2_correction} and \ref{sec:ecc_correction} become more significant, as $f_{\rm GW}$, $a_0$ and $e$ increases.
The approximation that $f_{\rm GW}$ coincides with the subband central frequency $\bar{f}$ grows worse as $a_0$ increases, hence requiring narrower subband widths to compensate, as highlighted in Section \ref{subsec:t_drift}.
In the longer run, it will be important to search for higher-frequency modes.
However, the computational cost of searching for continuous gravitational waves from neutron stars in a circular and eccentric binary orbit scales as $\propto f_{\rm GW}^3$ and $\propto f_{\rm GW}^5$ respectively \cite{leaciPrix2015}.
We emphasize that the main goal of this pilot search is to validate the enhancements introduced in Sections \ref{sec:a0^2_correction}--\ref{sec:search_implementation} in real data, so we choose to limit our search range to 60--200 Hz to meet this goal with the limited computational resources at our disposal.

\section{Propagation of $T_{\rm asc}$ and $P$ to the midpoint of O3}
\label{appendix:p_t_prop}

In this appendix, we justify propagating $T_{\rm asc}$ and $P$ to the midpoint of O3 by investigating the shear in the mismatch ellipse, defined by $\mu\leq \mu_{\rm max}$. 

In a single coherent timestep, the shear is proportional to $T_{\rm asc} - T_{\rm mid}$, where $T_{\rm mid}$ is the midpoint of the coherent timestep \cite{leaciPrix2015}. We first propagate $\bar{T}_{\rm asc, \, ref}$ and $\bar{P}_{\rm ref}$ to the start of O3, viz.\ $T_{\rm O3, \, start}$, and inject a signal into Gaussian noise using these propagated $T_{\rm asc}$ and $P$ values, with $f_{\rm GW}=100 \, {\rm Hz}$, $h_0=10^{-24}$, $\sqrt{S_n}=4\times 10^{-24}\, {\rm Hz^{-1/2}}$, $T_{\rm drift}=10\, {\rm days}$, $T_{\rm obs}=360\, {\rm days}$ and all other parameters taking the central values listed in Table \ref{tab:params}. We construct a two-dimensional $201 \times 201$ template grid, centered on the $\bar{T}_{\rm asc, \, ref}$ and $\bar{P}_{\rm ref}$ values propagated to $T_{\rm O3, \, start}$. The grid points are separated by $(\Delta T_{\rm asc}, \Delta P)/200$, calculated using Eqs. (\ref{eq:template_spacing_t}) and (\ref{eq:template_spacing_p}) with $\mu_{\rm max}=0.1$. We then execute the HMM algorithm, described in Section \ref{sec:hmm}, and calculate the mismatch $\mu$, defined by $1-\mathcal{L}/\mathcal{L}_{\rm inj}$ where $\mathcal{L}_{\rm inj}$ is the log likelihood obtained at the center of the grid, i.e.\ the injected values. We then repeat the exercise with $T_{\rm asc}$ propagated to the middle of O3, $T_{\rm O3, \, mid}$.

\begin{figure}[htbp!]
    \centering
    \begin{subfigure}[b]{0.99\linewidth}
        \centering
        \includegraphics[width=\linewidth]{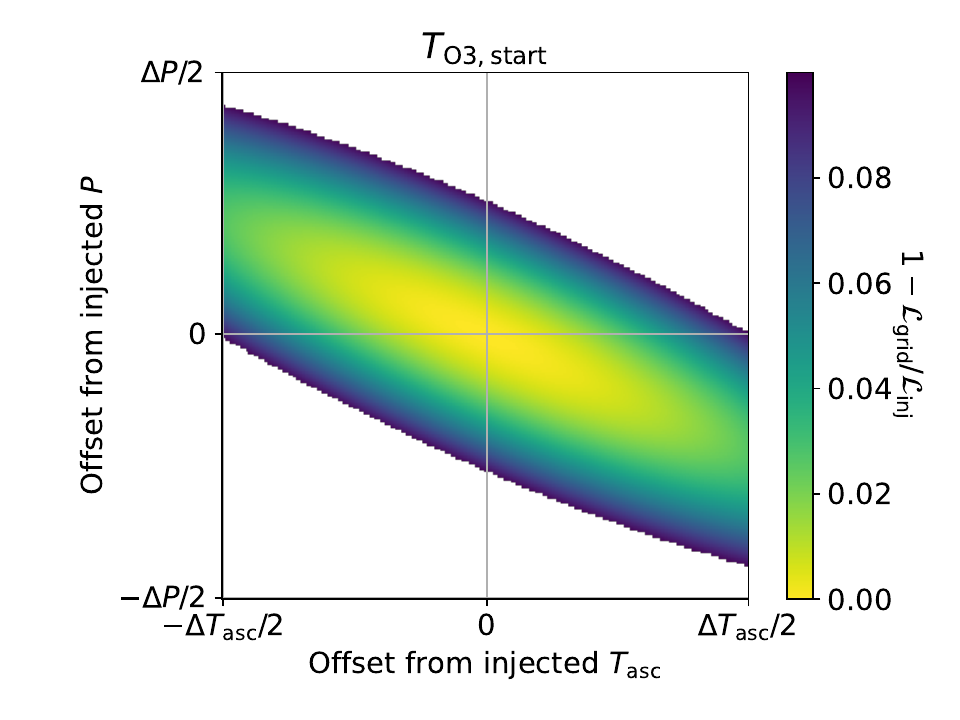}
        \caption{}
        \label{subfig:sheared_ellipse}
    \end{subfigure}
    \begin{subfigure}[b]{0.99\linewidth}
        \centering
        \includegraphics[width=\linewidth]{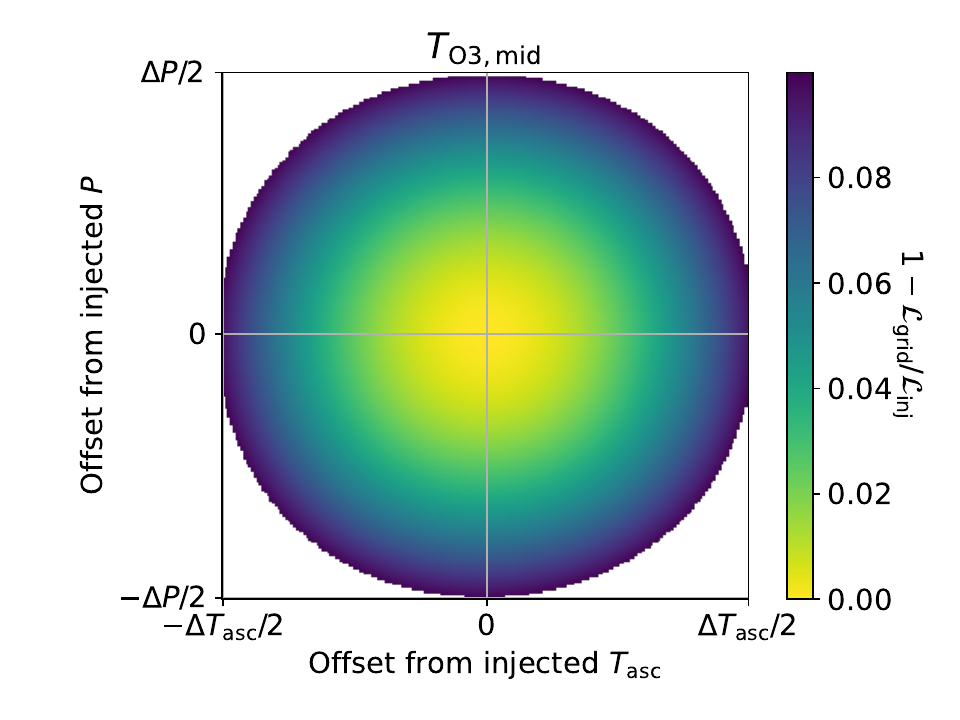}
        \caption{}
        \label{subfig:unsheared_ellipse}
    \end{subfigure}
    \caption{Heatmap of log likelihood mismatch $\mu$ as a function of the offset from the injected $T_{\rm asc}$ and $P$ values, if the central $T_{\rm asc}$ grid value is propagated to (a) $T_{\rm O3, \, start}$ or (b) $T_{\rm O3, \, mid}$. The uncolored white regions contain template sites with $\mu>\mu_{\rm max}$.}
    \label{fig:ellipse}
\end{figure}

Figs. \ref{subfig:sheared_ellipse} and \ref{subfig:unsheared_ellipse} show a heatmap of the mismatch as a function of the offset from the injected $T_{\rm asc}$ and $P$ values, for central search parameters propagated to $T_{\rm O3, \, start}$ and $T_{\rm O3, \, mid}$ respectively. The uncolored white regions mark grid points with $\mu > 0.1$. Fig. \ref{subfig:sheared_ellipse} (\ref{subfig:unsheared_ellipse}) shows that the mismatch ellipse is sheared (unsheared), if we construct a template grid from parameters propagated to $T_{\rm O3, \, start}$ ($T_{\rm O3, \, mid}$). Hence, to simplify template placement, we opt to propagate the orbital parameters to $T_{\rm O3, \, mid}$.

\section{Deriving the phase model}
\label{appendix:phase_model}

In this appendix, we derive the second-order phase model corrections $\propto a_0^2$ and $\propto a_0 e$ related to the semi-major axis and eccentricity. The corrections are given by Eqs. (\ref{eq:phase_second_order}) and (\ref{eq:phase_ecc}) respectively.

The gravitational wave phase from a neutron star in a binary, can be written in terms of the retarded time $t_{\rm r}$ as

\begin{equation}
\label{eq_app:phase_model_def}
    \Phi(t_{\rm r}) = 2\pi f_{\rm GW} t_{\rm r}.
\end{equation}

\noindent The gravitational wave front arrives at the Solar System barycenter at $t_{\rm s}$. Hence, $t_{\rm s}$ must be related to $t_{\rm r}$ through the light travel time to evaluate Eq. (\ref{eq_app:phase_model_def}). Ignoring relativistic corrections, one has

\begin{equation}
\label{eq_app:t_r_def}
    t_{\rm r}=t_{\rm s}-\frac{\left|\vec{x}_{\rm b}+\vec{x}_{\mathrm{ns}}(t_{\rm r})\right|}{c},
\end{equation}

\noindent where $\vec{x}_{\rm b}$ is the displacement from the Solar System barycenter to one focus of the ellipse describing the binary orbit, and $\vec{x}_{\mathrm{ns}}$ is the displacement from the focus to the neutron star. Assuming $|\vec{x}_{\mathrm{ns}}| \ll |\vec{x}_{\rm b}|$ and Taylor expanding, we obtain

\begin{align}
    \left|\vec{x}_{\rm b}+\vec{x}_{\mathrm{ns}}(t_{\rm r})\right| &\approx |\vec{x}_{\rm b}| \left[ 1+\frac{1}{2} \frac{ |\vec{x}_{\mathrm{ns}}|^2 + 2\vec{x}_{\rm b}\cdot \vec{x}_{\mathrm{ns}}}{|\vec{x}_{\rm b}|^2} \right. \notag \\
    & \quad - \left. \frac{1}{8} \left(\frac{ |\vec{x}_{\mathrm{ns}}|^2 + 2\vec{x}_{\rm b}\cdot \vec{x}_{\mathrm{ns}}}{|\vec{x}_{\rm b}|^2}  \right)^2 \right].  \label{eq_app:x_binom}
\end{align}

\noindent For a Keplerian orbit, one has \cite{Roy2004}


\begin{align}
    |\vec{x}_{\mathrm{ns}}| &= a(1-e\cos E), \label{eq_app:x_ns_norm} \\
    \vec{x}_{\rm b} \cdot \vec{x}_{\mathrm{ns}} &= a_0 D B(E)\label{eq_app:x_dot_norm} \\
    B(E)& = (\cos E-e)\sin\omega   \notag \\
    & \quad + \sqrt{1-e^{2}} \cos\omega\sin E, \label{eq_app:b_def} 
\end{align}

\noindent with $D=|\vec{x}_{\rm b}|$, where $a=a_0/\sin\iota$ is the semi-major axis of the binary and $E$, termed the eccentric anomaly, parametrizes the orbit. Substituting Eqs. (\ref{eq_app:x_ns_norm})--(\ref{eq_app:b_def}) into Eq. (\ref{eq_app:x_binom}) and discarding terms of order $D^{-2}$ and smaller, we find

\begin{align}
    \left|\vec{x}_{\rm b}+\vec{x}_{\mathrm{ns}}(t_{\rm r})\right| &\approx D + \frac{a^2\left(1-e \cos E\right)^2}{D} + a_0 B(E) \notag \\
    &\quad + \frac{a_0^2 B(E)^2}{2D}. \label{eq_app:x_substituted}
\end{align}

\noindent $E$ is an implicit function of $t_{\rm r}$ through Kepler's equation

\begin{align}
    M(t_{\rm r}) &= \Omega (t_{\rm r} - T_p) \\
    &= E - e \sin E, \label{eq_app:kepler}
\end{align}

\noindent where $T_p=T_{\rm asc} + \omega / \Omega$ is the time of periapsis, and $M(t_{\rm r})$ is termed the mean anomaly. Eq. (\ref{eq_app:kepler}) can be expanded as \cite{Brumberg2012}

\begin{equation}
\label{eq_app:E_series}
    E=M+2\sum_{n=1}^{\infty}\frac{J_n(ne)}{n}\sin(nM),
\end{equation}

\noindent where $J_n$ is the Bessel function of order $n$ of the first kind \cite{mathSpecialFunction}.

We are now ready to Taylor expand Eqs. (\ref{eq_app:x_substituted})--(\ref{eq_app:E_series}) to second order in $e\ll 1$. Before doing so, we note that Eq. (\ref{eq_app:x_substituted}) is Taylor expanded already to second order in $a/D \ll 1$. The expansions in $e$ and $a/D$ are independent, so we carry out both of them independently to second order, meaning that we keep cubic and quartic terms $\propto e a_0^2/D$ and $e^2 a_0^2/D$ (among others) until the end of the calculation, when we compare their relative magnitudes\footnote{Recall that the expansion in $a/D$ in Eq.(\ref{eq_app:x_binom}) is performed on a term of order $D\sqrt{1+a/D}$ $\propto a^2/D$ to second order.}. For a typical source like Her X-1, the magnitudes are ordered according to $a_0/c \gg a_0^2 \Omega /c^2 \sim ea_0/c \gg a_0^2/cD$, as discussed below.

Eq. (\ref{eq_app:E_series}) expands to second order in $e$ as

\begin{equation}
\label{eq_app:E_approx}
    E \approx M + e \sin M + \frac{e^2 \sin (2M)}{2}.
\end{equation}

\noindent Substituting Eq. (\ref{eq_app:E_approx}) into (\ref{eq_app:x_substituted}), Taylor expanding to second order in $e$ and discarding constant terms (which are equivalent to shifting the time origin), we obtain

\begin{align}
    & \left|\vec{x}_{\rm b}+\vec{x}_{\mathrm{ns}}(t_{\rm r})\right| \notag \\
    &\approx a_0 \sin (M+\omega ) -\frac{a_0^2 \cos (2 M+ 2\omega )}{4 D} \notag \\
    & \quad + e \left[\frac{1}{2} a_0 \sin (2
   M+\omega )-\frac{2 a^2 \cos (M)}{D}-\frac{a_0^2 \cos (M)}{2 D} \right. \notag \\
   & \qquad \left. -\frac{a_0^2 \cos (3 M+2 \omega )}{4
   D} +\frac{3 a_0^2 \cos (M+2 \omega )}{4 D}\right] \notag \\
   &\quad +
    e^2 \left[\frac{3}{8} a_0 \sin (3 M+\omega ) -\frac{1}{8} a_0 \sin (M-\omega ) \right. \notag \\
   & \qquad -\frac{1}{2} a_0 \sin (M+\omega
   ) -\frac{a^2 \cos (2 M)}{2 D}-\frac{a_0^2 \cos (2 M)}{8 D}  \notag \\
   & \qquad \left. +\frac{5 a_0^2 \cos (2M+2\omega )}{8 D} -\frac{a_0^2 \cos (4 M+2\omega )}{4
   D}  \right]. 
\end{align}

\noindent Writing Eq. (\ref{eq_app:t_r_def}) as

\begin{equation}
    t_{\rm r}-t_{\rm s}= T(t_{\rm r}),
\end{equation}

\noindent one can expand $T(t_{\rm r})$ to second order in a Taylor series around $t_{\rm r} = t_{\rm s}$ as

\begin{equation}
\label{eq_app:tr_ts_taylor}
    t_{\rm r}-t_{\rm s} \approx T(t_{\rm s})+T'(t_{\rm s})(t_{\rm r}-t_{\rm s})+\frac{1}{2}T''(t_{\rm s})(t_{\rm r}-t_{\rm s})^2.
\end{equation}

\noindent Eq. (\ref{eq_app:tr_ts_taylor}) is a quadratic in $t_{\rm r}-t_{\rm s}$. Taking the negative root\footnote{The expansion of the positive root contains divergent terms such as $1/\cos(M+\omega)$.} and keeping all terms up to order $e^p a_0^q/D^{q-1}$, with $0 \leq p, q \leq 2$, we obtain

\begin{align}
    &t_{\rm r}-t_{\rm s} \notag \\
    &\approx -\frac{a_0 \sin (M+\omega )}{c}+\frac{a_0^2 \Omega  \sin (2M+2\omega )}{2 c^2} \notag\\
   &\quad+\frac{a_0^2 \cos (2M+2\omega
   )}{4 c D} \notag \\
   &\quad +e \left[ -\frac{a_0 \sin (2
   M+\omega )}{2 c} +\frac{2 a^2 \cos (M)}{c D } \right. \notag \\
   &\qquad + \frac{3 a_0^2
   \Omega  \sin (3 M+2 \omega )}{4 c^2}-\frac{a_0^2 \Omega
    \sin (M)}{4 c^2} \notag \\
    &\qquad -\frac{3 a_0^2 \cos (M+2 \omega )}{4 c
   D} +\frac{a_0^2 \cos (M)}{2 c D} \notag \\
   &\qquad \left. +\frac{a_0^2 \cos (3 M+2 \omega )}{4 c
   D} \right] \notag \\
   &\quad +e^2 \left[\frac{a_0 \sin (M-\omega )}{8
   c} +\frac{a_0 \sin (M+\omega )}{2 c} \right. \notag \\
   &\qquad -\frac{3 a_0
   \sin (3 M+\omega )}{8 c} + \frac{a_0^2 \cos (2
   M)}{8 c D} \notag \\
   & \qquad -\frac{a_0^2 \Omega  \sin (2M+2\omega )}{2
   c^2}+\frac{a_0^2 \Omega  \sin ( 4 M+2\omega
   )}{c^2} \notag \\
   &\qquad -\frac{a_0^2 \Omega  \sin (2 M)}{2
   c^2}+\frac{a^2 \cos (2 M)}{2 c D } \notag \\
   &\qquad -\frac{5
   a_0^2 \cos (2M+2\omega )}{8 c D}+\frac{a_0^2
   \cos (4 M+2\omega )}{4 c D} \notag \\ 
   &\qquad  . \label{eq_app:tr_ts_root}
\end{align}

\noindent In Eq. (\ref{eq_app:tr_ts_root}), $M$ is now a function of $t_{\rm s}$. 

For \herx[,] the $a_0^2$ and $a_0 e$ terms are not negligible. One has $a_0/c \sim 10\, {\rm s}$, $\Omega \sim 10^{-5} \, {\rm s}^{-1}$, $e \sim 10^{-4}$ and $ a_0/D\sim 10^{-12}$. Note that every term in Eq. (\ref{eq_app:tr_ts_root}) is a sinusoid multiplied by some power of $e$ and either $a_0/c \sim 10\, {\rm s}$, $a_0^2 \Omega / c^2 \sim 10^{-3} \, {\rm s}$ or $a_0^2/cD \sim 10^{-11} \, {\rm s}$. For $f_{\rm GW}\sim 10^2\,$Hz, if the maximum acceptable phase error is 0.01 cycles, the surviving terms in Eq. (\ref{eq_app:tr_ts_root}) are

\begin{align}
    t_{r}-t_{\rm s}&\approx-\frac{a_0 \sin (M+\omega
   )}{c} + \frac{a_0^2 \Omega  \sin (2
   M+2\omega )}{2 c^2} \notag \\
   &\quad- \frac{a_0 e \sin (2 M+\omega )}{2
   c}. \label{eq_app:final_phase}
\end{align}

\noindent The first, second and third terms on the right-hand side of Eq. (\ref{eq_app:final_phase}) are Eqs. (\ref{eq:phase_first_order}), (\ref{eq:phase_second_order}) and (\ref{eq:phase_ecc}) respectively after multiplying by the factor $2\pi f_{\rm GW}$.

\texttt{lalpulsar\_Makefakedata\_v5} \cite{lalsuite} is the standard LALSuite tool used to simulate \cw signals. Hence, it is important that Eq. (\ref{eq_app:final_phase}) is consistent with the phase model assumed in \texttt{Makefakedata}, which assumes a slightly different equation for the time delay $t_{\rm r}-t_{\rm s}$, viz.

\begin{equation}
\label{eq_app:mfd_time_delay}
    t_{\rm r} - t_{\rm s} = -a_0 B(E).
\end{equation}

\noindent Eq. (\ref{eq_app:mfd_time_delay}) matches Eq. (\ref{eq_app:x_substituted}), when all terms involving $D$ are dropped. \texttt{Makefakedata} numerically inverts Eq. (\ref{eq_app:kepler}) to obtain $E$ as a function of $t_{\rm r}$, instead of expanding as in Eq. (\ref{eq_app:E_series}). If the procedure of going from Eq. (\ref{eq_app:x_substituted}) to (\ref{eq_app:tr_ts_root}) follows Eq. (\ref{eq_app:mfd_time_delay}) instead, the same expression as Eq. (\ref{eq_app:tr_ts_root}) is obtained without the $D^{-1}$ terms. The $D^{-1}$ terms are negligible in the \herx regime, so the phase model assumed in \texttt{Makefakedata} is consistent with Eq. (\ref{eq_app:final_phase}) for the purposes of this paper.

\section{Truncating the Bessel sum in the \jstat filter}
\label{appendix:bessel_sum}

In this appendix, we discuss to what order to truncate the Bessel sum in the \jstat[,] for the second-order ($\propto a_0^2$) and leading-order ($\propto a_0 e$) eccentricity corrections discussed in Sections \ref{sec:a0^2_correction} and \ref{sec:ecc_correction} respectively. With the inclusion of Eqs.\ (\ref{eq:phase_second_order}) and (\ref{eq:phase_ecc}) into the phase model, the \jstat contains a product of three sums after expanding using the Jacobi-Anger identity, viz.\ \cite{jstat}

\begin{align}
    \mathcal{J} \propto & \sum_{s_1=-M_1}^{M_1} J_{s_1}\left(\frac{2\pi \bar{f} a_0}{c} \right) \sum_{s_2=-M_2}^{M_2} J_{s_2}\left(\frac{\pi \bar{f} a_0^2\Omega}{c^2}\right) \notag \\ 
    & \times \sum_{s_e=-M_e}^{M_e} J_{s_e} \left(\frac{\pi \bar{f} a_0 e}{c} \right). \label{eq_app:bessel}
\end{align}

\noindent Traditionally, the leading-order ($\propto a_0$) Bessel sum is truncated to $M_1={\rm ceil} (2\pi \bar{f} a_0/c)$ \cite{jstat}. Hence, one might expect the corresponding truncation orders of the $a_0^2$ and $a_0 e$ Bessel sums to be $M_2={\rm ceil} (\pi \bar{f} a_0^2\Omega/c^2)$ and $M_e={\rm ceil} (\pi \bar{f} a_0 e/c)$ respectively. However, the situation is more complicated, as discussed below.

For $60 \leq f_{\rm GW}/{\rm Hz} \leq 200$, we have $2\leq M_2,M_e \leq 5$, cf.\ $M_1 \gtrsim 10^3$. Consider a scenario where $M_2$ is on the verge of incrementing to the next integer, e.g. $\bar{f}=85.63\, {\rm Hz}$ and the central \herx parameters listed in Table \ref{tab:params}. With these values, we obtain $M_2=2$. However, if we recompute $M_2$ assuming $\bar{a}_0+\Delta a_0/2$, with $\Delta a_0=1.6507\times 10^{-3}\,$lt-s computed from Eq.\ (\ref{eq:template_spacing_first}), we obtain $M_2=3$. We investigate the mismatch $1-\mathcal{L}/\mathcal{L}_{\rm inj}$, where $\mathcal{L}_{\rm inj}$ is the log likelihood obtained using the injected parameters, as we step from $\bar{a}_0$ to $\bar{a}_0 + \Delta a_0 / 2$ by injecting a signal into Gaussian noise using central \herx parameters, $f_{\rm GW}=85.63 \, {\rm Hz}$, $h_0=10^{-24}$, $\sqrt{S_n}=4\times 10^{-24}\, {\rm Hz^{-1/2}}$ and $T_{\rm drift}=T_{\rm obs}=10\, {\rm days}$.

\begin{figure}[htbp!]
    \centering
    \includegraphics[width=\linewidth]{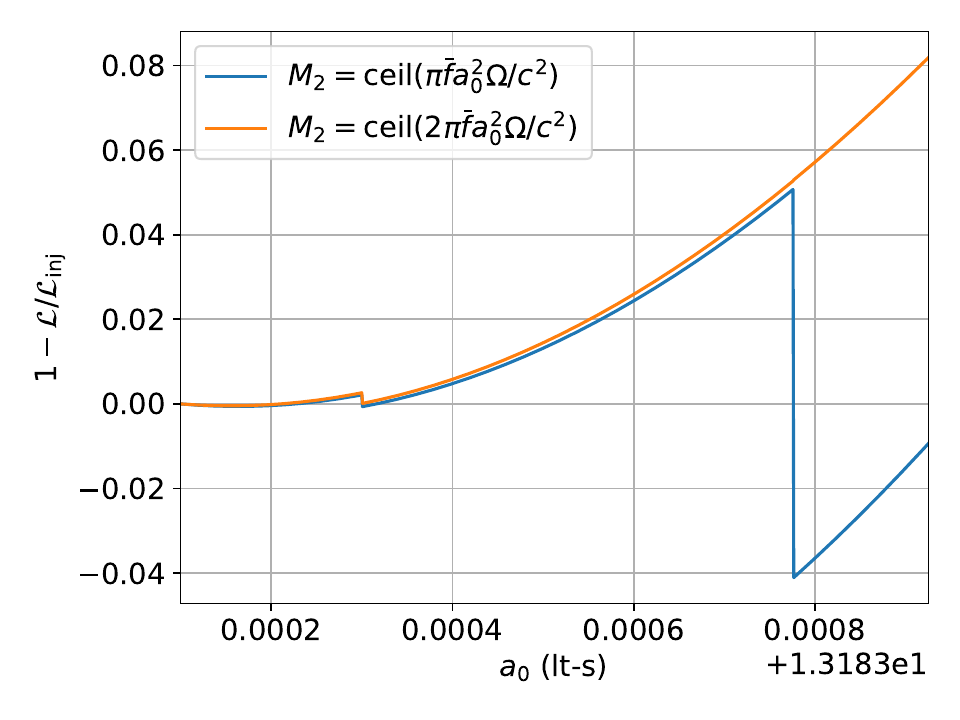}
    \caption{Log likelihood mismatch as a function of $a_0$, ranging from $\bar{a}_0$ to $\bar{a}_0+\Delta a_0/2$. The injected $a_0$ is $\bar{a}_0$. See Appendix \ref{appendix:bessel_sum} for a description of the aim and method of the mismatch test. The sudden drops at $a_0 = 13.1833$ and $13.1838 \,$lt-s arise from the interplay between ${\cal L}$ and the truncation order of the Bessel sums in Eq. (\ref{eq_app:bessel}).}
    \label{fig_app:teeth_mismatch}
\end{figure}

Fig.\ \ref{fig_app:teeth_mismatch} displays the mismatch as a function of $a_0$, by truncating at $M_2={\rm ceil} (\pi \bar{f} a_0^2\Omega/c^2)$ (blue curve) or doubling the truncation order to $M_2={\rm ceil} (2\pi \bar{f} a_0^2\Omega/c^2)$ (orange curve). Without doubling $M_2$, we see a noticeable drop in the mismatch at $a_0=13.1838\,$lt-s. This sharp drop coincides with the transition between $M_2=2$ and $M_2=3$, which implies that one achieves an appreciable gain in $\mathcal{L}$ by keeping higher orders in the Bessel sum. If we double $M_2$, the drop disappears. A similar behavior occurs for the eccentricity correction. Hence, we opt to double $M_2$ and $M_e$ for this paper.

Fig\ \ref{fig_app:teeth_mismatch} also reveals a smaller drop for both curves at $a_0=13.1833\,$lt-s, which coincides with the transition between $M_1=7093$ and $M_1=7094$. The drop is less significant compared to transitioning from $M_2=2$ to $M_2=3$, because adding a term, when the sum already contains many terms, makes a small change. Hence there is no need to double $M_1$ in this paper \cite{jstat}.

\section{Mismatch when the signal is not in the center of the subband}
\label{appendix:mismatch_center_band}

As explained in Section \ref{subsec:t_drift}, if a gravitational wave signal lies $\delta f$ away from the center of the subband $\bar{f}$, then there is an error amounting to $\sim a_o \delta f / c$ cycles in the phase model. In this appendix, we quantify the mismatch, $\mufreq= 1 -\mathcal{L}/\mathcal{L}_{\rm mid}$,  with ${\mathcal L}_{\rm mid} = {\mathcal L}(\delta f = 0)$, as a function of $a_0 \delta f/c$.

\begin{figure}[!htbp]
    \centering
    \includegraphics[width=0.95\linewidth]{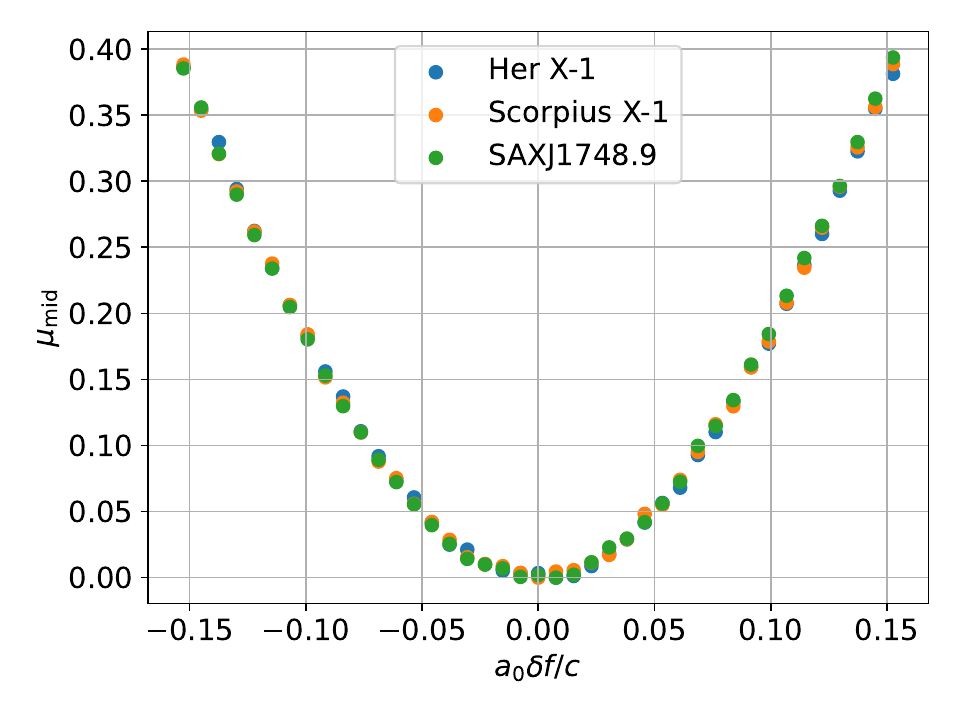}
    \caption{Mismatch $\mufreq$ as a function of $a_0 \delta f/c$ for \herx[,] Scorpius X-1 and SAX J1748.9–2021 (blue, orange and green dots respectively). See Table \ref{tab_app:mismatch_targets} and Appendix \ref{appendix:mismatch_center_band} for a description of the aim and method of the mismatch test.}
    \label{fig_app:mismatch_freq_center}
\end{figure}

We repeat the test in Fig. \ref{fig:mismatch_freq_center} using $\bar{f}=200\, {\rm Hz}$ and the central parameters of Scorpius X-1 \cite{PEGSIV_scox1} and SAX J1748.9--2021 \cite{SAXJ1748.9}, as tabulated in Table \ref{tab_app:mismatch_targets}. These targets are chosen, because their $a_0$  values are smaller than for \herx[,] and they have been targeted previously in HMM searches using the \jstat[] \cite{scox1, o3_amxp}. For each target, we inject signals at the same set of $a_0 \delta f / c$ values. Fig. \ref{fig_app:mismatch_freq_center} shows the mismatch for \herx[,] Scorpius X-1 and SAX J1748.9--2021 as a function of $a_0 \delta f / c$. The three objects all display similar trends. We find that $\mufreq=0.1$ corresponds to $|a_0 \delta f / c| \approx 0.07$ for all three targets, or $\delta f\approx5.3$, 30 and 181 mHz for \herx[,] Scorpius X-1 and SAX J1748.9–2021 respectively.

\begin{table}[htbp!]
    \centering
    \caption{Central (without error bars) orbital parameters of Scorpius X-1 \cite{PEGSIV_scox1} and SAX J1748.9--2021 \cite{SAXJ1748.9}. Both targets are assumed to have circular orbits in the mismatch test in Appendix \ref{appendix:mismatch_center_band}.}
    \begin{tabular}{@{}ccc@{}}
        \toprule
        Parameter & Scorpius X-1 & SAX J1748.9--2021 \\ \midrule
        $\alpha$ & 16h19m55.085s & 17h48m52.161s \\
        $\delta$ & $-15\degree$ 38$'$ 24.9$''$ & $-20\degree$ 21$'$ 32.406$''$ \\
        $a_0$ (lt-s) & 2.35 & 0.38757 \\
        $P$ (s) & 68023.919 & 31555.300 \\
        $T_{\rm asc}$ (GPS time) & 1078153676 & 1109500772.5
        \\
        \bottomrule
    \end{tabular}
    \label{tab_app:mismatch_targets}
\end{table}

\section{Interpolating the threshold as a function of frequency}
\label{appendix:threshold}

In this paper, we search 14766 frequency subbands. Setting a threshold $\mathcal{L}_{\rm th}$ at every subband is computationally expensive. Instead, we opt to set $\mathcal{L}_{\rm th}$ at a sample of 29 subbands, whose central frequencies $\bar{f}$ are separated by approximately $5 \, {\rm Hz}$, and then interpolate between them. For simplicity, we interpolate linearly between subbands. We ensure that each subband does not intersect any known lines \cite{O3lines}. We generate 2000 random off-target noise realizations for each subband and estimate $\mathcal{L}_{\rm th}$ using the percentile method, according to the procedure described in Appendix A in Ref.\ \cite{o3_amxp}.

Fig.\ \ref{fig_app:threshold} displays as points the sample of 29 $\bar{f}$ values and the corresponding percentiles. 
Because the exact percentile falls between two $\mathcal{L}$ values, we take the smaller and larger $\mathcal{L}$ values as the lower (blue crosses) and upper (orange crosses) limits respectively of $\mathcal{L}_{\rm th}$. 
As done in Ref.\ \cite{o3_amxp}, we use the lower limit as the threshold in this search. 
If we use the upper limit instead, we only recover the $109.47\, {\rm Hz}$ in Table \ref{tab:candidates} after applying all the vetoes described in Section \ref{subsec:vetoes}.

None of the subbands used to explicitly set thresholds contain any of the candidates in Table \ref{tab:candidates}.

\begin{figure}[htbp!]
    \centering
    \includegraphics[width=\linewidth]{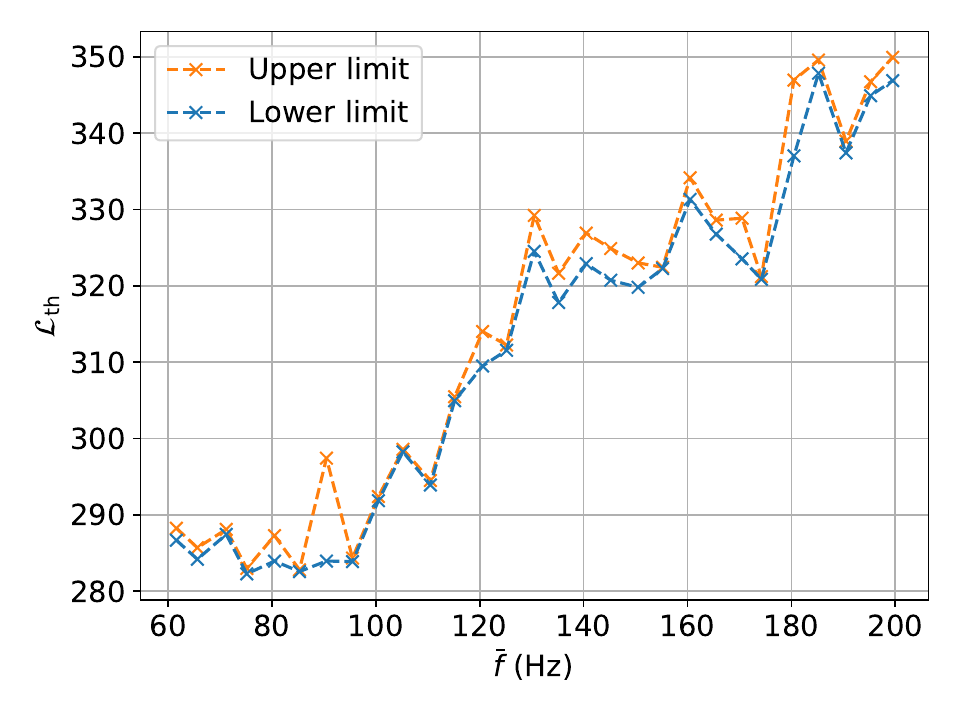}%
    \caption{Lower (blue crosses) and upper (orange crosses) limits of the threshold $\mathcal{L}_{\rm th}$ versus subband central frequency $\bar{f}$, computed using 2000 off-target noise realizations for each of the 29 subbands, and the percentile method.}
    \label{fig_app:threshold}
\end{figure}

\section{Comparison with the binary \fstat[]}
\label{appendix:binary_fstat}

In this appendix, we verify that the \jstat with $T_{\rm SFT}=1800 \, {\rm s}$ produces qualitatively similar $\mathcal{L}$ values as the binary \fstat \cite{fstat, lalsuite}. 

At $f_{\rm GW}=200 \, {\rm Hz}$, a representative test frequency, Eq.\ (\ref{eq:max_sft}) shows that the binary \fstat requires $T_{\rm SFT} \leq 444 \, {\rm s}$ for the \herx[-]like parameters listed in Table \ref{tab:params}.
We repeat the test shown in Fig.\ \ref{fig:ecc_correction}, except with $f_{\rm GW}=200 \, {\rm Hz}$.
We run the HMM algorithm, described in Section \ref{sec:hmm}, with both the $a_0^2$ and $a_0 e$ corrections presented in Sections \ref{sec:a0^2_correction} and \ref{sec:ecc_correction} respectively, once using $T_{\rm SFT}=1800 \, {\rm s}$ and once using $T_{\rm SFT}=400 \, {\rm s}$, with associated log-likelihoods $\mathcal{L}_{\rm 1800s,\mathcal{J}}$ and $\mathcal{L}_{\rm 400s,\mathcal{J}}$ respectively.
We also run the HMM algorithm a third time with $T_{\rm SFT}=400 \, {\rm s}$, using the binary \fstat[,] and record the log likelihood as $\mathcal{L}_{\rm 400s,\mathcal{F}}$.

\begin{figure}[htbp!]
    \centering

\includegraphics[width=0.99\linewidth]{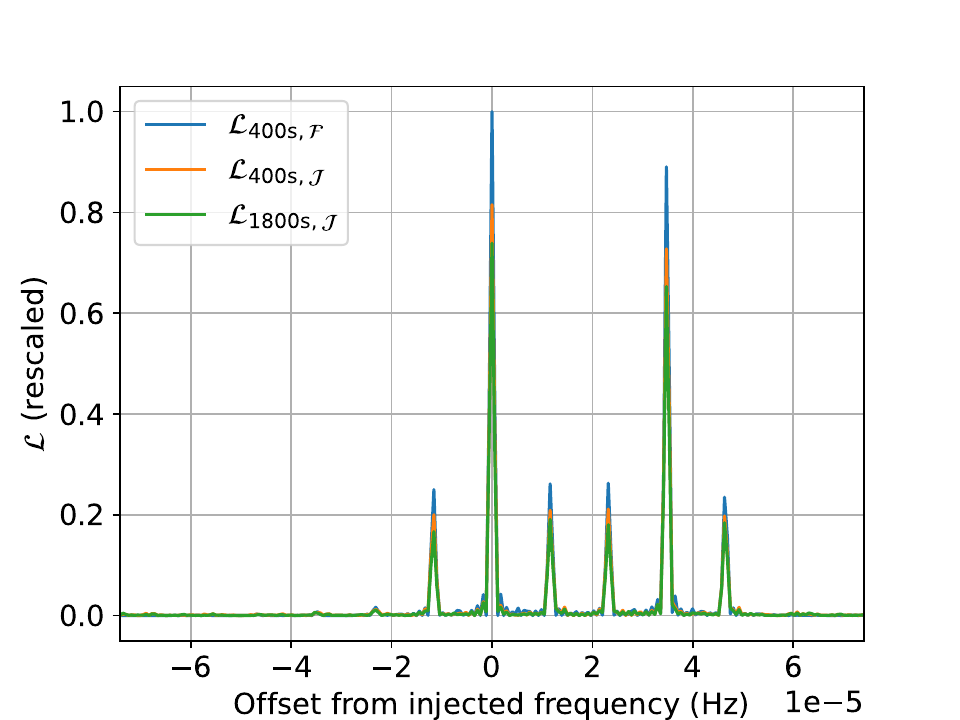}
    \caption{Log likelihood of the most likely path, $\mathcal{L}$, as a function of search frequency offset (units: $10^{-5} \, {\rm Hz}$), using $T_{\rm SFT}=1800 \, {\rm s}$ and the \jstat (green curve), $T_{\rm SFT}=400 \, {\rm s}$ and the \jstat (orange curve), and $T_{\rm SFT}=400 \, {\rm s}$ and the binary \fstat (blue curve). 
    The injection parameters are identical to Fig.\ \ref{fig:ecc_correction} of the paper, except for $f_{\rm GW}=200\, {\rm Hz}$.}
    \label{fig_app:sft_length}
\end{figure}

Fig.\ \ref{fig_app:sft_length} shows $\mathcal{L}_{\rm 1800s,\mathcal{J}}$, $\mathcal{L}_{\rm 400s,\mathcal{J}}$ and $\mathcal{L}_{\rm 400s,\mathcal{F}}$ as functions of the offset from the injected frequency. 
We find that $\mathcal{L}_{\rm 1800s,\mathcal{J}}$, $\mathcal{L}_{\rm 400s,\mathcal{J}}$ and $\mathcal{L}_{\rm 400s,\mathcal{F}}$ all achieve their maximum at the same frequency bin, which coincides with the injected frequency.
All three log-likelihoods have six visible peaks that align at the same frequency bins. 
We obtain $\max(\mathcal{L}_{\rm 1800s,\mathcal{J}})/\max(\mathcal{L}_{\rm 400s,\mathcal{J}})=0.91$,  $\max(\mathcal{L}_{\rm 400s,\mathcal{J}})/\max(\mathcal{L}_{\rm 400s,\mathcal{F}})=0.82$ and $\max(\mathcal{L}_{\rm 1800s,\mathcal{J}}) / \max(\mathcal{L}_{\rm 400s,\mathcal{F}})=0.74$. 
In other words, we tolerate a sensitivity penalty of $\lesssim 26\%$ when using the \jstat (with respect to the binary \fstat[)] as a trade-off to access computational savings.

\section{Manual inspection of the PSD near $T_{\rm obs}/2$ veto survivors}
\label{appendix:psd_inspection}

\begin{figure*}[htbp!]
    \centering

\includegraphics[width=0.45\linewidth]{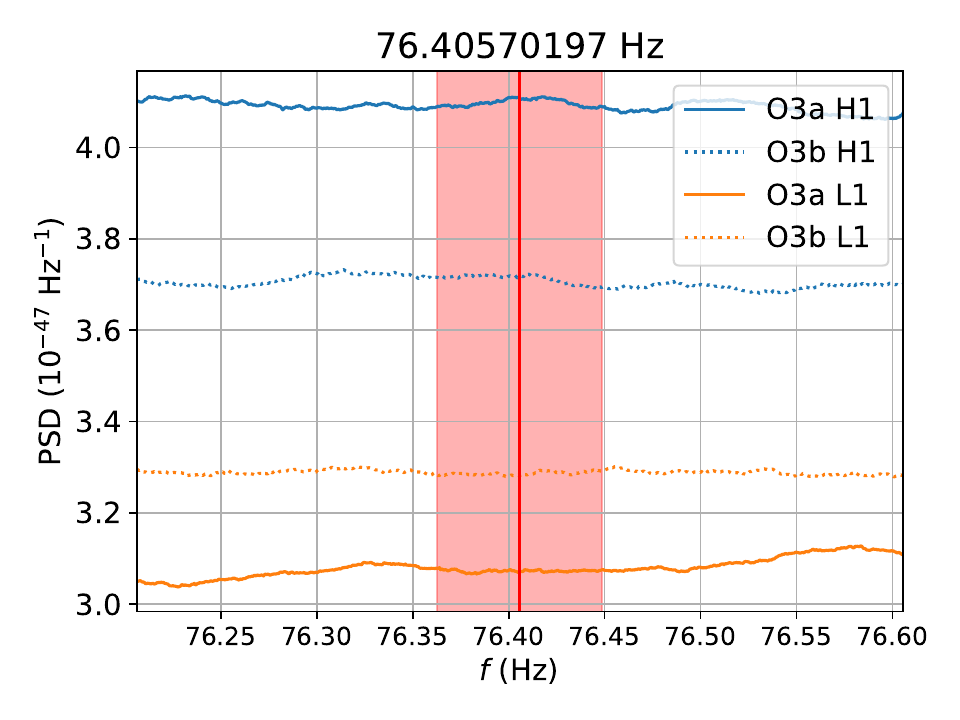}
\includegraphics[width=0.45\linewidth]{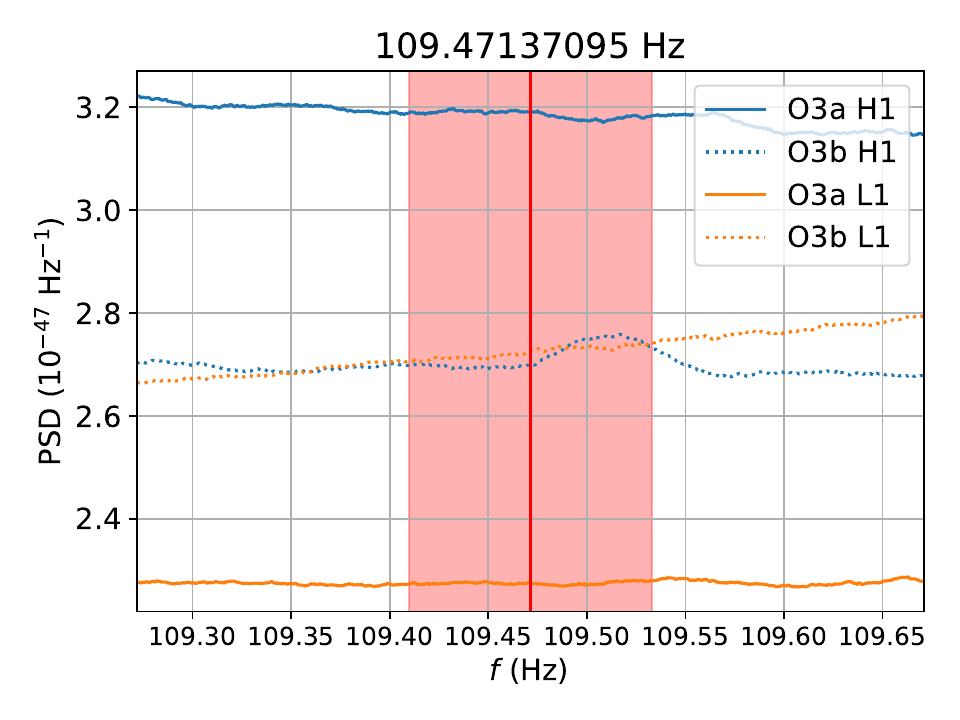}
\includegraphics[width=0.45\linewidth]{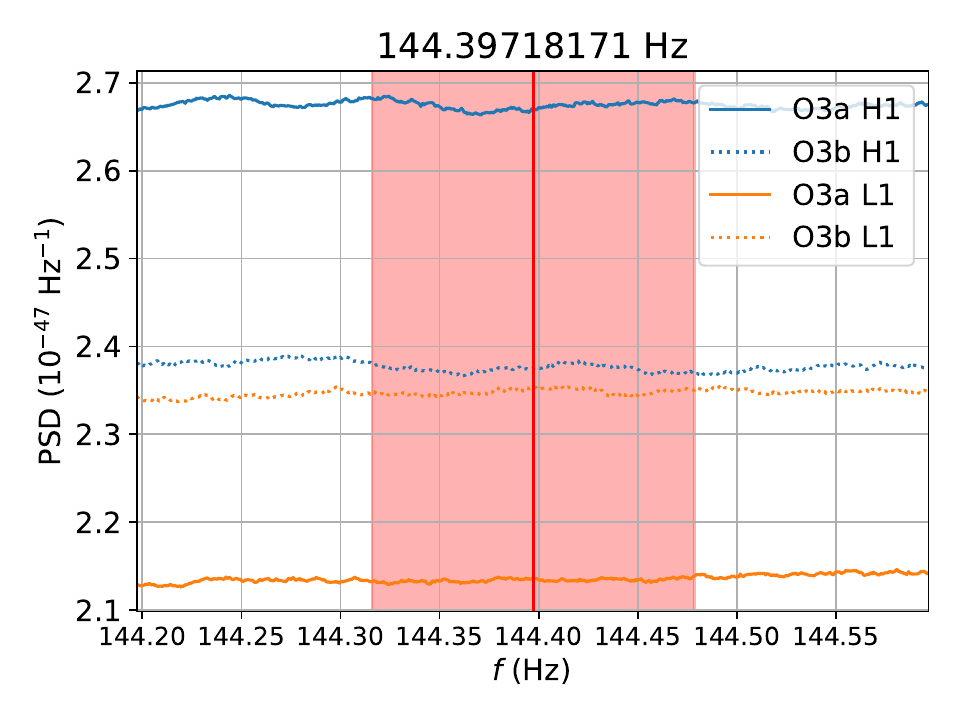}
\includegraphics[width=0.45\linewidth]{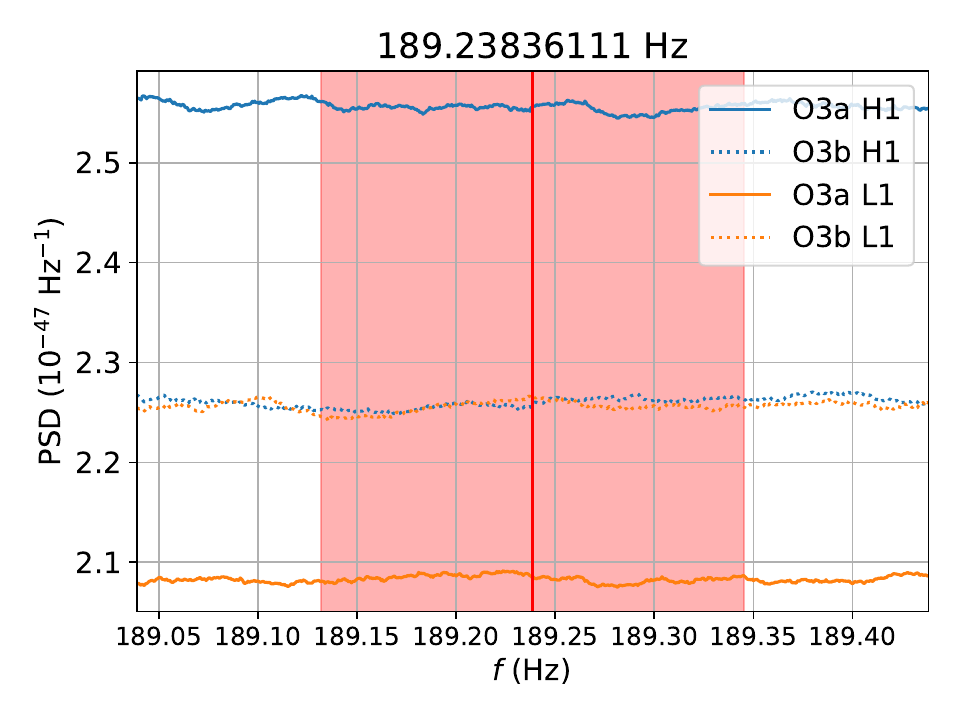}
    \caption{PSDs near each candidate that survives the $T_{\rm obs}/2$ veto. The blue and orange curves represent the O3a and O3b PSDs respectively, while solid and dotted curves represent the H1 and L1 PSDs respectively. The solid vertical red line marks the terminating frequency of the candidate, and the red band shows the extent of the \jstat filter (see Section \ref{subsec:t_drift}). The title of each subplot is the terminating frequency of the candidate.}
    \label{fig_app:psd}
\end{figure*}

In this appendix, we display the time-averaged PSDs of the Hanford (H1) and Livingston (L1) detectors during O3a and O3b, separately. The PSD is computed with the \texttt{lalpulsar\_ComputePSD} program in \texttt{LALSuite} \cite{lalsuite}, using the \texttt{--PSDmthopSFTs=harmsum} and \texttt{--normalizeByTotalNumSFTs=true} arguments, consistent with the definition of Ref.\ \cite{DreissigackerEtAl2018}. The PSDs near each candidate that survives the $T_{\rm obs}/2$ veto are plotted in Fig.\ \ref{fig_app:psd}. The O3a and O3b PSDs are plotted in blue and orange respectively, while the H1 and L1 PSDs are plotted as solid and dotted curves respectively. The solid vertical red line marks the terminating frequency of the candidate, and the red band is the extent of the \jstat filter (see Section \ref{subsec:t_drift}).

The candidate at $109.47\, {\rm Hz}$ appears contaminated in O3b H1.
The PSDs near the other candidates do not show visible contamination.

\section{Orbital parameters of above-threshold candidates}
\label{appendix:candidate_params}

For reproducibility, we list, in Table \ref{tab_app:candidate_params}, the orbital parameters of the 11 candidates in Table \ref{tab:candidates}. All of these candidates pass the known lines veto.

\begin{table*}
\caption{Orbital parameters of each of the 11 candidates listed in Table \ref{tab:candidates}, together with $\max \mathcal{L}$ and the threshold $\mathcal{L}_{\rm th}$ in the subband.}
\label{tab_app:candidate_params}
\begin{tabular}{cccccccc}
\toprule
Terminating frequency (Hz) & $\max \mathcal{L}$ & $\mathcal{L}_{\rm th}$ & $a_0$ (lt-s) &       $P$ (s) & $T_{\rm asc}$ (GPS s) & $e/10^{-4}$ & $\omega$ (rad) \\
\midrule
               64.23285822 &             304.33 &                 285.04 &      13.1825 & 146894.425806 &        1253782016.213 &         3.0 &            2.0 \\
               64.33650174 &             289.20 &                 284.97 &      13.1825 & 146894.428912 &        1253782027.406 &         3.0 &            2.0 \\
               64.36863600 &             301.25 &                 284.96 &      13.1825 & 146894.425806 &        1253782016.213 &         3.0 &            2.0 \\
               64.60612674 &             284.91 &                 284.81 &      13.1837 & 146894.427877 &        1253782023.675 &         3.0 &            1.3 \\
               64.67312153 &             296.22 &                 284.77 &      13.1825 & 146894.425806 &        1253782016.213 &         3.0 &            2.0 \\
               64.67559201 &             292.96 &                 284.76 &      13.1837 & 146894.424865 &        1253782012.482 &         3.0 &            2.0 \\
               76.40570197 &             283.23 &                 282.64 &      13.1837 & 146894.431167 &        1253782036.303 &         3.0 &            2.0 \\
              109.47137095 &             301.03 &                 294.74 &      13.1825 & 146894.426295 &        1253782017.975 &         5.8 &            1.3 \\
              114.70074190 &             305.37 &                 303.83 &      13.1825 & 146894.423603 &        1253782006.395 &         5.8 &            1.3 \\
              144.39718171 &             324.57 &                 321.10 &      13.1831 & 146894.432216 &        1253782041.630 &         4.8 &            1.7 \\
              189.23836111 &             341.30 &                 339.92 &      13.1834 & 146894.432296 &        1253782042.209 &         6.1 &            1.5 \\
\bottomrule
\end{tabular}
\end{table*}

\bibliography{ref}

@ARTICLE{tananbaum1972,
       author = {{Tananbaum}, H. and {Gursky}, H. and {Kellogg}, E.~M. and {Levinson}, R. and {Schreier}, E. and {Giacconi}, R.},
        title = "{Discovery of a Periodic Pulsating Binary X-Ray Source in Hercules from UHURU}",
      journal = {The Astrophysical Journal},
         year = 1972,
        month = jun,
       volume = {174},
        pages = {L143},
          doi = {10.1086/180968},
       adsurl = {https://ui.adsabs.harvard.edu/abs/1972ApJ...174L.143T}
}

@article{Staubert_2009,
   title={Updating the orbital ephemeris of Hercules X-1; rate of decay  and 
 eccentricity of the orbit},
   volume={500},
   ISSN={1432-0746},
   url={http://dx.doi.org/10.1051/0004-6361/200911690},
   DOI={10.1051/0004-6361/200911690},
   number={2},
   journal={Astronomy \& Astrophysics},
   publisher={EDP Sciences},
   author={Staubert, R. and Klochkov, D. and Wilms, J.},
   year={2009},
   month=apr, pages={883–889} }

@dataset{herx1_simbad,
  author={{Gaia Collaboration}},
  title = {{Gaia EDR3}},
  publisher = { Centre de Donnees astronomique de Strasbourg (CDS) },
  doi = {10.26093/cds/vizier.1350},
year=2020
}

@article{wette2023,
title = {Searches for continuous gravitational waves from neutron stars: A twenty-year retrospective},
journal = {Astroparticle Physics},
volume = {153},
pages = {102880},
year = {2023},
issn = {0927-6505},
doi = {https://doi.org/10.1016/j.astropartphys.2023.102880},
url = {https://www.sciencedirect.com/science/article/pii/S092765052300066X},
author = {Karl Wette},
}

@article{Gittins_2024,
doi = {10.1088/1361-6382/ad1c35},
url = {https://dx.doi.org/10.1088/1361-6382/ad1c35},
year = {2024},
month = {jan},
publisher = {IOP Publishing},
volume = {41},
number = {4},
pages = {043001},
author = {Fabian Gittins},
title = {Gravitational waves from neutron-star mountains},
journal = {Classical and Quantum Gravity},
}

@article{o3_amxp,
  title = {Search for continuous gravitational waves from 20 accreting millisecond x-ray pulsars in O3 LIGO data},
  author = {Abbott, R and Abbott, TD and Acernese, F and Ackley, K and Adams, C and Adhikari, N and Adhikari, RX and Adya, VB and Affeldt, C and Agarwal, D and others},
  collaboration = {LIGO Scientific, Virgo, KAGRA},
  journal = {Physical Review D},
  volume = {105},
  issue = {2},
  pages = {022002},
  numpages = {42},
  year = {2022},
  month = {Jan},
  publisher = {American Physical Society},
  doi = {10.1103/PhysRevD.105.022002}
}

@article{ligo,
doi = {10.1088/0264-9381/32/7/074001},
url = {https://dx.doi.org/10.1088/0264-9381/32/7/074001},
year = {2015},
month = {mar},
publisher = {IOP Publishing},
volume = {32},
number = {7},
pages = {074001},
author = {J Aasi and B P Abbott and R Abbott and T Abbott and M R Abernathy and K Ackley and C Adams and T Adams and P Addesso and R X Adhikari and others},
collaboration = {LIGO Scientific},
title = {Advanced LIGO},
journal = {Classical and Quantum Gravity}
}

@article{virgo,
doi = {10.1088/0264-9381/32/2/024001},
url = {https://dx.doi.org/10.1088/0264-9381/32/2/024001},
year = {2015},
publisher = {IOP Publishing},
volume = {32},
number = {2},
pages = {024001},
author = {F Acernese and M Agathos and K Agatsuma and D Aisa and N Allemandou and A Allocca and J Amarni and P Astone and G Balestri and G Ballardin and others},
collaboration={Virgo Collaboration},
title = {Advanced Virgo: a second-generation interferometric gravitational wave detector},
journal = {Classical and Quantum Gravity},
}

@article{kagra,
    author = {Akutsu, T and Ando, M and Arai, K and Arai, Y and Araki, S and Araya, A and Aritomi, N and Aso, Y and Bae, S and Bae, Y },
collaboration = {KAGRA},
    title = "{Overview of KAGRA: Detector design and construction history}",
    journal = {Progress of Theoretical and Experimental Physics},
    volume = {2021},
    number = {5},
    pages = {05A101},
    year = {2021},
    month = {05},
    issn = {2050-3911},
    doi = {10.1093/ptep/ptaa125},
    url = {https://doi.org/10.1093/ptep/ptaa125}
}

@article{HMM,
  title = {Hidden Markov model tracking of continuous gravitational waves from a neutron star with wandering spin},
  author = {Suvorova, S. and Sun, L. and Melatos, A. and Moran, W. and Evans, R. J.},
  journal = {Physical Review D},
  volume = {93},
  issue = {12},
  pages = {123009},
  numpages = {17},
  year = {2016},
  month = {Jun},
  publisher = {American Physical Society},
  doi = {10.1103/PhysRevD.93.123009},
  url = {https://link.aps.org/doi/10.1103/PhysRevD.93.123009}
}

@article{jstat,
  title = {Hidden Markov model tracking of continuous gravitational waves from a binary neutron star with wandering spin. II. Binary orbital phase tracking},
  author = {Suvorova, S. and Clearwater, P. and Melatos, A. and Sun, L. and Moran, W. and Evans, R. J.},
  journal = {Physical Review D},
  volume = {96},
  issue = {10},
  pages = {102006},
  numpages = {24},
  year = {2017},
  month = {Nov},
  publisher = {American Physical Society},
  doi = {10.1103/PhysRevD.96.102006},
  url = {https://link.aps.org/doi/10.1103/PhysRevD.96.102006}
}

@article{herx1CompanionMassAndDistance,
doi = {10.1088/0004-637X/793/2/79},
url = {https://dx.doi.org/10.1088/0004-637X/793/2/79},
year = {2014},
month = {sep},
publisher = {The American Astronomical Society},
volume = {793},
number = {2},
pages = {79},
author = {D. A. Leahy and M. H. Abdallah},
title = {HZ HER: STELLAR RADIUS FROM X-RAY ECLIPSE OBSERVATIONS, EVOLUTIONARY STATE, AND A NEW DISTANCE},
journal = {The Astrophysical Journal}
}

@article{herx1Edd1,
    author = {Kosec, P and Fabian, A C and Pinto, C and Walton, D J and Dyda, S and Reynolds, C S},
    title = "{An ionized accretion disc wind in Hercules X-1}",
    journal = {Monthly Notices of the Royal Astronomical Society},
    volume = {491},
    number = {3},
    pages = {3730-3750},
    year = {2019},
    month = {11},
    issn = {0035-8711},
    doi = {10.1093/mnras/stz3200},
    url = {https://doi.org/10.1093/mnras/stz3200}
}

@article{herx1Edd2,
doi = {10.3847/1538-4357/ac897e},
url = {https://dx.doi.org/10.3847/1538-4357/ac897e},
year = {2022},
month = {sep},
publisher = {The American Astronomical Society},
volume = {936},
number = {2},
pages = {185},
author = {P. Kosec and E. Kara and A. C. Fabian and F. Fürst and C. Pinto and I. Psaradaki and C. S. Reynolds and D. Rogantini and D. J. Walton and R. Ballhausen and C. Canizares and S. Dyda and R. Staubert and J. Wilms},
title = {The Long Stare at Hercules X-1. I. Emission Lines from the Outer Disk, the Magnetosphere Boundary, and the Accretion Curtain},
journal = {The Astrophysical Journal},
}

@article{scox1,
  title = {Search for gravitational waves from Scorpius X-1 with a hidden Markov model in O3 LIGO data},
  author = {Abbott, R. and Abe, H. and Acernese, F. and Ackley, K. and Adhikari, N. and Adhikari, R. X. and Adkins, V. K. and Adya, V. B. and Affeldt, C. and Agarwal, D. and Agathos, M. and others},
  collaboration = {LIGO Scientific, Virgo, KAGRA},
  journal = {Physical Review D},
  volume = {106},
  issue = {6},
  pages = {062002},
  numpages = {23},
  year = {2022},
  month = {Sep},
  publisher = {American Physical Society},
  doi = {10.1103/PhysRevD.106.062002},
  url = {https://link.aps.org/doi/10.1103/PhysRevD.106.062002}
}

@article{leaciPrix2015,
  title = {Directed searches for continuous gravitational waves from binary systems: Parameter-space metrics and optimal Scorpius X-1 sensitivity},
  author = {Leaci, Paola and Prix, Reinhard},
  journal = {Physical Review D},
  volume = {91},
  issue = {10},
  pages = {102003},
  numpages = {26},
  year = {2015},
  month = {May},
  publisher = {American Physical Society},
  doi = {10.1103/PhysRevD.91.102003},
  url = {https://link.aps.org/doi/10.1103/PhysRevD.91.102003}
}

@article{psrJ0437andres,
  title = {Search for continuous gravitational waves from PSR $\mathrm{J}0437\text{\ensuremath{-}}4715$ with a hidden Markov model in O3 LIGO data},
  author = {Vargas, Andr\'es F. and Melatos, Andrew},
  journal = {Physical Review D},
  volume = {107},
  issue = {6},
  pages = {064062},
  numpages = {16},
  year = {2023},
  month = {Mar},
  publisher = {American Physical Society},
  doi = {10.1103/PhysRevD.107.064062},
  url = {https://link.aps.org/doi/10.1103/PhysRevD.107.064062}
}

@article{caride2019,
  title = {How to search for gravitational waves from $r$-modes of known pulsars},
  author = {Caride, Santiago and Inta, Ra and Owen, Benjamin J. and Rajbhandari, Binod},
  journal = {Physical Review D},
  volume = {100},
  issue = {6},
  pages = {064013},
  numpages = {10},
  year = {2019},
  month = {Sep},
  publisher = {American Physical Society},
  doi = {10.1103/PhysRevD.100.064013},
  url = {https://link.aps.org/doi/10.1103/PhysRevD.100.064013}
}

@Article{Riles2023,
author={Riles, Keith},
title={Searches for continuous-wave gravitational radiation},
journal={Living Reviews in Relativity},
year={2023},
month={Apr},
day={17},
volume={26},
number={1},
pages={3},
issn={1433-8351},
doi={10.1007/s41114-023-00044-3},
url={https://doi.org/10.1007/s41114-023-00044-3}
}

@article{o3LIGOsensitivity,
  title = {Sensitivity and performance of the Advanced LIGO detectors in the third observing run},
  author = {Buikema, A. and Cahillane, C. and Mansell, G. L. and Blair, C. D. and Abbott, R. and Adams, C. and Adhikari, R. X. and Ananyeva, A. and Appert, S. and Arai, K. and others },
  journal = {Physical Review D},
  volume = {102},
  issue = {6},
  pages = {062003},
  numpages = {27},
  year = {2020},
  month = {Sep},
  publisher = {American Physical Society},
  doi = {10.1103/PhysRevD.102.062003},
  url = {https://link.aps.org/doi/10.1103/PhysRevD.102.062003}
}

@article{O3LIGOsensitivty2,
  title = {Quantum-Enhanced Advanced LIGO Detectors in the Era of Gravitational-Wave Astronomy},
  author = {Tse, M. and Yu, Haocun and Kijbunchoo, N. and Fernandez-Galiana, A. and Dupej, P. and Barsotti, L. and Blair, C. D. and Brown, D. D. and Dwyer, S. E. and Effler, A. and others},
  journal = {Physical Review Letters},
  volume = {123},
  issue = {23},
  pages = {231107},
  numpages = {8},
  year = {2019},
  month = {Dec},
  publisher = {American Physical Society},
  doi = {10.1103/PhysRevLett.123.231107},
  url = {https://link.aps.org/doi/10.1103/PhysRevLett.123.231107}
}

@book{aerts2010asteroseismology,
  title={Asteroseismology},
  author={Aerts, Conny and Christensen-Dalsgaard, J{\o}rgen and Kurtz, Donald W},
  year={2010},
  publisher={Springer Science \& Business Media}
}

@article{keer2014starquake,
    author = {Keer, L. and Jones, D. I.},
    title = {Developing a model for neutron star oscillations following starquakes},
    journal = {Monthly Notices of the Royal Astronomical Society},
    volume = {446},
    number = {1},
    pages = {865-891},
    year = {2014},
    month = {11},
    issn = {0035-8711},
    doi = {10.1093/mnras/stu2123},
    url = {https://doi.org/10.1093/mnras/stu2123}
}

@article{Giliberti2022starquake,
    author = {Giliberti, E and Cambiotti, G},
    title = {Starquakes in millisecond pulsars and gravitational waves emission},
    journal = {Monthly Notices of the Royal Astronomical Society},
    volume = {511},
    number = {3},
    pages = {3365-3376},
    year = {2022},
    month = {01},
    issn = {0035-8711},
    doi = {10.1093/mnras/stac245},
    url = {https://doi.org/10.1093/mnras/stac245},
}

@Article{Andersson2021,
AUTHOR = {Andersson, Nils},
TITLE = {A Gravitational-Wave Perspective on Neutron-Star Seismology},
JOURNAL = {Universe},
VOLUME = {7},
YEAR = {2021},
NUMBER = {4},
ARTICLE-NUMBER = {97},
URL = {https://www.mdpi.com/2218-1997/7/4/97},
ISSN = {2218-1997},
DOI = {10.3390/universe7040097}
}

@article{Baym1971starquake,
title = {Neutron starquakes and pulsar speedup},
journal = {Annals of Physics},
volume = {66},
number = {2},
pages = {816-835},
year = {1971},
issn = {0003-4916},
doi = {https://doi.org/10.1016/0003-4916(71)90084-4},
url = {https://www.sciencedirect.com/science/article/pii/0003491671900844},
author = {Gordon Baym and David Pines},
}

@article{Stergioulas2004splitting,
    author = {Stergioulas, Nikolaos and Apostolatos, Theocharis A. and Font, José A.},
    title = {Non-linear pulsations in differentially rotating neutron stars: mass-shedding-induced damping and splitting of the fundamental mode},
    journal = {Monthly Notices of the Royal Astronomical Society},
    volume = {352},
    number = {4},
    pages = {1089-1101},
    year = {2004},
    month = {08},
    issn = {0035-8711},
    doi = {10.1111/j.1365-2966.2004.07973.x},
    url = {https://doi.org/10.1111/j.1365-2966.2004.07973.x},
}

@ARTICLE{kruger2021splitting,
AUTHOR={Krüger, Christian J.  and Kokkotas, Kostas D.  and Manoharan, Praveen  and Völkel, Sebastian H. },
TITLE={Fast Rotating Neutron Stars: Oscillations and Instabilities},
JOURNAL={Frontiers in Astronomy and Space Sciences},
VOLUME={8},
YEAR={2021},
URL={https://www.frontiersin.org/journals/astronomy-and-space-sciences/articles/10.3389/fspas.2021.736918},
DOI={10.3389/fspas.2021.736918},
ISSN={2296-987X},}

@article{dong2024,
    author = {Dong, Wenhao and Melatos, Andrew},
    title = {Gravitational waves from non-radial oscillations of stochastically accreting neutron stars},
    journal = {Monthly Notices of the Royal Astronomical Society},
    volume = {530},
    number = {3},
    pages = {2822-2839},
    year = {2024},
    month = {04},
    issn = {0035-8711},
    doi = {10.1093/mnras/stae1028},
    url = {https://doi.org/10.1093/mnras/stae1028},
}

@article{dong2025,
    author = {Dong, Wenhao and Melatos, Andrew},
    title = {Gravitational waves from r-mode oscillations of stochastically accreting neutron stars},
    journal = {Monthly Notices of the Royal Astronomical Society},
    volume = {537},
    number = {2},
    pages = {650-660},
    year = {2025},
    month = {01},
    issn = {0035-8711},
    doi = {10.1093/mnras/staf033},
    url = {https://doi.org/10.1093/mnras/staf033}
}

@article{nagar2004accretion,
  title = {Accretion-driven gravitational radiation from nonrotating compact objects: Infalling quadrupolar shells},
  author = {Nagar, Alessandro and D\'{\i}az, Guillermo and Pons, Jos\'e A. and Font, Jos\'e A.},
  journal = {Physical Review D},
  volume = {69},
  issue = {12},
  pages = {124028},
  numpages = {21},
  year = {2004},
  month = {Jun},
  publisher = {American Physical Society},
  doi = {10.1103/PhysRevD.69.124028},
  url = {https://link.aps.org/doi/10.1103/PhysRevD.69.124028}
}

@article{Gaertig2009,
  title = {Relativistic g-modes in rapidly rotating neutron stars},
  author = {Gaertig, Erich and Kokkotas, Kostas D.},
  journal = {Physical Review D},
  volume = {80},
  issue = {6},
  pages = {064026},
  numpages = {13},
  year = {2009},
  month = {Sep},
  publisher = {American Physical Society},
  doi = {10.1103/PhysRevD.80.064026},
  url = {https://link.aps.org/doi/10.1103/PhysRevD.80.064026}
}

@article{andersson1998,
    author = {Andersson, Nils and Kokkotas, Kostas D.},
    title = {Towards gravitational wave asteroseismology},
    journal = {Monthly Notices of the Royal Astronomical Society},
    volume = {299},
    number = {4},
    pages = {1059-1068},
    year = {1998},
    month = {10},
    issn = {0035-8711},
    doi = {10.1046/j.1365-8711.1998.01840.x},
    url = {https://doi.org/10.1046/j.1365-8711.1998.01840.x},
}

@article{Benhar2004,
  title = {Gravitational wave asteroseismology reexamined},
  author = {Benhar, Omar and Ferrari, Valeria and Gualtieri, Leonardo},
  journal = {Physical Review D},
  volume = {70},
  issue = {12},
  pages = {124015},
  numpages = {9},
  year = {2004},
  month = {Dec},
  publisher = {American Physical Society},
  doi = {10.1103/PhysRevD.70.124015},
  url = {https://link.aps.org/doi/10.1103/PhysRevD.70.124015}
}

@article{Ashok_2021,
doi = {10.3847/1538-4357/ac2582},
url = {https://dx.doi.org/10.3847/1538-4357/ac2582},
year = {2021},
month = {dec},
publisher = {The American Astronomical Society},
volume = {923},
number = {1},
pages = {85},
author = {Ashok, A. and Beheshtipour, B. and Papa, M. A. and Freire, P. C. C. and Steltner, B. and Machenschalk, B. and Behnke, O. and Allen, B. and Prix, R.},
title = {New Searches for Continuous Gravitational Waves from Seven Fast Pulsars},
journal = {The Astrophysical Journal}
}

@article{knownPulsars,
doi = {10.3847/1538-4357/ac6acf},
url = {https://dx.doi.org/10.3847/1538-4357/ac6acf},
year = {2022},
month = {aug},
publisher = {The American Astronomical Society},
volume = {935},
number = {1},
pages = {1},
author = {Abbott, R. and Abe, H. and Acernese, F. and Ackley, K. and Adhikari, N. and Adhikari, R. X. and Adkins, V. K. and Adya, V. B. and Affeldt, C. and Agarwal, D. and others},
title = {Searches for Gravitational Waves from Known Pulsars at Two Harmonics in the Second and Third LIGO-Virgo Observing Runs},
collaboration = {LIGO Scientific, Virgo, KAGRA},
journal = {The Astrophysical Journal}
}

@article{knownPulsars2,
doi = {10.3847/1538-4357/ac6ad0},
url = {https://dx.doi.org/10.3847/1538-4357/ac6ad0},
year = {2022},
month = {jun},
publisher = {The American Astronomical Society},
volume = {932},
number = {2},
pages = {133},
author = {Abbott, R. and Abbott, T. D. and Acernese, F. and Ackley, K. and Adams, C. and Adhikari, N. and Adhikari, R. X. and Adya, V. B. and Affeldt, C. and Agarwal, D. and others},
title = {Narrowband Searches for Continuous and Long-duration Transient Gravitational Waves from Known Pulsars in the LIGO-Virgo Third Observing Run},
journal = {The Astrophysical Journal},
collaboration = {LIGO Scientific, Virgo, KAGRA },
}

@ARTICLE{viterbi,
  author={Viterbi, A.},
  journal={IEEE Transactions on Information Theory}, 
  title={Error bounds for convolutional codes and an asymptotically optimum decoding algorithm}, 
  year={1967},
  volume={13},
  number={2},
  pages={260-269},
  doi={10.1109/TIT.1967.1054010}}

@article{Leahy_2020_inclination,
doi = {10.3847/1538-4357/ab65f9},
url = {https://dx.doi.org/10.3847/1538-4357/ab65f9},
year = {2020},
month = {jan},
publisher = {The American Astronomical Society},
volume = {889},
number = {2},
pages = {131},
author = {Leahy, D. A. and Postma, J. and Chen, Y.},
title = {AstroSat UVIT Observations of Her X-1},
journal = {The Astrophysical Journal}
}

@article{fstat,
  title = {Data analysis of gravitational-wave signals from spinning neutron stars: The signal and its detection},
  author = {Jaranowski, Piotr and Kr\'olak, Andrzej and Schutz, Bernard F.},
  journal = {Physical Review D},
  volume = {58},
  issue = {6},
  pages = {063001},
  numpages = {24},
  year = {1998},
  month = {Aug},
  publisher = {American Physical Society},
  doi = {10.1103/PhysRevD.58.063001},
  url = {https://link.aps.org/doi/10.1103/PhysRevD.58.063001}
}

@article{GBMpulsars,
doi = {10.3847/1538-4357/ab855c},
url = {https://dx.doi.org/10.3847/1538-4357/ab855c},
year = {2020},
month = {jun},
publisher = {The American Astronomical Society},
volume = {896},
number = {1},
pages = {90},
author = {Malacaria, C. and Jenke, P. and Roberts, O. J. and Wilson-Hodge, C. A. and Cleveland, W. H. and Mailyan, B. and on behalf of the GBM Accreting Pulsars Program Team},
title = {The Ups and Downs of Accreting X-Ray Pulsars: Decade-long Observations with the Fermi Gamma-Ray Burst Monitor},
journal = {The Astrophysical Journal}
}

@misc{herx1GBM,
  title = {Her X-1},
  howpublished = {\url{https://gammaray.nsstc.nasa.gov/gbm/science/pulsars/lightcurves/herx1.html}},
  note = {Accessed: 2025-02-20},
author = {Peter Jenke}
}

@ARTICLE{McDermott1988,
       author = {{McDermott}, P.~N. and {van Horn}, H.~M. and {Hansen}, C.~J.},
        title = "{Nonradial Oscillations of Neutron Stars}",
      journal = {\apj},
         year = 1988,
        month = feb,
       volume = {325},
        pages = {725},
          doi = {10.1086/166044},
       adsurl = {https://ui.adsabs.harvard.edu/abs/1988ApJ...325..725M}
}

@article{sidery2010,
    author = {Sidery, T. and Passamonti, A. and Andersson, N.},
    title = {The dynamics of pulsar glitches: contrasting phenomenology with numerical evolutions},
    journal = {Monthly Notices of the Royal Astronomical Society},
    volume = {405},
    number = {2},
    pages = {1061-1074},
    year = {2010},
    month = {06},
    issn = {0035-8711},
    doi = {10.1111/j.1365-2966.2010.16497.x},
    url = {https://doi.org/10.1111/j.1365-2966.2010.16497.x},
}

@article{Mirasola2024,
  title = {New semicoherent targeted search for continuous gravitational waves from pulsars in binary systems},
  author = {Mirasola, L. and Leaci, P. and Astone, P. and D'Onofrio, L. and Dall'Osso, S. and De Falco, A. and Lai, M. and Mastrogiovanni, S. and Palomba, C. and Riggio, A. and Sanna, A.},
  journal = {Physical Review D},
  volume = {110},
  issue = {12},
  pages = {123043},
  numpages = {16},
  year = {2024},
  month = {Dec},
  publisher = {American Physical Society},
  doi = {10.1103/PhysRevD.110.123043},
  url = {https://link.aps.org/doi/10.1103/PhysRevD.110.123043}
}

@book{Roy2004,
  title = {Orbital Motion},
  ISBN = {9780367806620},
  url = {http://dx.doi.org/10.1201/9780367806620},
  DOI = {10.1201/9780367806620},
  publisher = {CRC Press},
  author = {Roy,  A.E.},
  year = {2004},
  edition=4,
chapter=4
}

@book{Brumberg2012,
  title = {Analytical Techniques of Celestial Mechanics},
  ISBN = {9783642794544},
  url = {https://doi.org/10.1007/978-3-642-79454-4},
  DOI = {10.1007/978-3-642-79454-4},
  publisher = {Springer Berlin Heidelberg},
  author = {Victor A. Brumberg},
  year = {2012},
  edition=1,
chapter=2
}

@misc{lalsuite,
       author         = "{LIGO Scientific Collaboration} and {Virgo Collaboration} and {KAGRA Collaboration}",
       title          = "{LVK} {A}lgorithm {L}ibrary - {LALS}uite",
       howpublished   = "Free software (GPL)",
       doi            = "10.7935/GT1W-FZ16",
       year           = "2018"
 }

@book{mathSpecialFunction,
  author = {Frank Olver and Daniel Lozier and Ronald Boisvert and Charles Clark},
  title = {The NIST Handbook of Mathematical Functions},
  year = {2010},
  month = {2010-05-17},
  publisher = {Cambridge University Press, New York, NY},
edition={1}
}

@article{PEGSIV_scox1,
    author = {Killestein, T L and Mould, M and Steeghs, D and Casares, J and Galloway, D K and Whelan, J T},
    title = {Precision Ephemerides for Gravitational-wave Searches – IV. Corrected and refined ephemeris for Scorpius X-1},
    journal = {Monthly Notices of the Royal Astronomical Society},
    volume = {520},
    number = {4},
    pages = {5317-5330},
    year = {2023},
    month = {02},
    issn = {0035-8711},
    doi = {10.1093/mnras/stad366},
}

@article{SAXJ1748.9,
    author = {Sanna, A. and Burderi, L. and Riggio, A. and Pintore, F. and Di Salvo, T. and Gambino, A. F. and Iaria, R. and Matranga, M. and Scarano, F.},
    title = {Timing of the accreting millisecond pulsar SAX J1748.9--2021 during its 2015 outburst},
    journal = {Monthly Notices of the Royal Astronomical Society},
    volume = {459},
    number = {2},
    pages = {1340-1349},
    year = {2016},
    month = {04},
    issn = {0035-8711},
    doi = {10.1093/mnras/stw740},
}

@article{crabVelaCW,
  title = {First search for long-duration transient gravitational waves after glitches in the Vela and Crab pulsars},
  author = {Keitel, David and Woan, Graham and Pitkin, Matthew and Schumacher, Courtney and Pearlstone, Brynley and Riles, Keith and Lyne, Andrew G. and Palfreyman, Jim and Stappers, Benjamin and Weltevrede, Patrick},
  journal = {Physical Review D},
  volume = {100},
  issue = {6},
  pages = {064058},
  numpages = {10},
  year = {2019},
  month = {Sep},
  publisher = {American Physical Society},
  doi = {10.1103/PhysRevD.100.064058}
}

@article{firstVelaCW,
  title = {Search for gravitational waves associated with the August 2006 timing glitch of the Vela pulsar},
  author = {Abadie, J. and Abbott, B. P. and Abbott, R. and Adhikari, R. and Ajith, P. and Allen, B. and Allen, G. and Amador Ceron, E. and Amin, R. S. and Anderson, S. B. and others},
  collaboration = {LIGO Scientific },
  journal = {Physical Review D},
  volume = {83},
  issue = {4},
  pages = {042001},
  numpages = {13},
  year = {2011},
  month = {Feb},
  publisher = {American Physical Society},
  doi = {10.1103/PhysRevD.83.042001}
}

@article{tempo2,
    author = {Hobbs, G. B. and Edwards, R. T. and Manchester, R. N.},
    title = {tempo2, a new pulsar-timing package – I. An overview},
    journal = {Monthly Notices of the Royal Astronomical Society},
    volume = {369},
    number = {2},
    pages = {655-672},
    year = {2006},
    month = {05},
    issn = {0035-8711},
    doi = {10.1111/j.1365-2966.2006.10302.x}
}

@article{Wagner_2022,
doi = {10.1088/1361-6382/ac5012},
year = {2022},
month = {mar},
publisher = {IOP Publishing},
volume = {39},
number = {7},
pages = {075013},
author = {Wagner, Katelyn J and Whelan, John T and Wofford, Jared K and Wette, Karl},
title = {Template lattices for a cross-correlation search for gravitational waves from Scorpius X-1},
journal = {Classical and Quantum Gravity}
}

@article{Whelan_2023,
doi = {10.3847/1538-4357/acc8d7},
year = {2023},
month = {jun},
publisher = {The American Astronomical Society},
volume = {949},
number = {2},
pages = {117},
author = {Whelan, John T. and Tenorio, Rodrigo and Wofford, Jared K. and Clark, James A. and Daw, Edward J. and Goetz, Evan and Keitel, David and Neunzert, Ansel and Sintes, Alicia M. and Wagner, Katelyn J. and Woan, Graham and Killestein, Thomas L. and Steeghs, Danny},
title = {Search for Gravitational Waves from Scorpius X-1 in LIGO O3 Data with Corrected Orbital Ephemeris},
journal = {The Astrophysical Journal}
}

@article{CrossCorrScox1,
doi = {10.3847/2041-8213/aca1b0},
url = {https://dx.doi.org/10.3847/2041-8213/aca1b0},
year = {2022},
month = {dec},
publisher = {The American Astronomical Society},
volume = {941},
number = {2},
pages = {L30},
author = {Abbott, R. and Abe, H. and Acernese, F. and Ackley, K. and Adhicary, S. and Adhikari, N. and Adhikari, R. X. and Adkins, V. K. and Adya, V. B. and Affeldt, C. and others},
title = {Model-based Cross-correlation Search for Gravitational Waves from the Low-mass X-Ray Binary Scorpius X-1 in LIGO O3 Data},
journal = {The Astrophysical Journal Letters},
collaboration = {LIGO Scientific, Virgo, KAGRA }
}

@misc{CalibrationO3,
       author = {{Zweizig}, J. and {Riles}, K.},
        title = "{LIGO-T2000384-v4: Information on self-gating of h(t) used in O3 continuous-wave and stochastic searches.}",
       adsurl = {https://dcc.ligo.org/LIGO-T2000384/public}
       }

@article{CalibrationO3_60Hz,
  title = {Machine-learning nonstationary noise out of gravitational-wave detectors},
  author = {Vajente, G. and Huang, Y. and Isi, M. and Driggers, J. C. and Kissel, J. S. and Szczepa\ifmmode \acute{n}\else \'{n}\fi{}czyk, M. J. and Vitale, S.},
  journal = {Physical Review D},
  volume = {101},
  issue = {4},
  pages = {042003},
  numpages = {12},
  year = {2020},
  month = {Feb},
  publisher = {American Physical Society},
  doi = {10.1103/PhysRevD.101.042003},
  url = {https://link.aps.org/doi/10.1103/PhysRevD.101.042003}
}

@misc{O3lines,
       author = {{Goetz}, E. and {Neunzert}, A. and {Riles}, K. and {Matas}, A. and {Kandhasamy}, S. and {Tasson}, J. and {Barschaw}, C. and {Middleton}, H. and {Hughey}, S. and {Mueller}, L. and {Heinzel}, J. and {Carlin}, J. and {Vargas}, A. and {Hollows}, I.},
        title = "{LIGO-T2100200-v2: O3 lines and combs in found in self-gated C01 data.}",
       url = {https://dcc.ligo.org/T2100200/public}
       }

@misc{O3segments,
    author = {Evan Goetz and Ketih Riles},
    title = "{LIGO-T2300068-v1: Segments used for creating standard SFTs in O3 data}",
    adsurl = {https://dcc.ligo.org/T2300068/public}
}

@article{Prix_template,
   title={Template-based searches for gravitational waves: efficient lattice covering of flat parameter spaces},
   volume={24},
   ISSN={1361-6382},
   url={http://dx.doi.org/10.1088/0264-9381/24/19/S11},
   DOI={10.1088/0264-9381/24/19/s11},
   number={19},
   journal={Classical and Quantum Gravity},
   publisher={IOP Publishing},
   author={Prix, Reinhard},
   year={2007},
   month=sep, pages={S481–S490} }

@article{Wette_template,
   title={Lattice template placement for coherent all-sky searches for gravitational-wave pulsars},
   volume={90},
   ISSN={1550-2368},
   url={http://dx.doi.org/10.1103/PhysRevD.90.122010},
   DOI={10.1103/physrevd.90.122010},
   number={12},
   journal={Physical Review D},
   publisher={American Physical Society (APS)},
   author={Wette, Karl},
   year={2014},
   month=dec }

@article{viterbiO2_AMXP,
  title = {Search for gravitational waves from five low mass x-ray binaries in the second Advanced LIGO observing run with an improved hidden Markov model},
  author = {Middleton, Hannah and Clearwater, Patrick and Melatos, Andrew and Dunn, Liam},
  journal = {Physical Review D},
  volume = {102},
  issue = {2},
  pages = {023006},
  numpages = {23},
  year = {2020},
  month = {Jul},
  publisher = {American Physical Society},
  doi = {10.1103/PhysRevD.102.023006},
  url = {https://link.aps.org/doi/10.1103/PhysRevD.102.023006}
}

@article{viterbiO1_scox1,
  title = {Search for gravitational waves from Scorpius X-1 in the first Advanced LIGO observing run with a hidden Markov model},
  author = {Abbott, B. P. and Abbott, R. and Abbott, T. D. and Acernese, F. and Ackley, K. and Adams, C. and Adams, T. and Addesso, P. and Adhikari, R. X. and Adya, V. B. and others},
  collaboration = {LIGO Scientific, Virgo},
  journal = {Physical Review D},
  volume = {95},
  issue = {12},
  pages = {122003},
  numpages = {20},
  year = {2017},
  month = {Jun},
  publisher = {American Physical Society},
  doi = {10.1103/PhysRevD.95.122003},
  url = {https://link.aps.org/doi/10.1103/PhysRevD.95.122003}
}

@article{viterbiO2_scox1,
  title = {Search for gravitational waves from Scorpius X-1 in the second Advanced LIGO observing run with an improved hidden Markov model},
  author = {Abbott, B. P. and Abbott, R. and Abbott, T. D. and Abraham, S. and Acernese, F. and Ackley, K. and Adams, C. and Adhikari, R. X. and Adya, V. B. and Affeldt, C. and others},
  collaboration = {LIGO Scientific, Virgo},
  journal = {Physical Review D},
  volume = {100},
  issue = {12},
  pages = {122002},
  numpages = {21},
  year = {2019},
  month = {Dec},
  publisher = {American Physical Society},
  doi = {10.1103/PhysRevD.100.122002},
  url = {https://link.aps.org/doi/10.1103/PhysRevD.100.122002}
}

@article{viterbiO2_supernova,
  title = {Search for gravitational waves from 12 young supernova remnants with a hidden Markov model in Advanced LIGO's second observing run},
  author = {Millhouse, Margaret and Strang, Lucy and Melatos, Andrew},
  journal = {Physical Review D},
  volume = {102},
  issue = {8},
  pages = {083025},
  numpages = {14},
  year = {2020},
  month = {Oct},
  publisher = {American Physical Society},
  doi = {10.1103/PhysRevD.102.083025},
  url = {https://link.aps.org/doi/10.1103/PhysRevD.102.083025}
}

@article{viterbiO3_supernova,
doi = {10.3847/1538-4357/ac17ea},
url = {https://dx.doi.org/10.3847/1538-4357/ac17ea},
year = {2021},
month = {nov},
publisher = {The American Astronomical Society},
volume = {921},
number = {1},
pages = {80},
author = {Abbott, R. and Abbott, T. D. and Abraham, S. and Acernese, F. and Ackley, K. and Adams, A. and Adams, C. and Adhikari, R. X. and Adya, V. B. and Affeldt, C. and others},
title = {Searches for Continuous Gravitational Waves from Young Supernova Remnants in the Early Third Observing Run of Advanced LIGO and Virgo},
journal = {The Astrophysical Journal}
}

@article{dm_off_veto,
  title = {New veto for continuous gravitational wave searches},
  author = {Zhu, Sylvia J. and Papa, Maria Alessandra and Walsh, Sin\'ead},
  journal = {Physical Review D},
  volume = {96},
  issue = {12},
  pages = {124007},
  numpages = {9},
  year = {2017},
  month = {Dec},
  publisher = {American Physical Society},
  doi = {10.1103/PhysRevD.96.124007},
  url = {https://link.aps.org/doi/10.1103/PhysRevD.96.124007}
}

@article{Jones_2022,
   title={Validating continuous gravitational-wave candidates from a semicoherent search using Doppler modulation and an effective point spread function},
   volume={106},
   ISSN={2470-0029},
   url={http://dx.doi.org/10.1103/PhysRevD.106.123011},
   DOI={10.1103/physrevd.106.123011},
   number={12},
   journal={Physical Review D},
   publisher={American Physical Society (APS)},
   author={Jones, Dana and Sun, Ling and Carlin, Julian and Dunn, Liam and Millhouse, Meg and Middleton, Hannah and Meyers, Patrick and Clearwater, Patrick and Beniwal, Deeksha and Strang, Lucy and Vargas, Andrés and Melatos, Andrew},
   year={2022},
   month=dec }

@article{viterbiO2_Fomalhaut_b,
  title = {Search for continuous gravitational waves from Fomalhaut b in the second Advanced LIGO observing run with a hidden Markov model},
  author = {Jones, Dana and Sun, Ling},
  journal = {Physical Review D},
  volume = {103},
  issue = {2},
  pages = {023020},
  numpages = {13},
  year = {2021},
  month = {Jan},
  publisher = {American Physical Society},
  doi = {10.1103/PhysRevD.103.023020},
  url = {https://link.aps.org/doi/10.1103/PhysRevD.103.023020}
}

@article{DreissigackerEtAl2018,
  title = {Fast and accurate sensitivity estimation for continuous-gravitational-wave searches},
  author = {{Dreissigacker}, C. and {Prix}, R. and {Wette}, K.},
  journal = {Physical Review D},
  volume = {98},
  pages = {084058},
  numpages = {30},
  year = {2018},
  doi = {10.1103/PhysRevD.98.084058},
}

@article{Zhang_2021,
doi = {10.3847/2041-8213/abd256},
url = {https://doi.org/10.3847/2041-8213/abd256},
year = {2021},
month = {jan},
publisher = {The American Astronomical Society},
volume = {906},
number = {2},
pages = {L14},
author = {Zhang, Yuanhao and Papa, Maria Alessandra and Krishnan, Badri and Watts, Anna L.},
title = {Search for Continuous Gravitational Waves from Scorpius X-1 in LIGO O2 Data},
journal = {The Astrophysical Journal Letters}
}

@article{Ju_2000,
doi = {10.1088/0034-4885/63/9/201},
url = {https://doi.org/10.1088/0034-4885/63/9/201},
year = {2000},
month = {sep},
publisher = {},
volume = {63},
number = {9},
pages = {1317},
author = {L Ju and D G Blair and C Zhao},
title = {Detection of gravitational waves},
journal = {Reports on Progress in Physics}
}

@article{zhao2022,
  title = {Universal relations for neutron star $f$-mode and $g$-mode oscillations},
  author = {Zhao, Tianqi and Lattimer, James M.},
  journal = {Physical Review D},
  volume = {106},
  issue = {12},
  pages = {123002},
  numpages = {23},
  year = {2022},
  month = {Dec},
  publisher = {American Physical Society},
  doi = {10.1103/PhysRevD.106.123002},
  url = {https://link.aps.org/doi/10.1103/PhysRevD.106.123002}
}

@article{O3_open_data,
doi = {10.3847/1538-4365/acdc9f},
url = {https://doi.org/10.3847/1538-4365/acdc9f},
year = {2023},
month = {jul},
publisher = {The American Astronomical Society},
volume = {267},
number = {2},
pages = {29},
author = {Abbott, R. and Abe, H. and Acernese, F. and Ackley, K. and Adhicary, S. and Adhikari, N. and Adhikari, R. X. and Adkins, V. K. and Adya, V. B. and Affeldt, C. and others},
collaboration = {LIGO Scientific  Virgo, KAGRA},
title = {Open Data from the Third Observing Run of LIGO, Virgo, KAGRA, and GEO},
journal = {The Astrophysical Journal Supplement Series}
}

@article{Davis_2021,
doi = {10.1088/1361-6382/abfd85},
url = {https://doi.org/10.1088/1361-6382/abfd85},
year = {2021},
month = {jun},
publisher = {IOP Publishing},
volume = {38},
number = {13},
pages = {135014},
author = {Davis, D and Areeda, J S and Berger, B K and Bruntz, R and Effler, A and Essick, R C and Fisher, R P and Godwin, P and Goetz, E and Helmling-Cornell, A F and others},
title = {LIGO detector characterization in the second and third observing runs},
journal = {Classical and Quantum Gravity},
}

@article{Nieder_2020,
doi = {10.3847/1538-4357/abaf53},
url = {https://doi.org/10.3847/1538-4357/abaf53},
year = {2020},
month = {oct},
publisher = {The American Astronomical Society},
volume = {901},
number = {2},
pages = {156},
author = {Nieder, L. and Allen, B. and Clark, C. J. and Pletsch, H. J.},
title = {Exploiting Orbital Constraints from Optical Data to Detect Binary Gamma-Ray Pulsars},
journal = {The Astrophysical Journal},
}

@article{BehnkeEtAl2015,
  title = {Postprocessing methods used in the search for continuous gravitational-wave signals from the {G}alactic {C}enter},
  author = {{Behnke}, B. and {Papa}, M.~A. and {Prix}, R.},
  journal = {Physical Review D},
  volume = {91},
  pages = {064007},
  year = {2015},
  doi = {10.1103/PhysRevD.91.064007},
}

@ARTICLE{Reisenegger1992,
       author = {{Reisenegger}, Andreas and {Goldreich}, Peter},
        title = "{A New Class of g-Modes in Neutron Stars}",
      journal = {The Astrophysical Journal},
         year = 1992,
        month = aug,
       volume = {395},
        pages = {240},
          doi = {10.1086/171645},
       adsurl = {https://ui.adsabs.harvard.edu/abs/1992ApJ...395..240R}
}

@article{Gusakov2013,
  title = {Thermal g-modes and unexpected convection in superfluid neutron stars},
  author = {Gusakov, Mikhail E. and Kantor, Elena M.},
  journal = {Physical Review D},
  volume = {88},
  issue = {10},
  pages = {101302},
  numpages = {5},
  year = {2013},
  month = {Nov},
  publisher = {American Physical Society},
  doi = {10.1103/PhysRevD.88.101302},
  url = {https://link.aps.org/doi/10.1103/PhysRevD.88.101302}
}

@article{Kantor2014,
    author = {Kantor, E. M. and Gusakov, M. E.},
    title = {Composition temperature-dependent g modes in superfluid neutron stars},
    journal = {Monthly Notices of the Royal Astronomical Society: Letters},
    volume = {442},
    number = {1},
    pages = {L90-L94},
    year = {2014},
    month = {06},
    issn = {1745-3925},
    doi = {10.1093/mnrasl/slu061},
    url = {https://doi.org/10.1093/mnrasl/slu061}
}

@Article{Lozano2022,
AUTHOR = {Lozano, Nicholas and Tran, Vinh and Jaikumar, Prashanth},
TITLE = {Temperature Effects on Core g-Modes of Neutron Stars},
JOURNAL = {Galaxies},
VOLUME = {10},
YEAR = {2022},
NUMBER = {4},
ARTICLE-NUMBER = {79},
URL = {https://www.mdpi.com/2075-4434/10/4/79},
ISSN = {2075-4434},
DOI = {10.3390/galaxies10040079}
}

@article{Lai1999,
    author = {Lai, Dong},
    title = {Secular instability of g-modes in rotating neutron stars},
    journal = {Monthly Notices of the Royal Astronomical Society},
    volume = {307},
    number = {4},
    pages = {1001-1007},
    year = {1999},
    month = {08},
    issn = {0035-8711},
    doi = {10.1046/j.1365-8711.1999.02723.x},
    url = {https://doi.org/10.1046/j.1365-8711.1999.02723.x},
}

@article{Zhao_2025,
doi = {10.3847/1538-4357/ae0198},
url = {https://doi.org/10.3847/1538-4357/ae0198},
year = {2025},
month = {oct},
publisher = {The American Astronomical Society},
volume = {993},
number = {2},
pages = {161},
author = {Zhao, Tianqi and Rau, Peter B. and Haber, Alexander and Harris, Steven P. and Constantinou, Constantinos and Han, Sophia},
title = {Suppression of Composition g-modes in Chemically Equilibrating Warm Neutron Stars},
journal = {The Astrophysical Journal},
}

@article{Wei_2020,
doi = {10.3847/1538-4357/abbe02},
url = {https://doi.org/10.3847/1538-4357/abbe02},
year = {2020},
month = {dec},
publisher = {The American Astronomical Society},
volume = {904},
number = {2},
pages = {187},
author = {Wei, Wei and Salinas, Marc and Klähn, Thomas and Jaikumar, Prashanth and Barry, Megan},
title = {Lifting the Veil on Quark Matter in Compact Stars with Core g-mode Oscillations},
journal = {The Astrophysical Journal},
}

@article{Cheunchitra2024,
    author = {Cheunchitra, T and Melatos, A and Carlin, J B and Howitt, G},
    title = {Persistent gravitational radiation from glitching pulsars – II. Updated scaling with vortex number},
    journal = {Monthly Notices of the Royal Astronomical Society},
    volume = {528},
    number = {2},
    pages = {1360-1371},
    year = {2024},
    month = {01},
    issn = {0035-8711},
    doi = {10.1093/mnras/stae130},
    url = {https://doi.org/10.1093/mnras/stae130}
}

@article{Melatos2015,
doi = {10.1088/0004-637X/807/2/132},
url = {https://doi.org/10.1088/0004-637X/807/2/132},
year = {2015},
month = {jul},
publisher = {The American Astronomical Society},
volume = {807},
number = {2},
pages = {132},
author = {Melatos, A. and Douglass, J. A. and Simula, T. P.},
title = {PERSISTENT GRAVITATIONAL RADIATION FROM GLITCHING PULSARS},
journal = {The Astrophysical Journal},
}

@article{besedina2023,
  title = {Microseisms as a Tool for Geophysical Research. A Review},
  author = {A. N. Besedina and Ts. A. Tubanov},
  journal = {Journal of Volcanology and Seismology},
  year = {2023},
  volume = {17},
  number = {2},
  pages = {83--101},
  isbn = {1819-7108},
  doi = {10.1134/S0742046323700112},
  url = {https://doi.org/10.1134/S0742046323700112}
}

@article{Kulkarni2009,
    author = {Kulkarni, A. K. and Romanova, M. M.},
    title = {Possible quasi-periodic oscillations from unstable accretion: 3D magnetohydrodynamic simulations},
    journal = {Monthly Notices of the Royal Astronomical Society},
    volume = {398},
    number = {2},
    pages = {701-714},
    year = {2009},
    month = {09},
    issn = {0035-8711},
    doi = {10.1111/j.1365-2966.2009.15186.x},
    url = {https://doi.org/10.1111/j.1365-2966.2009.15186.x}
}

@Article{Romanova2015,
author={Romanova, Marina M.
and Owocki, Stanley P.},
title={Accretion, Outflows, and Winds of Magnetized Stars},
journal={Space Science Reviews},
year={2015},
month={Oct},
day={01},
volume={191},
number={1},
pages={339-389},
issn={1572-9672},
doi={10.1007/s11214-015-0200-9},
url={https://doi.org/10.1007/s11214-015-0200-9}
}

@ARTICLE{vanderKlis1984,
       author = {{van der Klis}, M. and {Bonnet-Bidaud}, J.~M.},
        title = "{The orbital parameters and the X-ray pulsation of VELA X-1 (4U 0900-40).}",
      journal = {Astronomy \& Astrophysics},
         year = 1984,
        month = jun,
       volume = {135},
        pages = {155-170},
       adsurl = {https://ui.adsabs.harvard.edu/abs/1984A&A...135..155V}
}

@article{Lange2001,
    author = {Lange, Ch. and Camilo, F. and Wex, N. and Kramer, M. and Backer, D.C. and Lyne, A.G. and Doroshenko, O.},
    title = {Precision timing measurements of PSR J1012+5307},
    journal = {Monthly Notices of the Royal Astronomical Society},
    volume = {326},
    number = {1},
    pages = {274-282},
    year = {2001},
    month = {09},
    issn = {0035-8711},
    doi = {10.1046/j.1365-8711.2001.04606.x},
    url = {https://doi.org/10.1046/j.1365-8711.2001.04606.x},
    eprint = {https://academic.oup.com/mnras/article-pdf/326/1/274/3408096/326-1-274.pdf},
}

\end{document}